\documentclass[journal]{IEEEtran}
\ifCLASSINFOpdf
\else
\fi

\usepackage{amsmath,graphicx,mathtools}
\usepackage{amsfonts,amssymb}
\usepackage{siunitx}
\usepackage{hyperref}
\usepackage{subcaption}
\usepackage{psfrag}
\usepackage{multirow}
\usepackage{tikz,pgfplots}
\usetikzlibrary{positioning}

\pgfplotsset{
    compat=newest,
    /pgfplots/legend image code/.code={%
        \draw[mark repeat=2,mark phase=2,#1] 
            plot coordinates {
                (0cm,0cm) 
                (0.1cm,0cm)
                (0.1cm,0cm)
                (0.1cm,0cm)
                (0.2cm,0cm)%
            };
    },
}

\usepackage{subfiles}
\begin{document}
%
\title{Learning Wavefront Coding for Extended Depth of Field Imaging}
%
%
%

\author{Ugur~Akpinar,
        Erdem~Sahin, 
        Monjurul~Meem,
        Rajesh~Menon,
        and~Atanas~Gotchev\thanks{U. Akpinar, E. Sahin, and A. Gotchev are with the Faculty of Information Technology and Communication Sciences, Tampere University, 33720 Tampere, Finland (e-mail: \href{mailto:ugur.akpinar@tuni.fi}{ugur.akpinar@tuni.fi})}\thanks{M. Meem and R. Menon are with the Department of Electrical and Computer Engineering, University of Utah, Salt Lake City, Utah 84102, USA}
        }
\maketitle

\begin{abstract}
Depth of field is an important factor of imaging systems that highly affects the quality of the acquired spatial information. Extended depth of field (EDoF) imaging is a challenging ill-posed problem and has been extensively addressed in the literature. We propose a computational imaging approach for EDoF, where we employ wavefront coding via a diffractive optical element (DOE) and we achieve deblurring through a convolutional neural network. Thanks to the end-to-end differentiable modeling of optical image formation and computational post-processing, we jointly optimize the optical design, i.e., DOE, and the deblurring through standard gradient descent methods. Based on the properties of the underlying refractive lens and the desired EDoF range, we provide an analytical expression for the search space of the DOE, which is instrumental in the convergence of the end-to-end network. We achieve superior EDoF imaging performance compared to the state of the art, where we demonstrate results with minimal artifacts in various scenarios, including deep 3D scenes and broadband imaging.
\end{abstract}


%
\IEEEpeerreviewmaketitle

\section{Introduction}


\IEEEPARstart{A}{} conventional imaging system can produce sharp images only for objects that are within the so-called depth of field (DoF). The DoF resides around the focused depth of the scene, which is determined by the effective focal length of the main lens and the lens-to-sensor distance, and its extent is dictated by the aperture size (more specifically, numerical aperture), i.e., the smaller the aperture, the larger the DoF. Although for some applications smaller DoFs may be preferred, it is mostly desirable to have as large DoF as possible to be able to acquire sharp images of deep scenes. This, however, cannot be achieved arbitrarily without sacrificing some other image quality factors. For instance, in microscopy, high numerical aperture objectives are especially critical to acquire fine details around the focused object depth. Since such objectives result in very shallow DoF, the microscopic imaging systems usually have to do depth scanning \cite{EDOFMicroscopy} to be able to cover the whole depth range of interest, which is typically much larger than the DoF. In photography, on the other hand, the imaging resolution at the focused depth is not usually a concern, since the features of interest are mostly much coarser than the diffraction-limited resolution. Thus, there is some flexibility of applying the naive approach of decreasing the aperture size to extend the DoF. By doing this, the intensity of the light reaching the sensor is reduced, which usually results in intolerable degradation in the image signal-to-noise ratio (SNR).

An extended DoF (EDoF) imaging approach should be able to provide larger DoF compared to such conventional imaging scenarios without significantly sacrificing the spatial resolution at the focused object depth. In general, one can express the forward imaging process mathematically as the convolution of the ideal all-in-focus (pinhole) image with the depth-dependent point spread function (PSF) of the imaging system. The optical defocus blur in conventional imaging systems with clear aperture is low-pass in nature. Therefore, the problem of defocus deblurring can be rarely solved via a single-image computational-only method, though several methods exist that rely on conventional clear aperture images \cite{DeblurringEM,DeblurringVariational,BlindDefDeblur,DeblurringSingleImage,GraphBlindDec}. This has led to the development of computational imaging approaches, for at least a few decades now, which essentially aim at engineering the PSF to make it more suitable for EDoF imaging \cite{LevinCA,ZhouCA,ZhouCAPairs,AnnularCA,DappledCA,MasiaCA,Cubic,tangent,sinusoidal,logarithmic,Ref-Diff,SpectralSweepEDoF,EDoFMobilePhone}. Such approaches can be mainly organized into two categories: coded-aperture and wavefront coding. 

In the coded-aperture imaging, the image is modified by means of a PSF coded with an optimized amplitude-mask at the aperture position \cite{LevinCA,ZhouCA,ZhouCAPairs,AnnularCA,DappledCA,MasiaCA}. Such mask basically modulates the intensity of the incident light, e.g., for a binary coded aperture the light is either transmitted or blocked at a given aperture position. Under the ray optics formalism (which is widely adopted in coded-aperture approaches), the depth-dependency of PSF can be described through the simple scaling operation, where the scale of the aperture is determined by the depth of the object point. Having estimated the depth information based on such (depth-dependent) defocus blur, i.e. defocus map, the problem of defocus deblurring is usually solved locally by applying (non-blind) deconvolution. Thus, such approaches need to find an optimal coded aperture that enables solving the monocular depth estimation and deblurring at the same time (blind deconvolution), which is a challenging problem due to contradictory requirements. An effective way for depth estimation, for instance, is to discriminate the depths based on the shifting behaviour of the zero-crossings in the modulation transfer functions (MTFs), which are the Fourier transforms of the corresponding PSFs \cite{LevinCA}. On the other hand, for efficient deblurring, the pass-bands of the MTFs should be as broad as possible to recover features at different spatial frequencies. Furthermore, depending on the amount of attenuation in the light intensity, a coded-aperture camera also suffers from degradations in the SNR.

The wavefront coding \cite{Cubic,tangent,sinusoidal,logarithmic,Ref-Diff,SpectralSweepEDoF,EDoFMobilePhone}, on the other hand, employs a (transparent) phase element at the aperture position, such as a diffractive optical element (DOE) or a refractive free-form lens. The main objective of the wavefront coding is to achieve a depth-invariant PSF, while still preserving the information at all spatial frequencies. Having a depth-invariant PSF, the deblurring problem is simplified to non-blind deconvolution without an explicit depth estimation. Unlike the coded-aperture methods, the wavefront coding approach does not attenuate the intensity of the incident light. That is, ideally, the image SNR is not degraded. Despite such advantages, the EDoF imaging capabilities of above-mentioned conventional wavefront coding techniques are limited, i.e., they are not able to provide satisfactory image qualities in the desired large EDoF ranges. There are few reasons for this. First, to the best of our knowledge, none of those approaches systematically analyse or derive the search space for the phase code that would serve well for the given EDoF setting. This is a critical aspect to address for the efficient convergence of the underlying optimization problem, which otherwise would converge to sub-optimal solutions. Second, all such approaches try to mainly optimize the optics, i.e., the phase code, based on the optimization criteria on the PSF being depth-independent and wide-band. The deconvolution then follows, just to digitally decode the optically-coded information, with a known code, that is assumed to be depth-invariant. However, it is usually not possible to achieve such a perfectly depth-invariant code. Furthermore, the employed phase element might cause dispersion of the chromatic components. This make the problem of achieving depth-independent PSF for all channels and for the whole continuous spectrum of interest more challenging, even if the channels are rigorously modeled and incorporated into the formulation of the optimization problem. 

Besides the above-mentioned categories of EDoF imaging, it is also worth mentioning the group of methods relying on light field \cite{DefPlenoptic,FocusedPlenoptic,EDOFPlenoptic,Latticefocal,LFDiffuser}. Plenoptic cameras \cite{DefPlenoptic,FocusedPlenoptic,EDOFPlenoptic} have originally aimed at acquiring the multi-dimensional light field information through spatio-angular multiplexing, which is implemented via a microlens array placed just in front of the sensor. Although plenoptic cameras can deliver significant EDoF imaging capabilities \cite{DefPlenoptic}, this extension in DoF comes at the expense of (significant) decrease in the spatial resolution due to the inherent spatio-angular trade-off in lens array based light-field imaging. Additional designs are focused on optimizing the imaging model for EDoF imaging, rather than capturing the light field \cite{Latticefocal,LFDiffuser}. The advantage of such methods is that in frequency domain, only a 3D subset of the 4D light field contributes to focus due to the dimensionality gap \cite{Latticefocal,FourierSlice}. The key concept of \cite{Latticefocal} is to maximize the EDoF imaging by concentrating the limited energy onto such subset. The PSF of their setup is still depth dependent, which requires a coarse depth estimation. In \cite{LFDiffuser}, an approximately depth-invariant PSF is achieved via radially symmetric diffuser encoding, which is analyzed in the light field space. 


Unlike the above-mentioned conventional coded-aperture and wavefront coding approaches that put the main emphasize on the optics, in this work, we solve the computational EDoF imaging problem via an end-to-end optimization framework, where we model the optics (phase-coded camera) and the post-processing (deblurring) algorithm as an end-to-end differentiable neural network. In particular, our computational EDoF camera employs the hybrid combination of a refractive lens and a DOE at the aperture position, where the DOE serves for the wavefront coding, and a convolutional neural network (CNN) as the deblurring algorithm. By training such end-to-end network we jointly optimize the DOE and the deblurring-CNN based on the characteristics of natural images. Regarding the optics, the hybrid refractive-diffractive optics is particularly useful for our approach. First, the refractive lens, which serves as the main lens, significantly relaxes the sampling requirements of the DOE compared to DOE-only optics, which consequently contributes to the fast convergence of the end-to-end network. Second, the opposite color dispersion characteristics of the refractive lens and DOE enables achieving broadband imaging with minimal dispersion. The above-mentioned factors together result in superior EDoF performance with respect to the state-of-the-art. 

Preliminary results of the proposed method have been presented in \cite{EDOFAkpinar}. In this article, we extend that work by providing a more detailed discussion and analysis of the algorithm, as well as demonstrating the validity and robustness of the approach through rigorous simulation results, e.g., covering broadband imaging and 3D scene scenarios. Furthermore, we demonstrate the EDoF imaging capabilities of our approach in the case of real-world imagery. In particular, we have fabricated the optimized DOE by using gray-scale lithography and employ it in a benchtop optical setup, used in the real-world imaging tests. The organization of the paper is as follows. In Sec.~\ref{sec:RelatedWork}, we review the latest end-to-end optimization methods, focusing on EDoF imaging approaches. In Sec.~\ref{sec:Problem}, the problem is discussed through wave optics based image formation and MTF analysis. In Sec.~\ref{sec:Method}, the proposed method is described in detail. In Sec.~\ref{sec:Simulations}, we summarize the simulation results. In Sec.~\ref{sec:Experiments}, the real-world experiments are presented and discussed. Finally, we conclude the work in Sec.~\ref{sec:Conclusion}. 

 

\section{Related Work}
\label{sec:RelatedWork}

The approach of end-to-end imaging based on supervised learning have been recently applied not only for EDoF imaging \cite{EDOFSitzmann,EDOFElmalem} but also for single image depth estimation, \cite{DepthEstPhase,CADepth}, light field imaging \cite{LFCALearning}, spectral imaging \cite{HSCNNJoint}, image classification \cite{OpticalCNN}, etc., resulting in notable improvements in each of these problems. The widely demonstrated success of CNNs in such inverse problems as well as the approach of co-design of the optics and the post processing set the basis of all these works.

In particular, the recently proposed EDoF methods \cite{EDOFSitzmann} and \cite{EDOFElmalem} are closely related to our co-design of the optics and the post-processing deblurring algorithm. The method in \cite{EDOFSitzmann} utilizes only a single phase element, either a DOE or a free form refractive lens placed at the aperture position. The post-processing deblurring is implemented via Wiener deconvolution, which is likely sub-optimal given the deviations from the perfectly depth-invariant PSF. The search spaces for the phase elements are chosen based on the Fourier basis or Zernike polynomials. However, there is no systematic analysis or validation of the employed optimization search space in connection with the device limitations, such as the range of EDoF, provided. On the other hand, the phase mask optimized by \cite{EDOFElmalem} consists of a DOE, which is assumed at the aperture position of the color aberration-corrected (compound) refractive lens system. The post-processing deblurring algorithm is based on a CNN. The search space of the DOE is chosen to be a single-ring binary pattern with two concentric regions (introducing either no phase shift or a uniform phase shift). Although multiple-ring versions are also tested, such choices are likely to generate sub-optimal mask patterns (indeed, we demontrate this in Sec. \ref{sec:Simulations}), as, again, there is no validation provided demonstrating that the optimal solution lies in such a search space. 

\section{Problem Formulation}
\label{sec:Problem}

\begin{figure}[htbp]
    \centering
    \includegraphics[width=0.9\columnwidth]{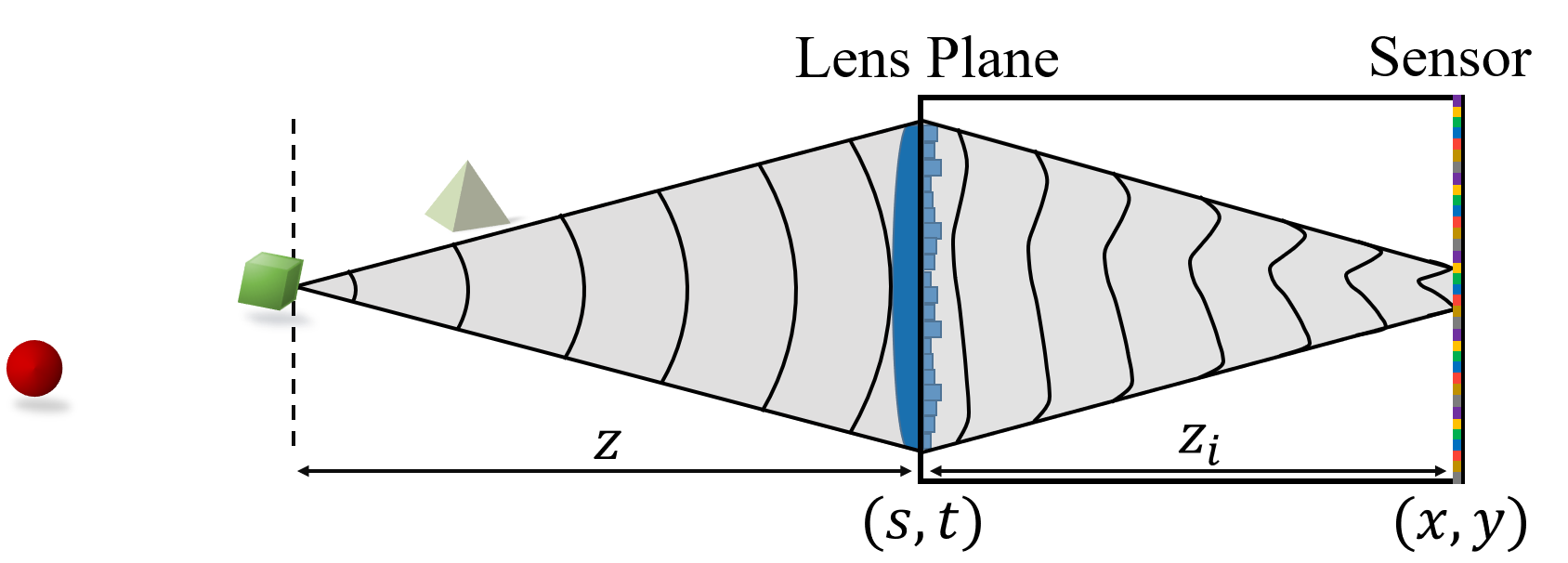}
    \caption{The optical setup of the imaging model, which consists of a refractive lens and a complex-valued coded mask.}
    \label{fig:CameraModel}
\end{figure}

In this section, we describe the problem through an image formation model derived based on wave optics. Fig.~\ref{fig:CameraModel} demonstrates the optical design of our camera. The sensor is placed at a distance $z_i$ from the lens plane. The spatial coordinates at the lens and the sensor planes are denoted as $(s,t)$ and $(x,y)$, respectively. Let us start with a scene illuminated by a monochromatic and spatially incoherent light source with wavelength $\lambda$, and each object point is defined by its intensity and distance from the lens plane. Ignoring the monocular parallax within the entrance pupil, such scene can be described via an image and depth map pair. The PSF of a point in 3D space, $h_{z,\lambda}(x,y)$, is both depth and wavelength dependent. The sensor image with respect to the points at $z$ is then the convolution between the ideal (pinhole) image $I_{z,\lambda}(x,y)$ and the PSF $h_{z,\lambda}(x,y)$,
\begin{equation}
    I^s_{z,\lambda}(x,y) = I_{z,\lambda}(x,y) \ast h_{z,\lambda}(x,y),
    \label{eq:Forwardmodel}
\end{equation}
where $I^s_{z,\lambda}$ is the sensor image and $\ast$ is the convolution operator. The final image recorded at the sensor, $I^s_\lambda$, is the integration of $I^s_z(x,y)$ over all possible depth values in the scene, accounting also for the sensor noise, 
\begin{equation}
    I^s_\lambda(x,y) = \int I^s_{z,\lambda}(x,y) dz + \eta_s,
    \label{eq:SensorImage}
\end{equation}
where $\eta_s \sim \mathcal{N}(0,\sigma_s^2)$ is zero-mean Gaussian noise with standard deviation $\sigma_s$.

The recovery of the sharp image depends on the characteristics of the PSF $h_{z,\lambda}(x,y)$. In particular, the reconstruction quality is dictated by the frequency support of the PSF. If, for instance, the corresponding MTF is a low-pass function, the problem becomes ill-posed with infinitely many solutions for the high frequency components of the original image. In the following, the mathematical model of the PSF is derived using wave optics, which is the basis for the design principles, as well as the proposed optimization scheme.

Consider an imaging system including a refractive lens and a complex coded mask as in Fig.~\ref{fig:CameraModel}, where the amplitude modulation and the phase shift by the mask are denoted by $A(s,t)$ and $\Phi_\lambda(s,t)$, respectively. For a transparent (refractive or diffractive) optical element, the wavelength-dependent phase shift $\Phi_\lambda(s,t)$ is manipulated by the thickness function of the element, $d(s,t)$ as 
\begin{equation}
    \Phi_\lambda(s,t) = k (n_\lambda-1) d(s,t), 
    \label{eq:dtophi}
\end{equation}
where $k=2\pi/\lambda$ is the wave number and $n_\lambda$ is the wavelength-dependent refractive index of the material. Within the paraxial optics limits and assuming thin lens model, the monochromatic incoherent PSF of such system is expressed as \cite{Goodman}
\begin{equation}
h_{\lambda,z}(x,y) \propto \biggl|\mathcal{F}\{Q_{\lambda,z}(s,t)\}|_{\bigl(\frac{x}{\lambda z_i},\frac{y}{\lambda z_i}\bigr)}\biggr|^2,
\label{eq:PSF}
\end{equation}
where $\mathcal{F}\{.\}$ is the Fourier transform operator, 
\begin{equation}
\begin{multlined}
Q_{\lambda,z}(s,t) = A(s,t)\exp(j\Phi_\lambda(s,t))\exp\biggl[j\Psi_{\lambda,z}\biggl(\frac{s^2 + t^2}{r^2}\biggr)\biggr]
\end{multlined}
\label{eq:Q}
\end{equation}
is the so-called generalized pupil function,
\begin{equation}
\Psi_{\lambda,z} = \frac{\pi}{\lambda}\biggl(\frac{1}{z}+\frac{1}{z_i}-\frac{1}{f_\lambda}\biggr)r^2
\label{eq:defocus}
\end{equation} 
is the defocus coefficient, $f_\lambda$ is the wavelength-dependent focal length, and $r$ is the radius of the pupil. Please refer to the Appendix for detailed derivation. 

\begin{figure}[htbp]
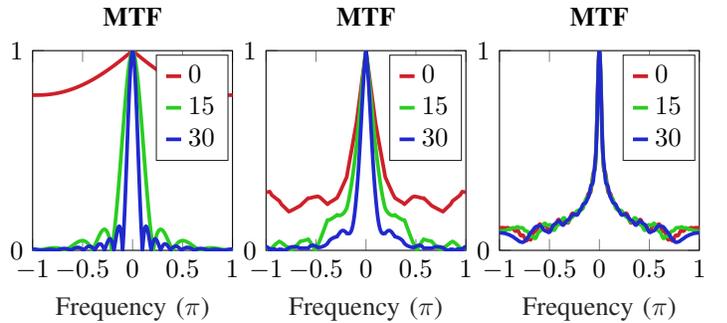
	
	\hspace{-4.5mm}
	\begin{subfigure}[t]{0.3\columnwidth}
	    \subfile{figures/MTF/MTF_OpenAperture}
	\end{subfigure}
	\hspace{2mm}
	\begin{subfigure}[t]{0.3\columnwidth}
		\subfile{figures/MTF/MTF_Levin}	
	\end{subfigure}
	\hspace{2mm}
	\begin{subfigure}[t]{0.3\columnwidth}
		\centering	
	    \subfile{figures/MTF/MTF_Cubic}
	    \label{sfig:MTFCubic}
	\end{subfigure}
	\caption{MTFs of three different imaging systems; conventional (left), coded-aperture \cite{LevinCA} (middle), and wavefront coding \cite{Cubic} (right), for three different values of the defocus coefficient $\Psi$.} 
	\label{fig:MTFDifApertures}
\end{figure}

Fig.~\ref{fig:MTFDifApertures} illustrates the MTFs for three different defocus values, $\Psi_{\lambda,z}$, where $\lambda = 543 nm$. The conventional imaging case with clear aperture (left) has two major problems. First, as the defocus scale is increased, the MTF quickly decreases to zero, which makes the problem severely ill-posed. Second, the MTF is strongly dependent on the defocus; therefore, the deconvolution with correct kernel requires a priori knowledge or estimation of the defocus, equivalently the object depth. 

The coded-aperture imaging (Fig.~\ref{fig:MTFDifApertures}, middle) can be thought as the modification of the lens amplitude function $A(s,t)$ of Eq.~\ref{eq:Q}. The above-mentioned issues related to the conventional cameras are addressed in the coded-aperture design. In \cite{LevinCA}, for instance, the change in the MTF due to the defocus is optimized such that at each defocus scale, the locations of the zeros in the MTFs are distinguishable. Such method eases the estimation of the depth, from which a non-blind deconvolution can be applied with estimated blur kernels. However, such coded-aperture camera fails to image details of spatial frequencies at or around the zero-crossings, where the image noise is also inevitably amplified during deblurring. Several other methods exist to overcome this and offer full-band (i.e., at all spatial frequencies) imaging via coded-aperture, which are reviewed in \cite{ZhouCA} in detail. However, in their comparisons, they assume that the depth is known \textit{a priori}, which is a strong assumption since alleviating the zero-crossings makes it harder to identify the defocus level with single camera.

The wavefront coding \cite{Cubic} (Fig.~\ref{fig:MTFDifApertures}, right) aims at altering the phase component $\Phi_\lambda(s,t)$ of the generalized pupil of Eq.~\ref{eq:Q}. As it can be seen in the figure, such approach provides few critical advantages in the EDoF problem compared to the coded-apertures and conventional cameras. First, the frequency response is wider in high defocus levels, which decreases the attenuation of the high frequency components making the problem less ill-posed. Second, the MTF is relatively independent with respect to the defocus level. The defocus deblurring can then be approximated as a non-blind deconvolution via an average blur kernel, which does not require any depth information. These factors form the basis of any wavefront-coding approaches. Our method also relies on them in achieving successful EDoF performance. However, unlike conventional wavefront-coding methods, our approach is able to achieve EDoF imaging in significantly larger depth ranges thanks to the co-design of optimal optics and subsequent deblurring. In the following section, the proposed method is described in details.

\section{Proposed Method}
\label{sec:Method}

\begin{figure}[htbp]
    \centering
    \includegraphics[clip, trim=180 125 80 160, width=0.8\columnwidth]{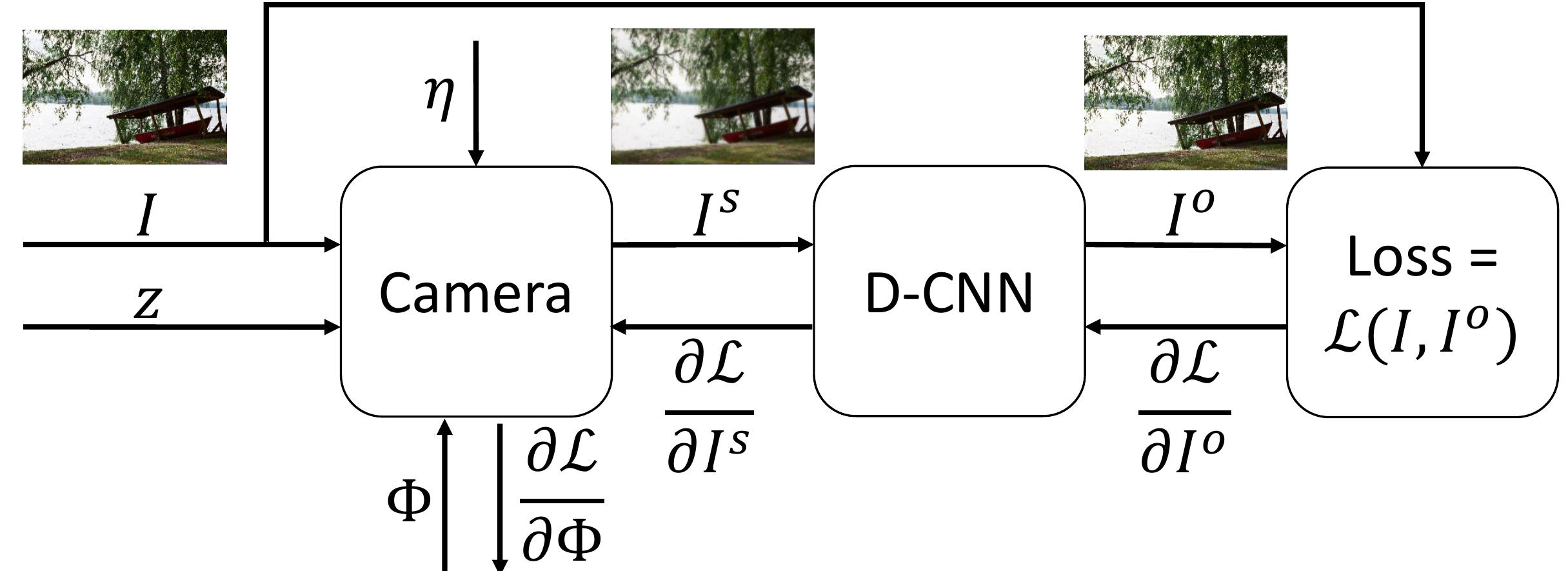}
    \caption{Overall representation of proposed setup.}
    \label{fig:Overall}
\end{figure}

Fig.~\ref{fig:Overall} illustrates the proposed end-to-end learning scheme. The inputs to the network are the patches of sharp natural images, and a single depth value for each image patch. The network is divided into two main parts, namely the computational camera model and the D-CNN. The computational camera model takes the input image and the depth values to simulate the sensor image based on the forward model described in Sec.~\ref{sec:Problem}. The post-processing network takes the resulting sensor image as the input, and estimates the sharp image as the output. The depths of the images are not given as input to the deblurring network; the post-processing algorithm is expected to obtain the sharp images without knowing the depth explicitly. In the following, more detailed description of the camera and D-CNN models are given.

\subsection{Computational Camera Model}

\begin{figure}[htbp]
    \centering
    \includegraphics[clip, trim=190 110 180 160, width=0.8\columnwidth]{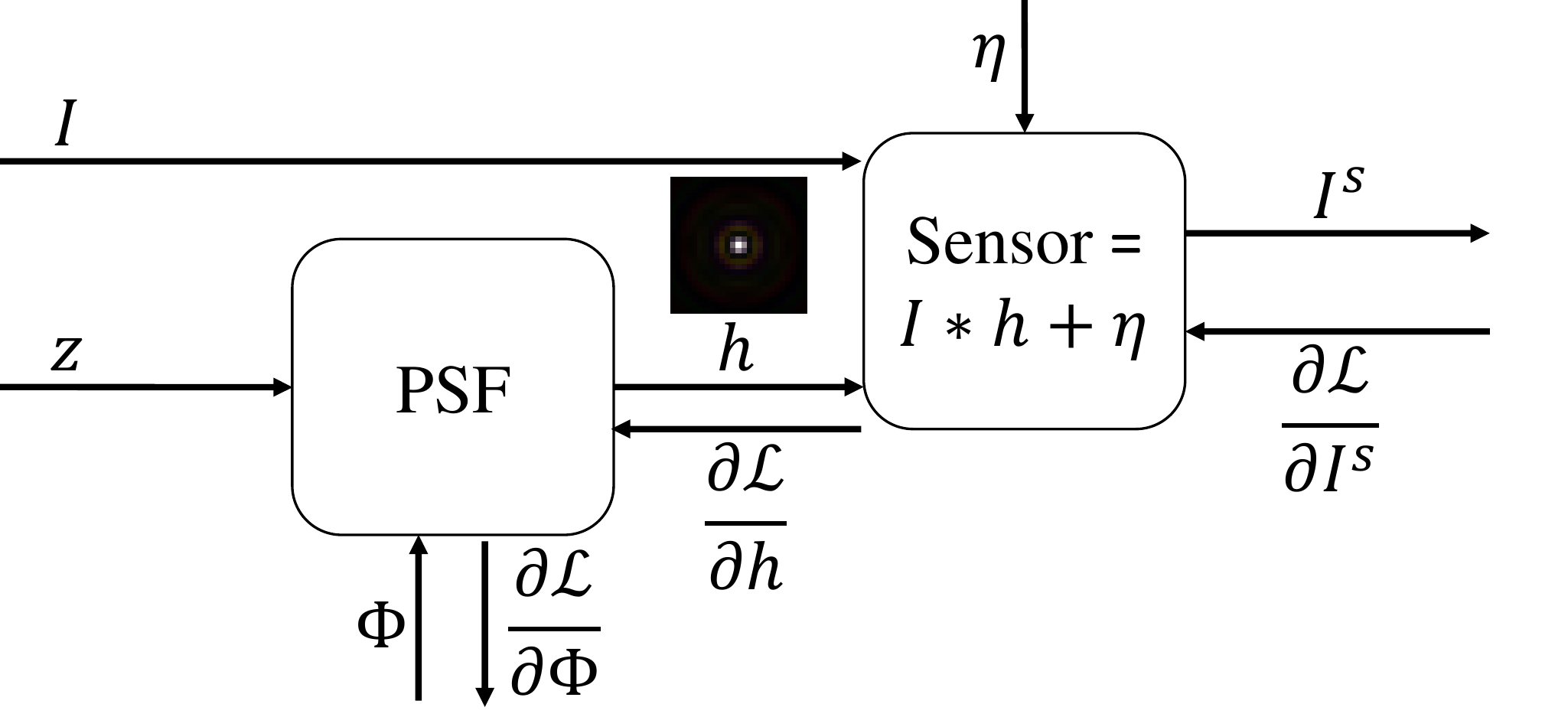}
    \caption{Computational camera model.}
    \label{fig:Camera_Net}
\end{figure}

The computational camera model takes the sharp image $I$ and a constant depth $z$ as inputs, and outputs the sensor image $I^s$. The PSF layer in Fig.~\ref{fig:Camera_Net} outputs the depth-dependent PSFs for three color channels based on wave-optics derivations. We utilize Eq.~\ref{eq:PSF} to calculate the PSF for each iteration. The amplitude $A(s,t)$ is kept fixed as the circular function within the aperture diameter. The camera parameters such as the phase function of the lens, the sensor-to-lens distance $z_i$, and the mask sampling grid $(s,t)$ are defined in advance and given as the layer properties. The phase $\Phi_\lambda(s,t)$ is defined as the parameter to be optimized (see Appendix for derivation of the derivative of error with respect to $\Phi_\lambda(s,t)$ that is used in the backward pass). 

Eq.~\ref{eq:dtophi} dictates a physical constraint between the thickness function and the phase delay of the DOE, which relates the phase delays at different wavelengths. For a nominal wavelength $\lambda_0$, such relation can be formulated as
\begin{equation}
    \Phi_\lambda(s,t) = \Phi_{\lambda_0}(s,t) \frac{\lambda_0(n_\lambda-1)}{\lambda(n_{\lambda_0}-1)}. 
    \label{eq:phi0tophi}
\end{equation}
During training, we optimize $\Phi_{\lambda_0}$, from which $\Phi_\lambda$ are obtained using Eq.~\ref{eq:phi0tophi} in the forward pass. Taking into account fabrication errors in the height profile of the DOE (correspondingly its phase), we also add random Gaussian noise to $\Phi_{\lambda_0}$ in each iteration to increase the robustness of the algorithm to such deviations in the phase. The standard deviation of the phase noise is derived via the production error of the DOE. In other words, the zero-mean production error is defined with the standard deviation $\sigma_d$ as $\eta_d \sim \mathcal{N}(0,\sigma_d^2)$, from which the phase error is calculated via Eq.~\ref{eq:dtophi}.

\subsubsection{Optimum Signal Space Analysis}

The object depth $z$ is chosen randomly from the uniform distribution $U(z^-,z^+)$, where $z^-$ and $z^+$ are the scene depth limits. In this section, we derive the optimum sampling strategy of the phase mask given the scene-depth limits, equivalently the defocus range. The sampling rate significantly affects the number of parameters to optimize within the mask, thus the training speed and convergence. On the other hand, since the PSF is the magnitude square of the Fourier transform of the defocused pupil function as in Eq.~\ref{eq:PSF}, the minimum sampling rate of the mask pattern should be derived in order to avoid aliasing due to under-sampling of the defocused pupil function, which would otherwise result in miscalculated PSF.

\begin{figure}[h!]
\hspace{-4mm}
\begin{subfigure}{0.48\columnwidth}
%
%
\definecolor{mycolor1}{rgb}{0.00000,0.44700,0.74100}%
\begin{tikzpicture}

\begin{axis}[%
width=0.7\columnwidth,
height=0.5\columnwidth,
at={(0\columnwidth,0\columnwidth)},
scale only axis,
unbounded coords=jump,
xmin=0,
xmax=100,
xlabel style={font=\color{white!15!black}},
xlabel={$\Psi$},
ymin=0,
ymax=10,
ylabel style={font=\color{white!15!black}},
ylabel={$B_D/2w_z$},
axis background/.style={fill=white},
xmajorgrids,
ymajorgrids,
title style={font=\bfseries},
title={Bandwidth},
legend style={legend cell align=left, align=left, draw=white!15!black}
]
\addplot [color=mycolor1, line width=1.2pt]
  table[row sep=crcr]{%
0	inf\\
1	90.8745797915231\\
2	45.4432042102621\\
3	30.3020374827072\\
4	22.7334190469871\\
5	18.1938155132031\\
6	15.1687146783091\\
7	13.0090389547296\\
8	11.3902487256586\\
9	10.1320434618224\\
10	9.12624216664929\\
11	8.31013974778369\\
12	7.64627326993347\\
13	7.08311001665928\\
14	6.58600446801525\\
15	6.15425074690734\\
16	5.77691024345982\\
17	5.44437499088065\\
18	5.14917171946406\\
19	4.88540008244822\\
20	4.65980980173306\\
21	4.46137853449382\\
22	4.26589066946318\\
23	4.08744905815304\\
24	3.92413427169116\\
25	3.77412531676339\\
26	3.64466434315671\\
27	3.52961221238432\\
28	3.4103157280529\\
29	3.29943448050266\\
30	3.19612128893751\\
31	3.10554917759691\\
32	3.02764349910547\\
33	2.94235913523101\\
34	2.86222879735741\\
35	2.78680514215611\\
36	2.73070557750623\\
37	2.66446405965179\\
38	2.60048978360185\\
39	2.53989431041186\\
40	2.49532956858642\\
41	2.44232256451093\\
42	2.39004333961688\\
43	2.34026917730985\\
44	2.30510020176439\\
45	2.26084611724267\\
46	2.21729322245009\\
47	2.17580828531578\\
48	2.14833906671013\\
49	2.10988077267297\\
50	2.07300170864306\\
51	2.0397366567906\\
52	2.01525869856492\\
53	1.98234466349826\\
54	1.95067490881015\\
55	1.93071655974813\\
56	1.90167850773754\\
57	1.87314959748515\\
58	1.84726104116347\\
59	1.82941488899763\\
60	1.80355471649894\\
61	1.77854686892333\\
62	1.76427455872676\\
63	1.74069911341463\\
64	1.71785850164729\\
65	1.7009507488323\\
66	1.68360921239336\\
67	1.6626403713239\\
68	1.64272502573077\\
69	1.63148122492553\\
70	1.6121382874954\\
71	1.59332424002843\\
72	1.58286935647088\\
73	1.56572187283555\\
74	1.54826289235968\\
75	1.53220517459238\\
76	1.52285962089553\\
77	1.50659078310676\\
78	1.490712511832\\
79	1.48258398957902\\
80	1.46787985782095\\
81	1.45300654922859\\
82	1.43909836521047\\
83	1.43176216678767\\
84	1.41777835514039\\
85	1.40408921944706\\
86	1.39239827643564\\
87	1.38472387971245\\
88	1.37179251447694\\
89	1.35911316642424\\
90	1.3498675199411\\
91	1.34131854389671\\
92	1.329286673759\\
93	1.31747146276167\\
94	1.30787643640924\\
95	1.30098463825073\\
96	1.28972290885583\\
97	1.27864833310388\\
98	1.26872213411921\\
99	1.26324164697843\\
100	1.25263856867321\\
};

\end{axis}

\end{tikzpicture}%
\end{subfigure}
\hspace{0mm}
\begin{subfigure}{0.48\columnwidth}
    \subfile{figures/DefocusChirp/Def_Psi40}
\end{subfigure}
    \caption{The frequency spectrum of the defocus chirp function and its theoretical and practical bandwidths. Left: The observed bandwidth of the chirp function with respect to the theoretical bandwith, as a function of changing defocus values. Right: An example defocus chirp function with $\Psi=40$ and $r=\SI{2.5}{\mm}$}
    \label{fig:DefPupil}
\end{figure}
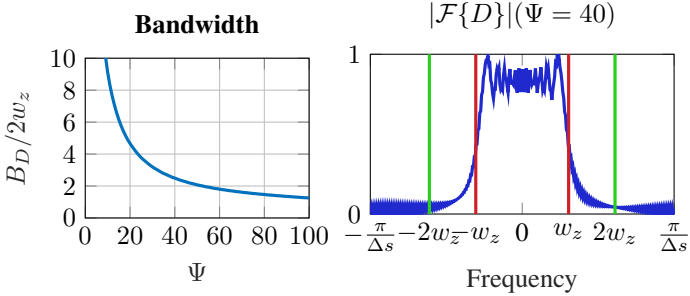

Let us separate Eq.~\ref{eq:Q} as the multiplication of two components, namely the phase delay by the DOE, $P(s,t) = \exp(j\Phi_\lambda(s,t))$, and the defocus, $D(s,t) = A(s,t)\exp(j\Psi_{\lambda,z}(s^2/r^2+t^2/r^2))$. The second term is a well-defined chirp function, of which the bandwidth can be approximated through the instantaneous frequency. In 1D case, the theoretical maximum frequency is given as
\begin{equation}
    w_{\lambda,z} = \frac{\partial (\Psi_{\lambda,z} s^2/r^2)}{\partial s}\bigg |_{s=r} = \frac{2 \Psi_{\lambda,z}}{r}.
    \label{eq:Bandwidth}
\end{equation}
In practice, however, the finite aperture size introduces oscillation and tails in the frequency spectrum, which increases the bandwidth. Fig.~\ref{fig:DefPupil} illustrates the ratio of the observed bandwidth of the finite chirp to the theoretical bandwidth of $2w_z$ with respect to the defocus, as well as an example chirp function in frequency domain where $\Psi_{\lambda,z} = 40$ and $r=\SI{2.5}{\mm}$. We observe from the figure that the ratio is inversely proportional to the defocus and assume that a bandwidth of $4w_z$ with bandlimits of $[-2w_z,2w_z]$ will be enough to account for the tails for higher defocus values. Furthermore, we assume that the bandlimit of the phase delay introduced by the DOE will be the same as the maximum target defocus bandlimit, as it will be enough to manipulate the PSF within the scene depth range. The defocused pupil function is then assumed to be bandlimited by $[-4w_z,4w_z]$. Using the Nyquist theorem, 
\begin{equation}
    \frac{\pi}{\Delta s} \geq 4|w_z|, 
    \label{eq:Nyquist}
\end{equation}
where $\Delta s$ is the mask sampling rate. Eq.~\ref{eq:Bandwidth} reveals that $w_z$ is linearly proportional to $\Psi_{\lambda,z}$. Therefore, the upper limit of $w_z$ can be derived via the maximum defocus value within the scene, $\Psi_{max} = \max\limits_\lambda(\max\{|\Psi_{\lambda,z^-}|,|\Psi_{\lambda,z^+}|\})$. Using Eq.~\ref{eq:Bandwidth} and~\ref{eq:Nyquist}, we conclude that the sampling rate for a given scene depth range should satisfy
\begin{equation}
    \Delta s \leq \frac{r\pi}{8\Psi_{max}}.
    \label{eq:MinSampling}
\end{equation}

\subsection{Deblurring CNN}

\begin{figure}
    \centering
    \includegraphics[width=0.99\columnwidth]{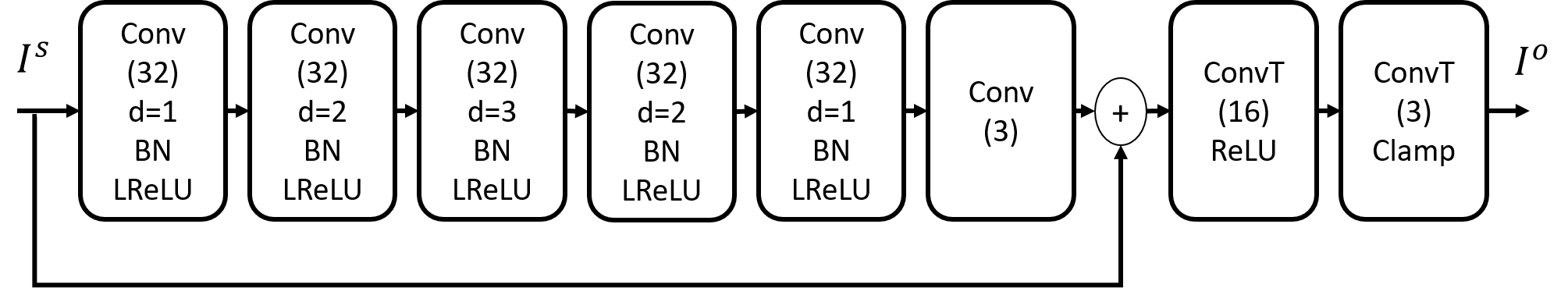}
    \caption{CNN-based deblurring network.}
    \label{fig:DCNN}
\end{figure}

Fig.~\ref{fig:DCNN} illustrates the proposed post-processing algorithm for image deblurring, which takes the RGB sensor output of the camera model, $I^s$, as input. Inspired by the recent works on image restoration \cite{EDOFElmalem,regularization,DebCNN,DebCNN2}, we propose a modified residual network for deblurring. The main advantage of such approach is to learn the residual image, i.e. the difference between the sharp image and the sensor output, instead of learning the original sharp image itself. For image deblurring, we expect to learn high frequency information, such as edges, through the residual network. Each of the first five blocks in the residual network consists of a convolution layer with 32 kernels of $3\times 3$ size, a batch normalization, and a leaky rectified linear unit (LReLU). The dilation in each convolution layer, denoted with d in the figure, is 1,2,3,2,1, respectively. A final convolution layer with 3 kernels of $3 \times 3$ size is used to make the residual output size equal to the original image size. The residual image is then added to the sensor output, and the resulting image is given to the transposed convolution layers, denoted as ConvT in the figure. A transposed convolution layer computes $y = C^\intercal x$, where C is the convolution matrix and $x$ and $y$ are the input and output vectors of the layer, respectively. We observe that the two transposed convolution layers after the residual network further enhance the image quality. Finally, the output of the transposed convolution layers is fed into the Clamp layer, which keeps the data in between 0 and 1, applying
\begin{equation}
    \operatorname{Clamp}(x) = \begin{cases}
    0, & \text{if $x<0$},\\
    x, & \text{if $0\leq x \leq 1$},\\
    1, & \text{if $x>1$}.
  \end{cases}
    \label{eq:Clip}
\end{equation}
We observe that it is crucial to introduce such layer to handle the dynamic range and channel normalization issues.

The loss function of the network is calculated accounting both for the data and the prior terms. Given a batch of $T$ ground-truth images $(I_1,..,I_T)$, and the network output $(I_1^o,..,I_T^o)$
\begin{equation}
    \mathcal{L}(I,I^o) = \frac{1}{T} \sum_{t=1}^T \mathcal{D}(I_t,I_t^o) + \alpha \mathcal{R}(I_t,I_t^o),
\label{eq:Loss}
\end{equation}
where $\mathcal{D}(I,I^o)$ is the data loss, $\mathcal{R}(I,I^o)$ is the regularization term, and $\alpha$ is the regularization weight. In the proposed method, we utilize L1-loss
\begin{equation}
    \mathcal{D}(I,I^o) = \sum_{\lambda} |I_\lambda - I_\lambda^o|_1,
\label{eq:L1}
\end{equation}
which is demonstrated to provide improvement compared to L2-loss for various image enhancement problems \cite{L1loss}. The sparse gradient prior \cite{LevinCA} is utilized for the regularization, which provides sharper details compared to Gaussian prior, while suppressing the artifacts. The regularization term is given as
\begin{equation}
    \mathcal{R}(I,I^o) = \sum_{\lambda,p,q} e^{(-\beta |\nabla_p I_\lambda|^\gamma)} |\nabla_p I_\lambda^o|^\gamma + e^{(-\beta |\nabla_q I_\lambda|^\gamma)} |\nabla_q I_\lambda^o|^\gamma,
\label{eq:Regularization}
\end{equation}
where $\nabla_p I = I_{p,q} - I_{p-1,q}, \nabla_q I = I_{p,q} - I_{p,q-1}$ are the vertical and horizontal derivatives, respectively. The exponential terms $e^{(-\beta |\nabla_p I_\lambda|^\gamma)}$ and $e^{(-\beta |\nabla_q I_\lambda|^\gamma)}$ are included to decrease the weight of the regularization term at the edges of the ground truth image, as suggested in \cite{regularization}. 

\section{Simulations}
\label{sec:Simulations}

The main optics of our EDoF camera is composed of a refractive lens that works in tandem with a wavefront coding DOE. The refractive lens is chosen to be an off-the-shelf plano-convex lens with the effective focal length of $f_s=\SI{36}{\mm}$ at the specification wavelength of $\lambda_s=\SI{587.6}{\nm}$ and aperture radius of $r=\SI{2.5}{\mm}$ (which is the same for the DOE). We model the red, green, and blue channels with three distinct wavelengths of \SI{611}{\nm}, \SI{543}{\nm}, and \SI{482}{\nm}, respectively. The material of the lens is silica with the refractive indices of 1.457, 1.460, and 1.463 at these distinct wavelengths, respectively. The focused distance at the specification wavelength is fixed as $z_{f_s}=\SI{1.50}{\meter}$ from the camera plane, in which case the sensor-to-lens distance is $z_i=\SI{36.9}{\mm}$. The resulting focus distances for each channel are $z_{f_R}=\SI{1.62}{\meter}, z_{f_G}=\SI{1.30}{\meter},$and $z_{f_B}=\SI{1.04}{\meter}$, respectively. However, please note that in practice the phase modification through the refractive lens is modeled via the thickness function using Eq.~\ref{eq:dtophi} and~\ref{eq:spherical}. The radius of curvature and the central thickness are taken from the lens data sheet as $R=\SI{16.51}{\mm}$ and $d_0=\SI{2}{\mm}$. The scene depth range is assumed to be $[\SI{0.5}{\meter}, \infty)$, which corresponds to the maximum defocus of $\Psi_{max} = 45$. From Eq.~\ref{eq:MinSampling}, taking the mask sampling rate as $\Delta s = \SI{21}{\um}$ will be enough to cover such depth range. The sensor resolution is taken to be $\Delta x = \SI{6}{\um}$.

The network is trained via an image data set provided by \cite{INRIADataset}, which contains high-resolution natural images. The training data is divided into patches of size 300$\times$300 pixels, where the batch size is set to 4. $10\%$ of the data is used for the validation. At each iteration, a random depth value from uniform distribution is assigned to each image patch within the scene depth limits, i.e. $z \sim U(0.5,\infty)$. To increase the robustness to different image noise levels, we choose the standard deviation of the Gaussian sensor noise from the uniform distribution $\sigma_s \sim U(0.001,0.015)$. The fabrication noise of the height map is also assumed to be Gaussian distributed, but with a fixed standard deviation of $\sigma_d=\SI{40}{\nm}$. The training is done in 73 epochs using Adam solver \cite{AdamOptimizer} as the optimization method, where the initial learning rate is 1e-3, first decay rate is 0.9, second decay rate is 0.999, and the weight decay is 1e-4.

\begin{figure}[htbp]
\hspace{-4.5mm}
\begin{subfigure}{0.40\columnwidth}
%
%
\begin{tikzpicture}

\begin{axis}[%
width=0.75\columnwidth,
height=0.75\columnwidth,
at={(0\columnwidth,0\columnwidth)},
scale only axis,
point meta min=0,
point meta max=1.13478324692551, 
axis on top,
xmin=-2.5095,
xmax=2.5095,
xtick={-2.5,  2.5},
xlabel style={font=\color{white!15!black}},
xlabel={$s (mm)$},
y dir=reverse,
ymin=-2.5095,
ymax=2.5095,
ytick={-2.5,  2.5},
axis background/.style={fill=white},
title style={font=\bfseries},
title={Height map},
legend style={legend cell align=left, align=left, draw=white!15!black},
colormap={mymap}{[1pt] rgb(0pt)=(0.2422,0.1504,0.6603); rgb(1pt)=(0.25039,0.164995,0.707614); rgb(2pt)=(0.257771,0.181781,0.751138); rgb(3pt)=(0.264729,0.197757,0.795214); rgb(4pt)=(0.270648,0.214676,0.836371); rgb(5pt)=(0.275114,0.234238,0.870986); rgb(6pt)=(0.2783,0.255871,0.899071); rgb(7pt)=(0.280333,0.278233,0.9221); rgb(8pt)=(0.281338,0.300595,0.941376); rgb(9pt)=(0.281014,0.322757,0.957886); rgb(10pt)=(0.279467,0.344671,0.971676); rgb(11pt)=(0.275971,0.366681,0.982905); rgb(12pt)=(0.269914,0.3892,0.9906); rgb(13pt)=(0.260243,0.412329,0.995157); rgb(14pt)=(0.244033,0.435833,0.998833); rgb(15pt)=(0.220643,0.460257,0.997286); rgb(16pt)=(0.196333,0.484719,0.989152); rgb(17pt)=(0.183405,0.507371,0.979795); rgb(18pt)=(0.178643,0.528857,0.968157); rgb(19pt)=(0.176438,0.549905,0.952019); rgb(20pt)=(0.168743,0.570262,0.935871); rgb(21pt)=(0.154,0.5902,0.9218); rgb(22pt)=(0.146029,0.609119,0.907857); rgb(23pt)=(0.138024,0.627629,0.89729); rgb(24pt)=(0.124814,0.645929,0.888343); rgb(25pt)=(0.111252,0.6635,0.876314); rgb(26pt)=(0.0952095,0.679829,0.859781); rgb(27pt)=(0.0688714,0.694771,0.839357); rgb(28pt)=(0.0296667,0.708167,0.816333); rgb(29pt)=(0.00357143,0.720267,0.7917); rgb(30pt)=(0.00665714,0.731214,0.766014); rgb(31pt)=(0.0433286,0.741095,0.73941); rgb(32pt)=(0.0963952,0.75,0.712038); rgb(33pt)=(0.140771,0.7584,0.684157); rgb(34pt)=(0.1717,0.766962,0.655443); rgb(35pt)=(0.193767,0.775767,0.6251); rgb(36pt)=(0.216086,0.7843,0.5923); rgb(37pt)=(0.246957,0.791795,0.556743); rgb(38pt)=(0.290614,0.79729,0.518829); rgb(39pt)=(0.340643,0.8008,0.478857); rgb(40pt)=(0.3909,0.802871,0.435448); rgb(41pt)=(0.445629,0.802419,0.390919); rgb(42pt)=(0.5044,0.7993,0.348); rgb(43pt)=(0.561562,0.794233,0.304481); rgb(44pt)=(0.617395,0.787619,0.261238); rgb(45pt)=(0.671986,0.779271,0.2227); rgb(46pt)=(0.7242,0.769843,0.191029); rgb(47pt)=(0.773833,0.759805,0.16461); rgb(48pt)=(0.820314,0.749814,0.153529); rgb(49pt)=(0.863433,0.7406,0.159633); rgb(50pt)=(0.903543,0.733029,0.177414); rgb(51pt)=(0.939257,0.728786,0.209957); rgb(52pt)=(0.972757,0.729771,0.239443); rgb(53pt)=(0.995648,0.743371,0.237148); rgb(54pt)=(0.996986,0.765857,0.219943); rgb(55pt)=(0.995205,0.789252,0.202762); rgb(56pt)=(0.9892,0.813567,0.188533); rgb(57pt)=(0.978629,0.838629,0.176557); rgb(58pt)=(0.967648,0.8639,0.16429); rgb(59pt)=(0.96101,0.889019,0.153676); rgb(60pt)=(0.959671,0.913457,0.142257); rgb(61pt)=(0.962795,0.937338,0.12651); rgb(62pt)=(0.969114,0.960629,0.106362); rgb(63pt)=(0.9769,0.9839,0.0805)},
colorbar,
colorbar style={
            at={(0.75\columnwidth,0.75\columnwidth)},
            anchor=north west,
            align=left,
            title style = {at = {(0.1\columnwidth,0.7\columnwidth)}},
            title={$\mu m$}},
colorbar/width=0.05\columnwidth
]
\addplot [forget plot] graphics [xmin=-2.5095, xmax=2.5095, ymin=-2.5095, ymax=2.5095] {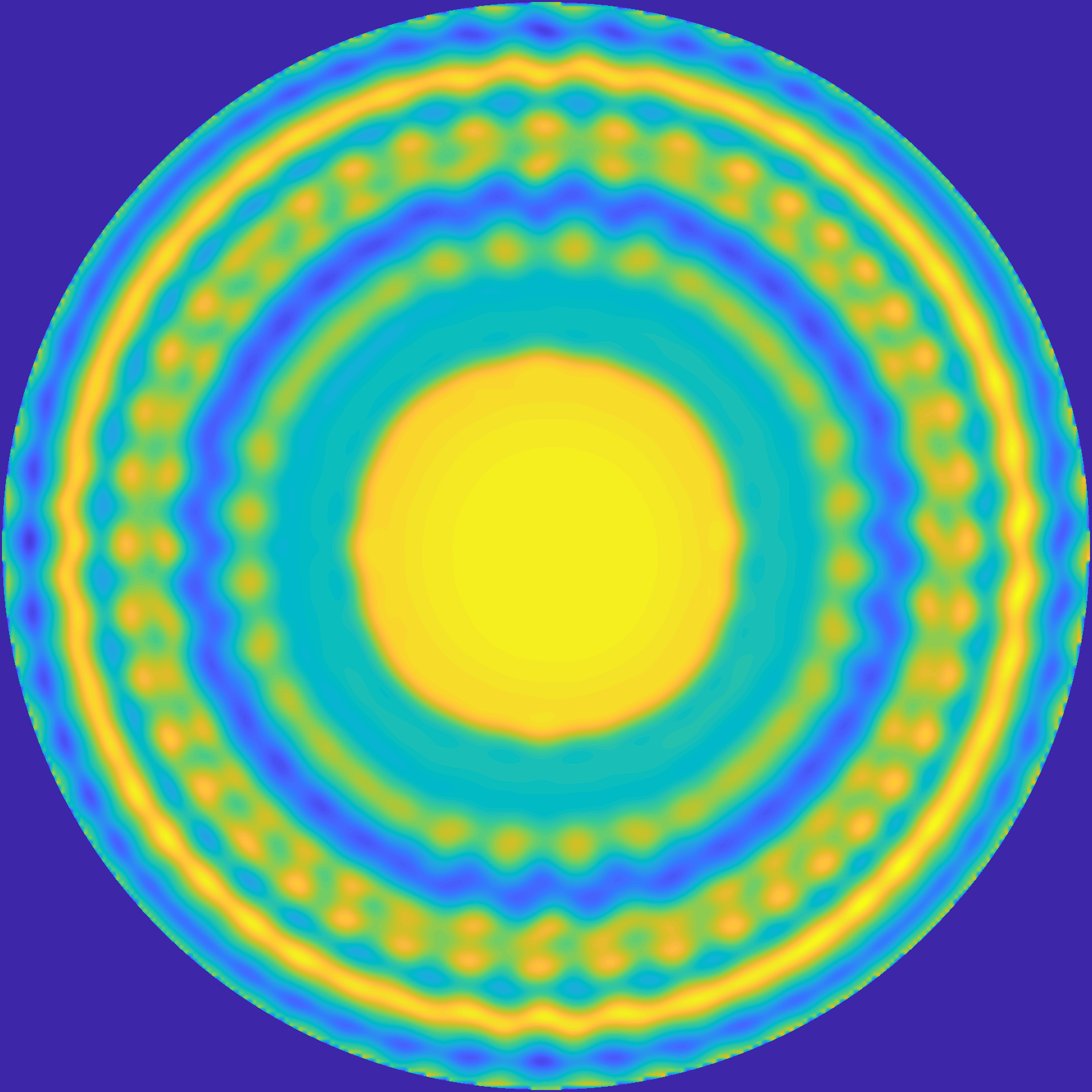};
\end{axis}

\put(-24,23.5){\rotatebox{90}{$t(mm)$}}

\end{tikzpicture}%
\end{subfigure}
\hspace{9mm}
\begin{subfigure}{0.40\columnwidth}
    \subfile{figures/MTF/MTF_Red}
\end{subfigure}

\hspace{-3mm}    
\begin{subfigure}{0.40\columnwidth}
    \subfile{figures/MTF/MTF_Green}
\end{subfigure}
\hspace{8mm}
\begin{subfigure}{0.40\columnwidth}
    \subfile{figures/MTF/MTF_Blue}
\end{subfigure}
\caption{The optimized height map (top left), and the modulation transfer functions (MTFs) at three different depths for red (\SI{611}{\nm}, top right), green (\SI{543}{\nm}, bottom left), and blue (\SI{482}{\nm}, bottom right) channels.} 
	\label{fig:MTF}
\end{figure}

After the optimization at the signal sampling rate of $\Delta s=\SI{21}{\um}$, we upsample the optimized DOE height map to the fabrication resolution at ${\Delta s}^f=\SI{3}{\um}$ by bicubic interpolation, which defines the actual physical DOE to be fabricated. Fig.~\ref{fig:MTF} illustrates the upsampled height map as well as the MTFs at three different depths for each color channel. As expected from the discussion in Sec.~\ref{sec:Problem}, the depth dependency of the MTFs decrease in the proposed method compared to the clear aperture case, easing the post-processing. However, please note that the proposed method does not provide a fully depth-invariant PSF, which is visible e.g. for the objects at infinity (blue curves in Fig.~\ref{fig:MTF}). That is due to the fact that our method is an end-to-end optimization scheme. The success of the algorithm is evaluated based on the final reconstruction image rather than the depth-invariancy of the PSF. Furthermore, our D-CNN is expected to be robust to such fluctuations in MTF. If, for instance, Wiener deconvolution is applied instead, these differences would decrease the image quality, since such method applies deconvolution with single (average) PSF. Additionally, we expect D-CNN to utilize the cross-channel features. The blue and the green channels (Fig.~\ref{fig:MTF}, bottom) have zero-crossings at infinity, while the red channel (Fig.~\ref{fig:MTF}, top right) has a wider response. Such diversity has been shown to improve the image quality and also been utilized as a key concept of \cite{EDOFElmalem}. In the following, further simulation results are discussed, considering the fabrication resolution of the mask.  

\subsubsection{Test Images}

\begin{figure*}[t!]	
	\centering
    \begin{subfigure}[t]{0.48\columnwidth}
	    \caption{Cubic \cite{Cubic}}
	    \centering
	\end{subfigure}
    \begin{subfigure}[t]{0.48\columnwidth}
        \centering
	    \caption{Elmalem et.al. \cite{EDOFElmalem}}
	\end{subfigure}
    \begin{subfigure}[t]{0.48\columnwidth}
        \centering
	    \caption{Sitzmann et.al. \cite{EDOFSitzmann}}
	\end{subfigure}
    \begin{subfigure}[t]{0.48\columnwidth}
        \centering
	    \caption{Proposed}
	\end{subfigure}
	\begin{subfigure}[t]{0.24\columnwidth}
		\caption{Sensor}
		\centering
		\includegraphics[width=\columnwidth]{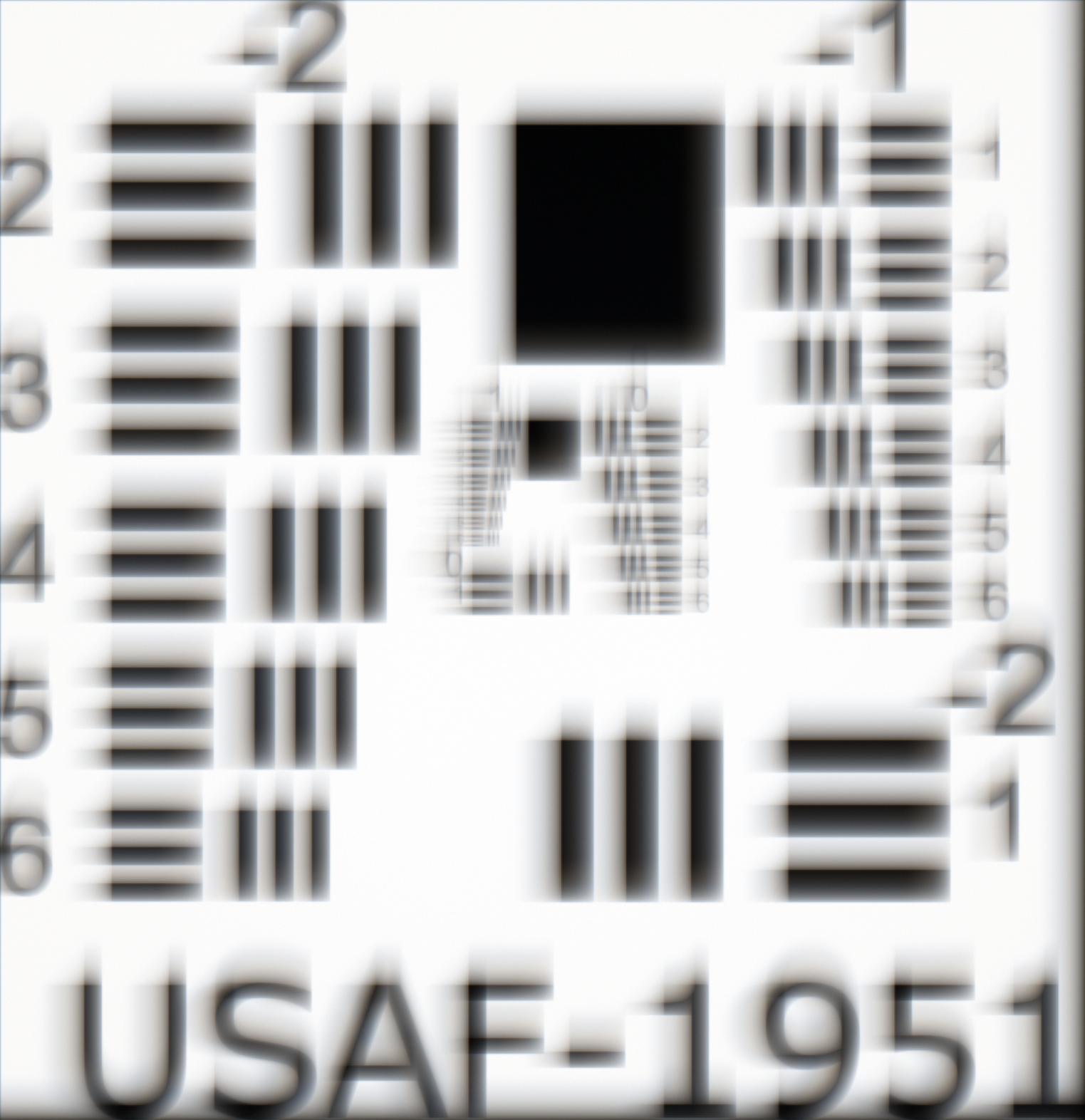}
	\end{subfigure}
	\vspace{1mm}
	\begin{subfigure}[t]{0.24\columnwidth}
		\caption{Output}
		\centering
		\includegraphics[width=\columnwidth]{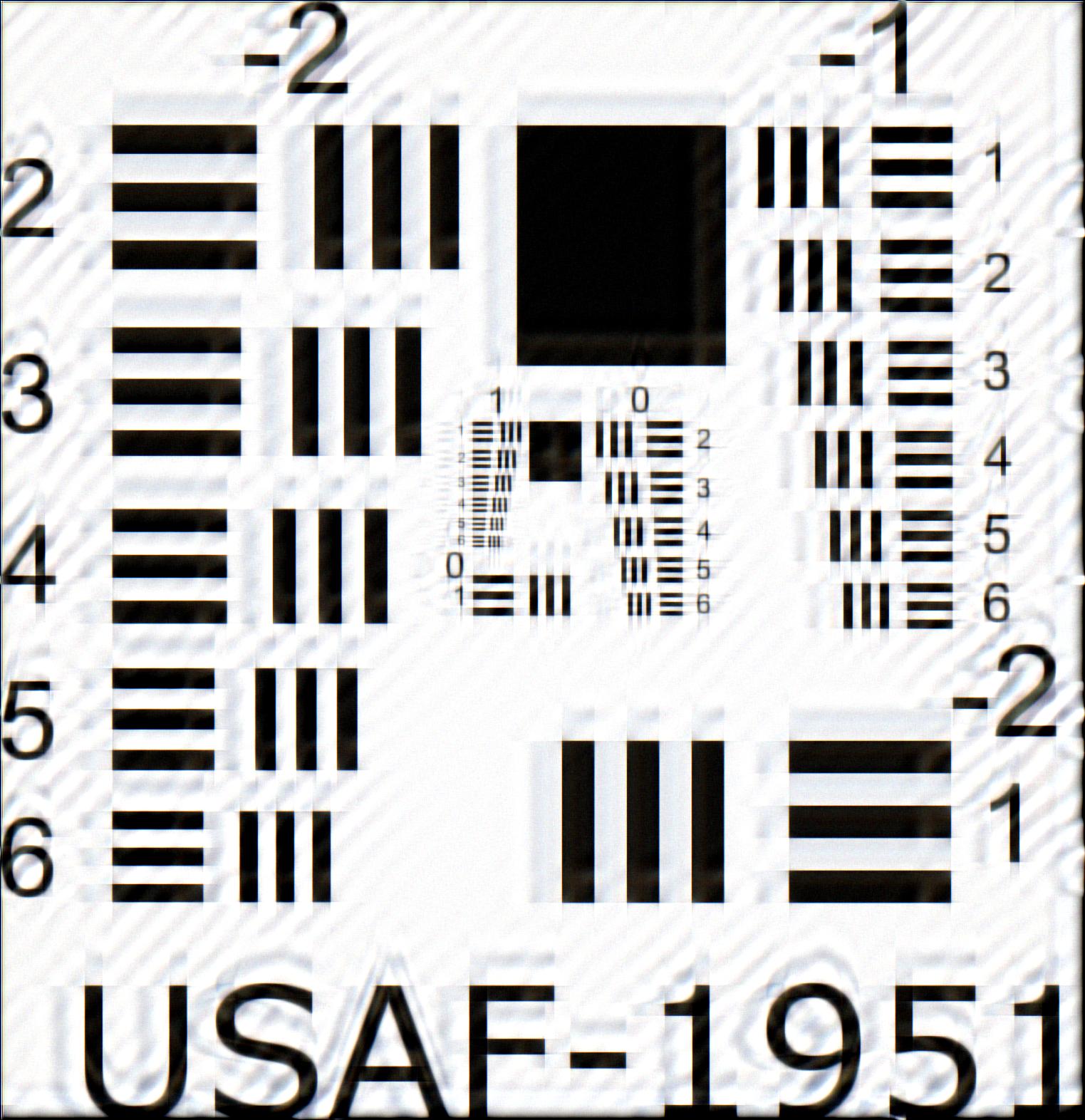}		
	\end{subfigure}
	\begin{subfigure}[t]{0.24\columnwidth}
	    \caption{Sensor}
		\centering
		\includegraphics[width=\columnwidth]{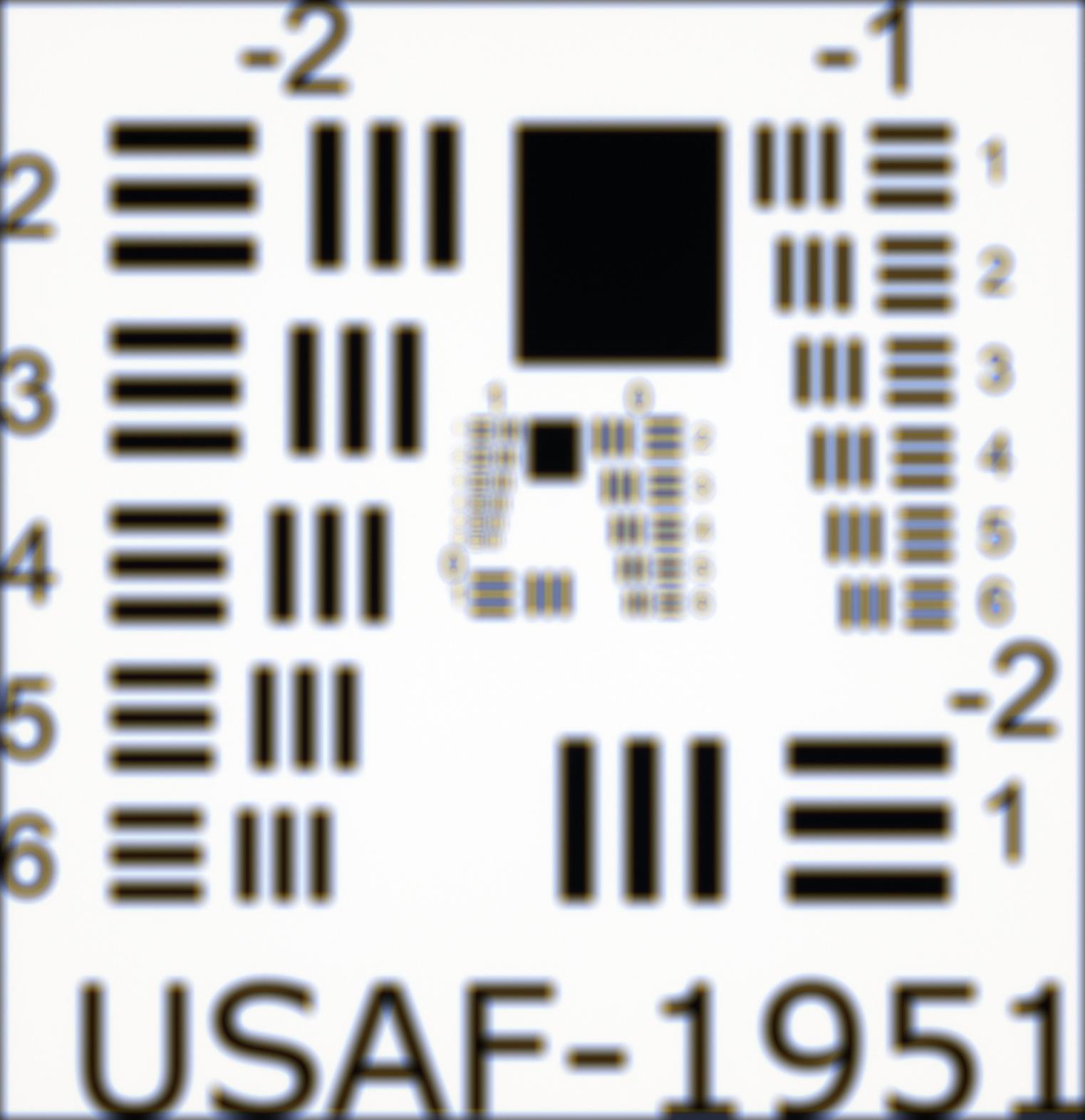}
	\end{subfigure}
	\begin{subfigure}[t]{0.24\columnwidth}
		\caption{Output}
		\centering
		\includegraphics[width=\columnwidth]{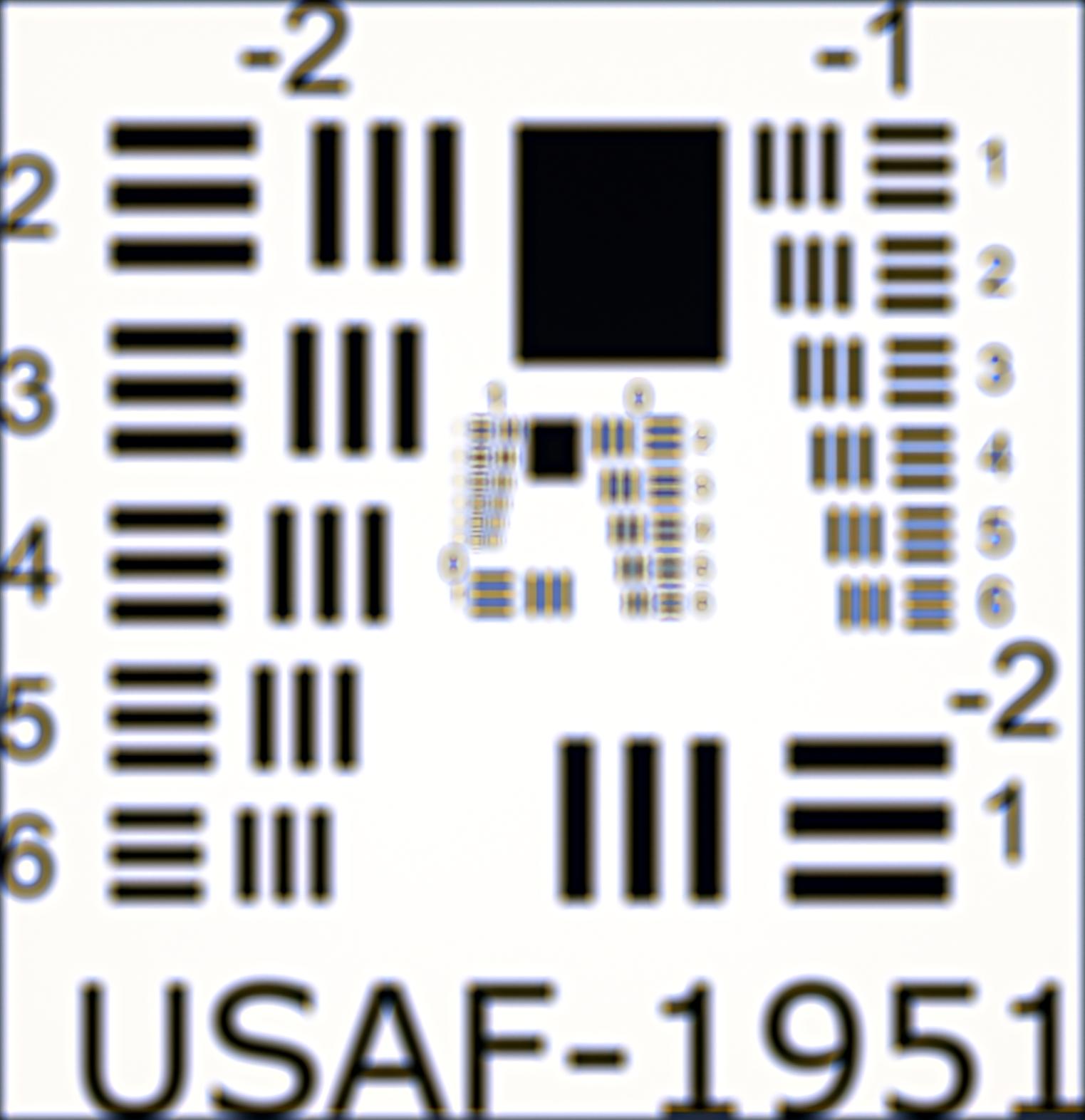}		
	\end{subfigure}
	\begin{subfigure}[t]{0.24\columnwidth}
		\caption{Sensor}
		\centering
		\includegraphics[width=\columnwidth]{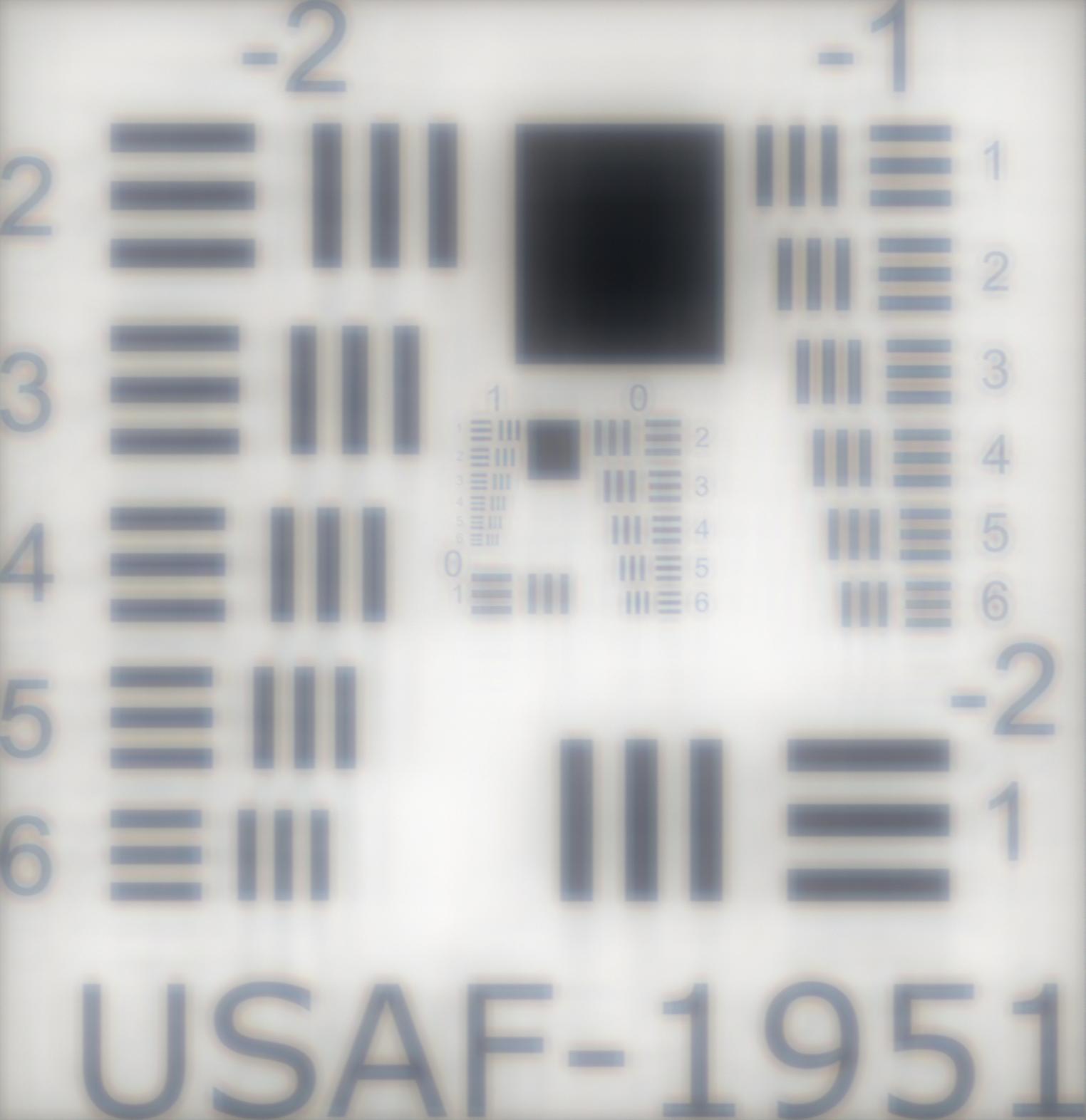}		
	\end{subfigure}
	\begin{subfigure}[t]{0.24\columnwidth}
		\caption{Output}
		\centering
		\includegraphics[width=\columnwidth]{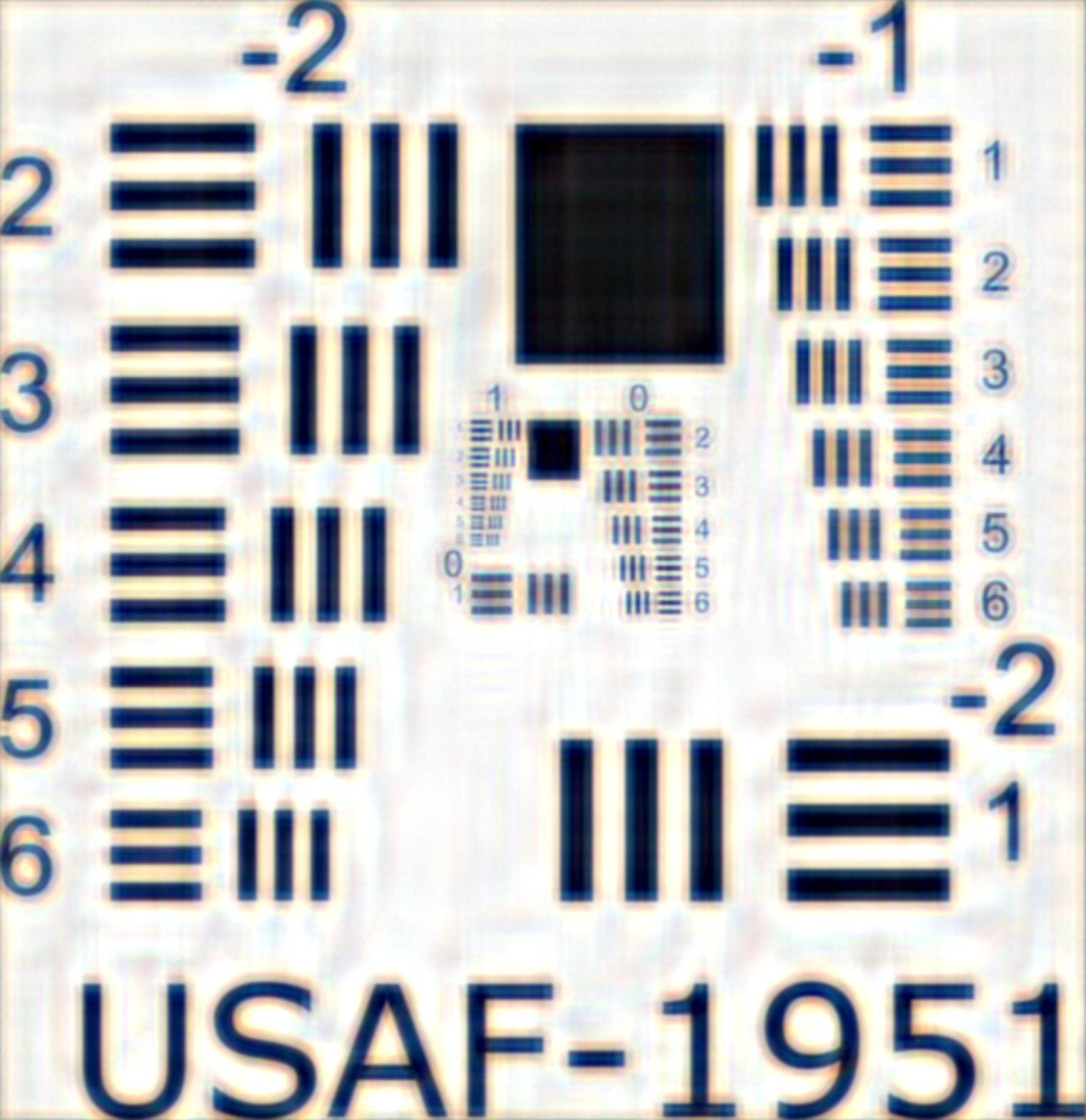}		
	\end{subfigure}
	\begin{subfigure}[t]{0.24\columnwidth}
		\caption{Sensor}
		\centering
		\includegraphics[width=\columnwidth]{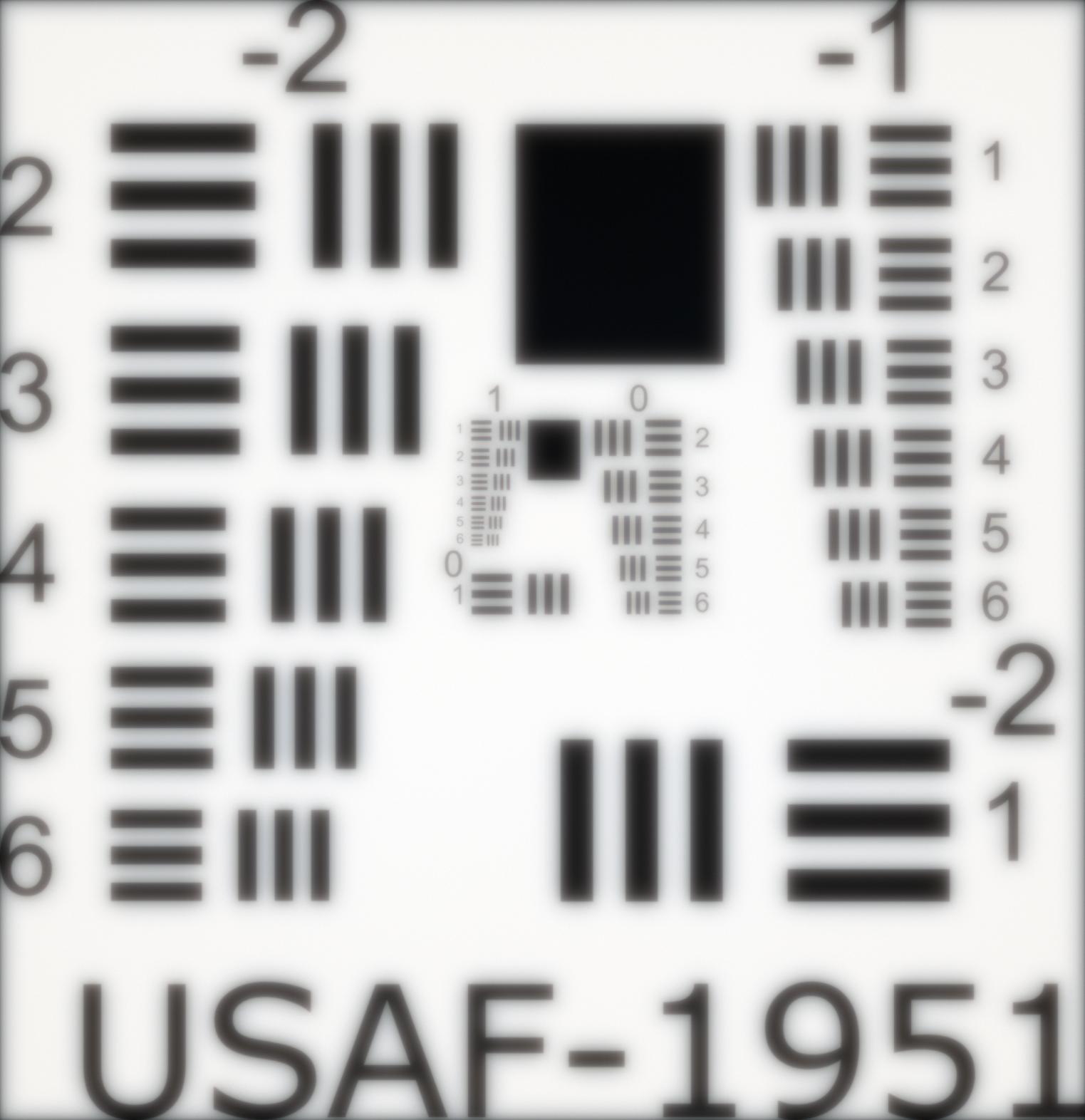}		
	\end{subfigure}
	\begin{subfigure}[t]{0.24\columnwidth}
		\caption{Output}
		\centering
		\includegraphics[width=\columnwidth]{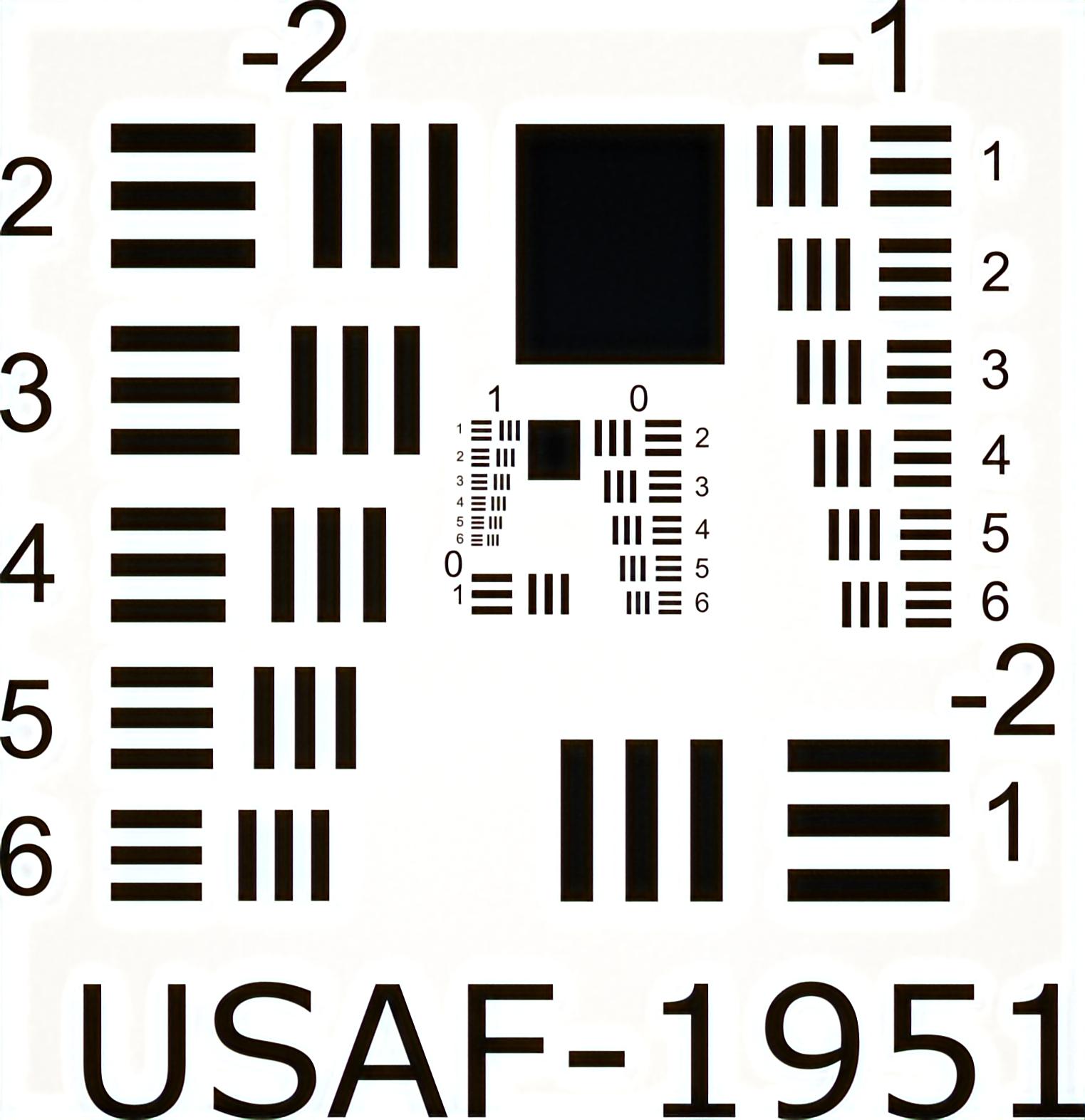}		
	\end{subfigure}
	\begin{subfigure}[t]{0.24\columnwidth}
		\centering
		\includegraphics[width=\columnwidth]{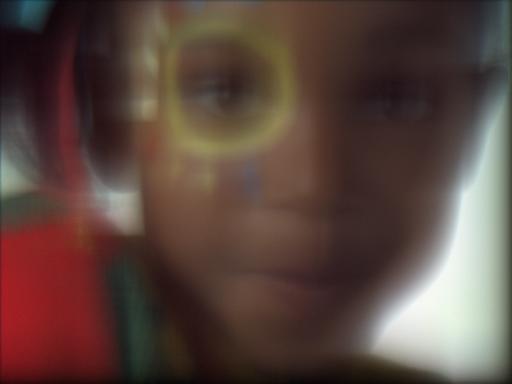}
		\put(-70,2){\rotatebox{90}{$z=\SI{0.5}{\meter}$}}		
	\end{subfigure}
	\vspace{1mm}
	\begin{subfigure}[t]{0.24\columnwidth}
		\centering
		\includegraphics[width=\columnwidth]{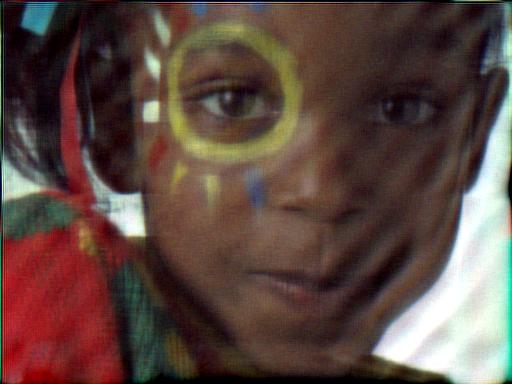}		
	\end{subfigure}
	\begin{subfigure}[t]{0.24\columnwidth}
		\centering
		\includegraphics[width=\columnwidth]{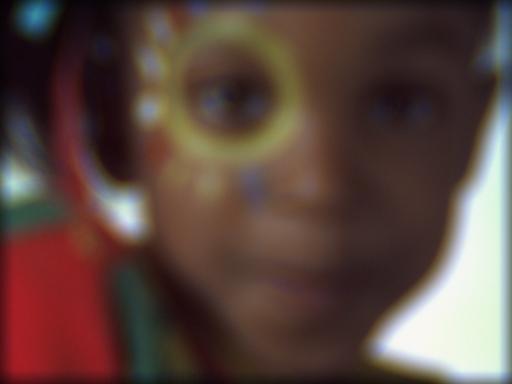}		
	\end{subfigure}
	\begin{subfigure}[t]{0.24\columnwidth}
		\centering
		\includegraphics[width=\columnwidth]{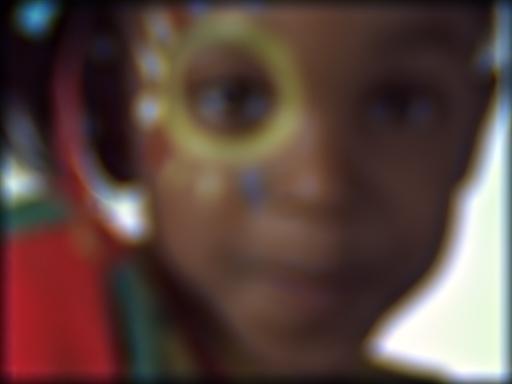}		
	\end{subfigure}
	\begin{subfigure}[t]{0.24\columnwidth}
		\centering
		\includegraphics[width=\columnwidth]{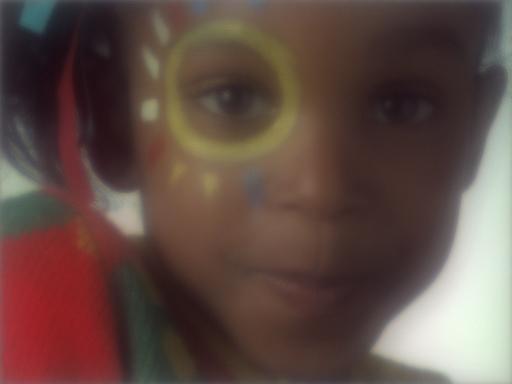}		
	\end{subfigure}
	\begin{subfigure}[t]{0.24\columnwidth}
		\centering
		\includegraphics[width=\columnwidth]{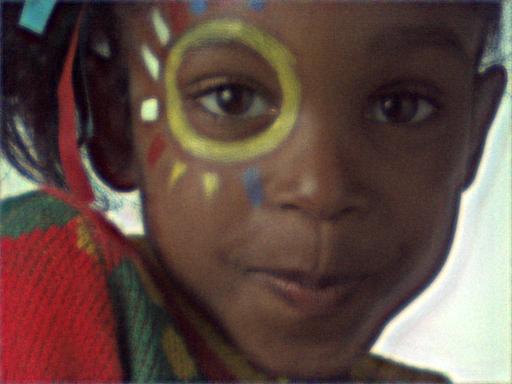}		
	\end{subfigure}
	\begin{subfigure}[t]{0.24\columnwidth}
		\centering
		\includegraphics[width=\columnwidth]{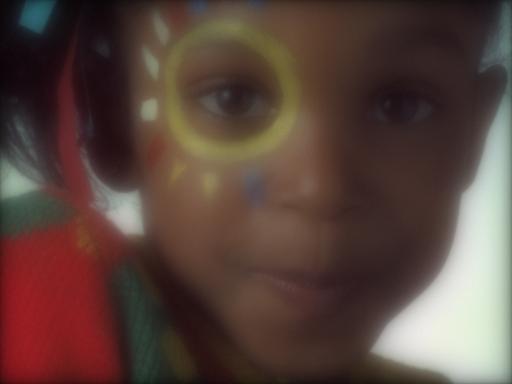}		
	\end{subfigure}
	\begin{subfigure}[t]{0.24\columnwidth}
		\centering
		\includegraphics[width=\columnwidth]{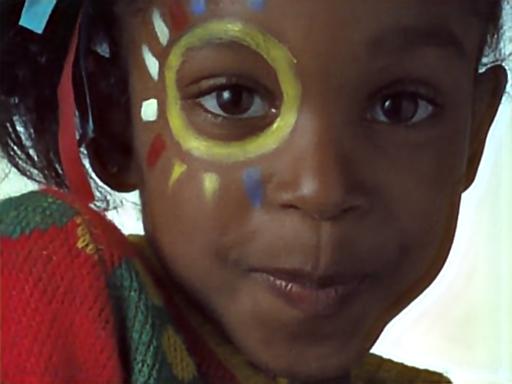}		
	\end{subfigure}
	\begin{subfigure}[t]{0.24\columnwidth}
		\centering
		\includegraphics[width=\columnwidth]{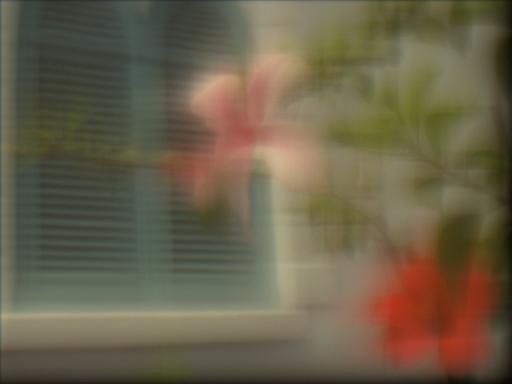}		
	\end{subfigure}
	\vspace{2mm}
	\begin{subfigure}[t]{0.24\columnwidth}
		\centering
		\includegraphics[width=\columnwidth]{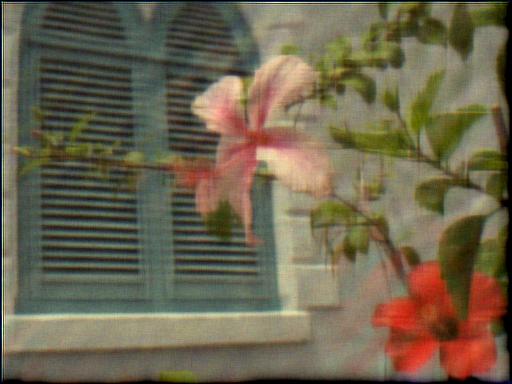}		
	\end{subfigure}
	\begin{subfigure}[t]{0.24\columnwidth}
		\centering
		\includegraphics[width=\columnwidth]{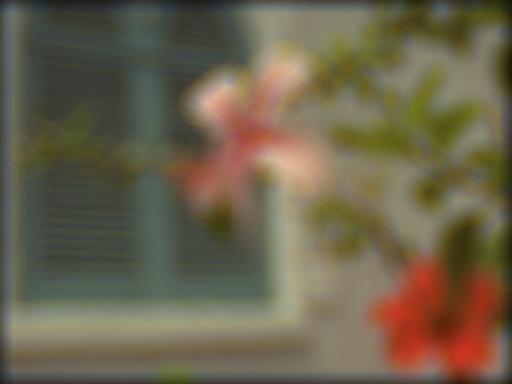}		
	\end{subfigure}
	\begin{subfigure}[t]{0.24\columnwidth}
		\centering
		\includegraphics[width=\columnwidth]{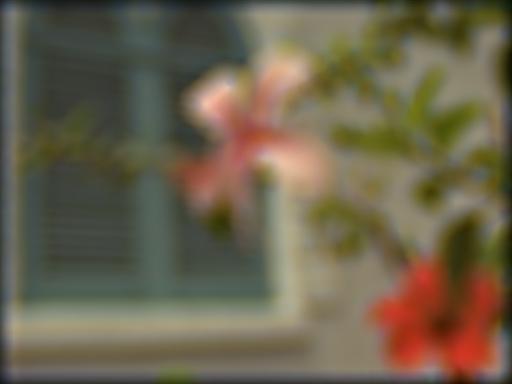}		
	\end{subfigure}
	\begin{subfigure}[t]{0.24\columnwidth}
		\centering
		\includegraphics[width=\columnwidth]{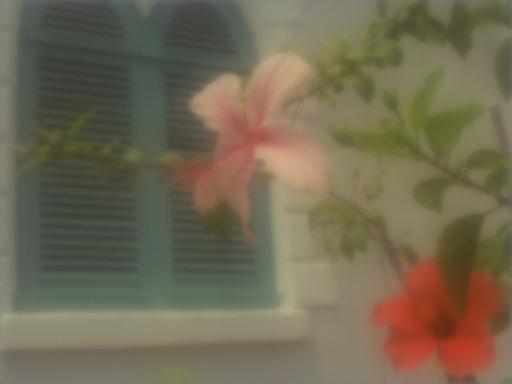}		
	\end{subfigure}
	\begin{subfigure}[t]{0.24\columnwidth}
		\centering
		\includegraphics[width=\columnwidth]{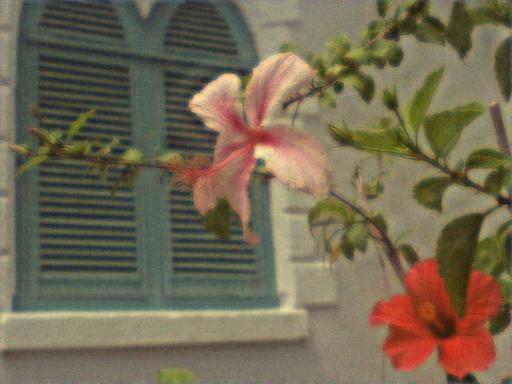}		
	\end{subfigure}
	\begin{subfigure}[t]{0.24\columnwidth}
		\centering
		\includegraphics[width=\columnwidth]{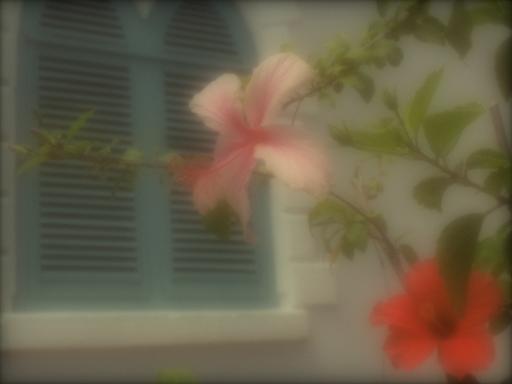}		
	\end{subfigure}
	\begin{subfigure}[t]{0.24\columnwidth}
		\centering
		\includegraphics[width=\columnwidth]{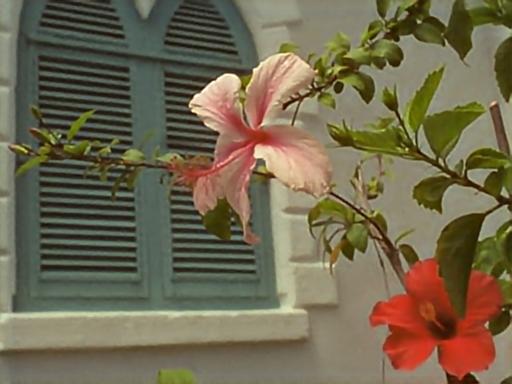}		
	\end{subfigure}
	\begin{subfigure}[t]{0.24\columnwidth}
		\centering
		\includegraphics[width=\columnwidth]{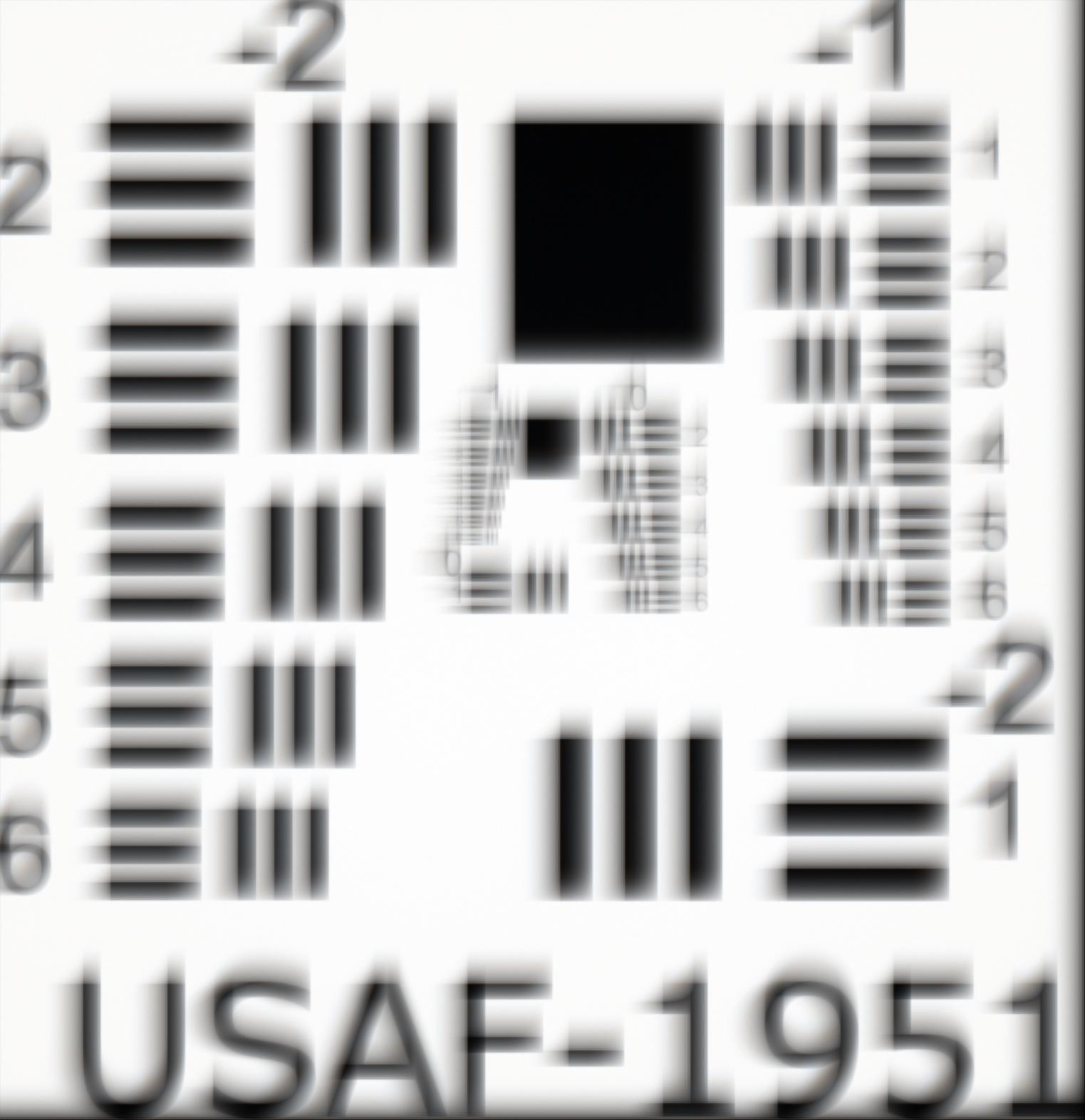}		
	\end{subfigure}
	\vspace{1mm}
	\begin{subfigure}[t]{0.24\columnwidth}
		\centering
		\includegraphics[width=\columnwidth]{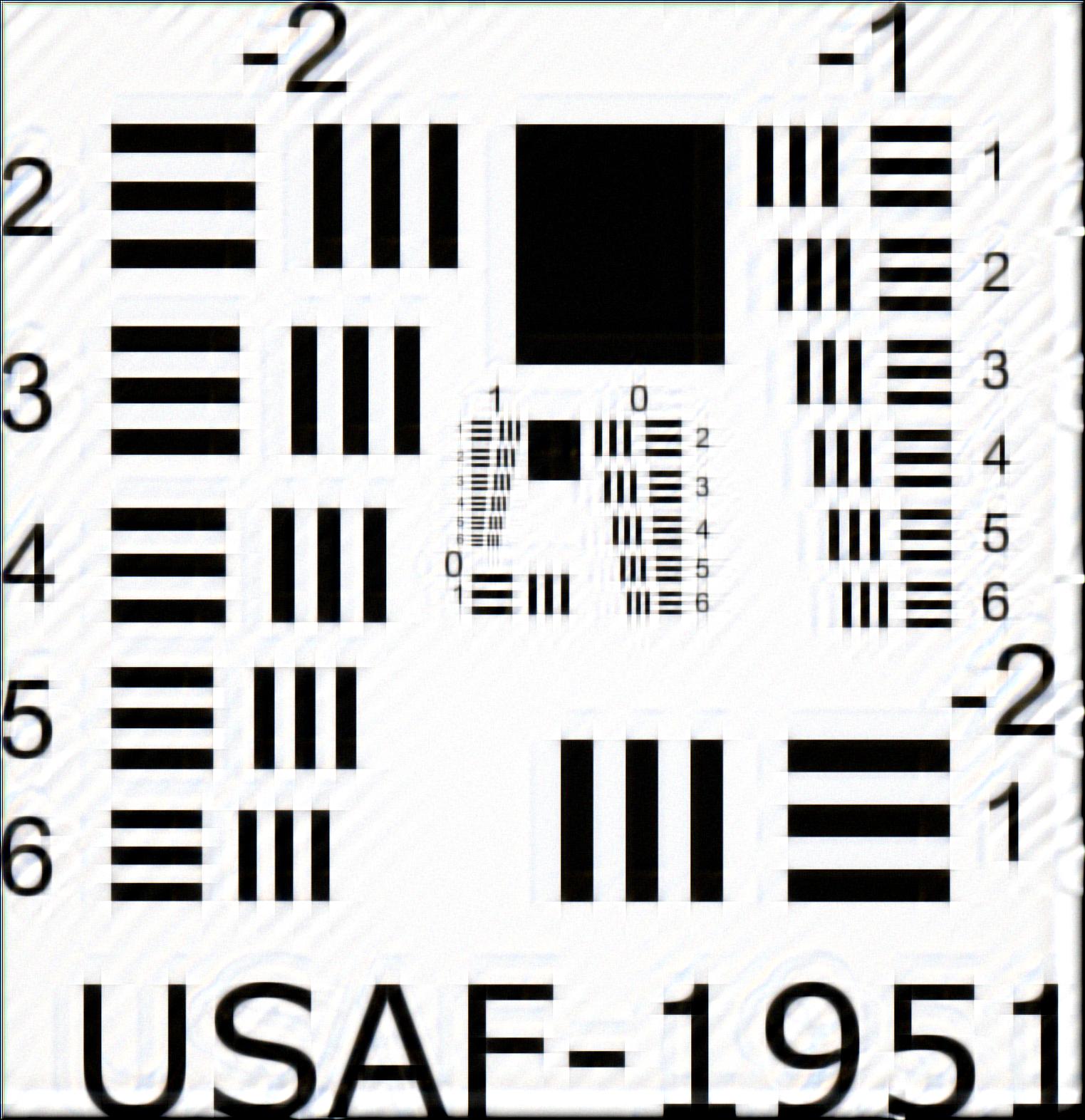}		
	\end{subfigure}
	\begin{subfigure}[t]{0.24\columnwidth}
		\centering
		\includegraphics[width=\columnwidth]{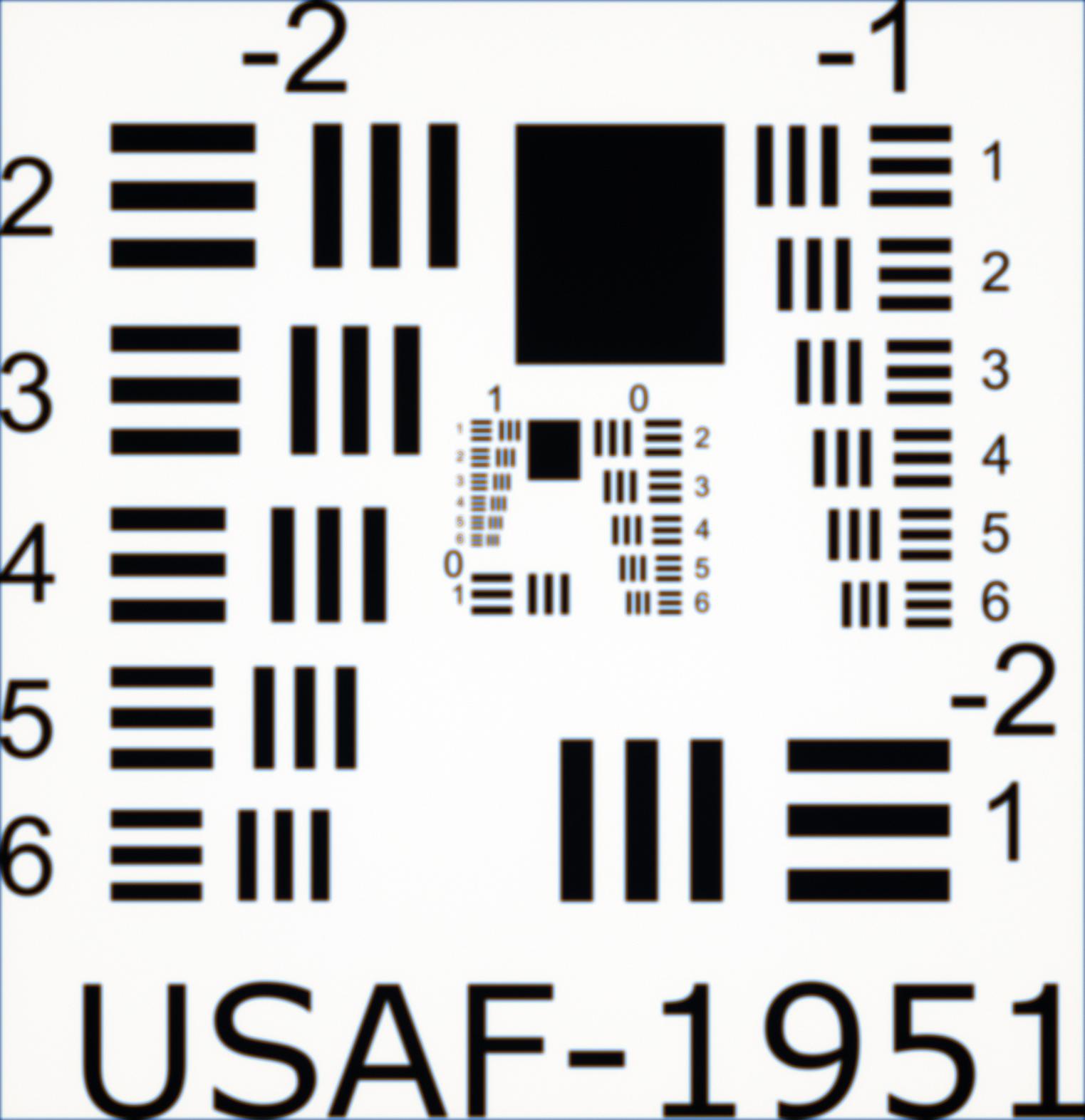}
	\end{subfigure}
	\begin{subfigure}[t]{0.24\columnwidth}
		\centering
		\includegraphics[width=\columnwidth]{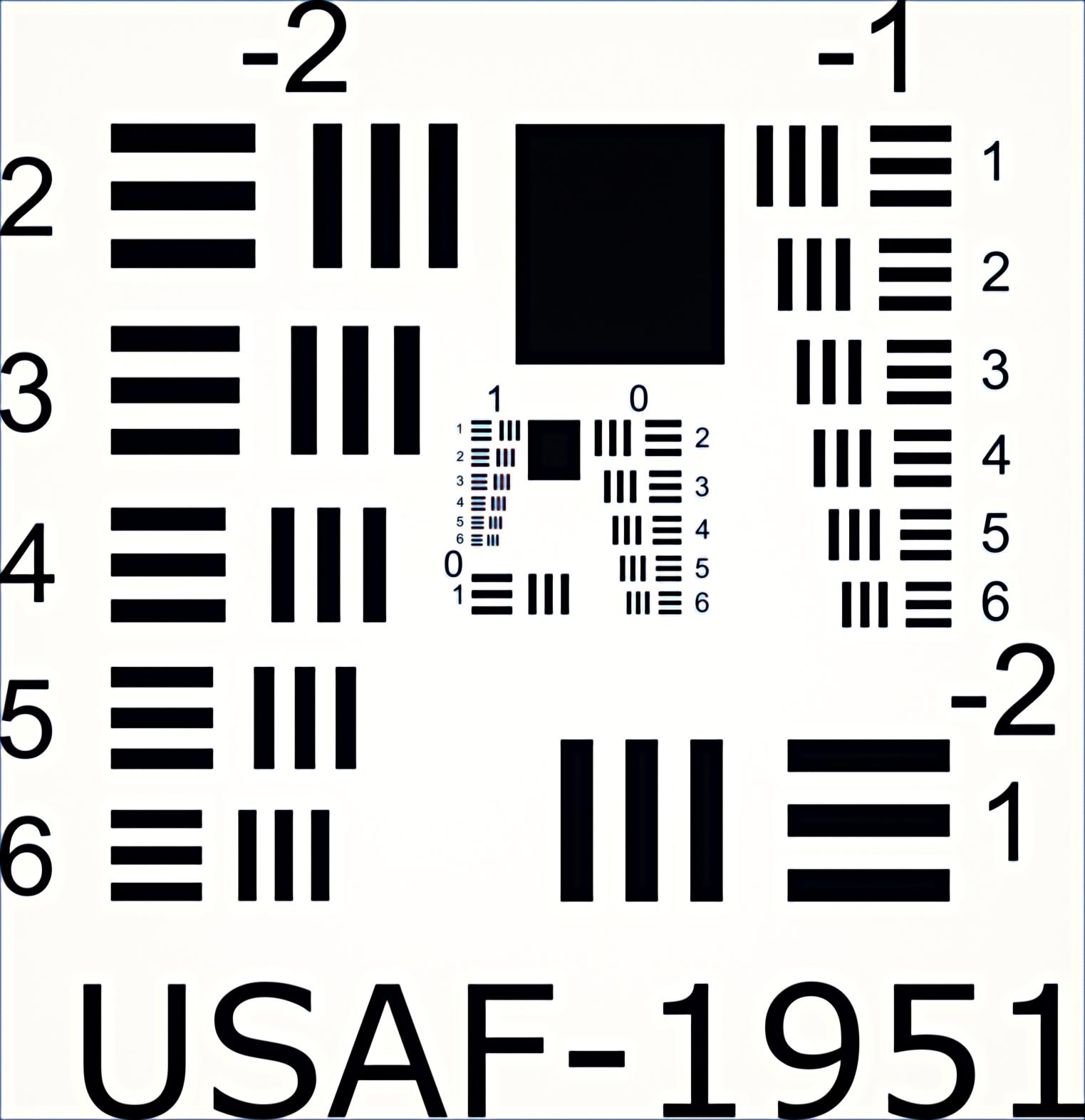}		
	\end{subfigure}
	\begin{subfigure}[t]{0.24\columnwidth}
		\centering
		\includegraphics[width=\columnwidth]{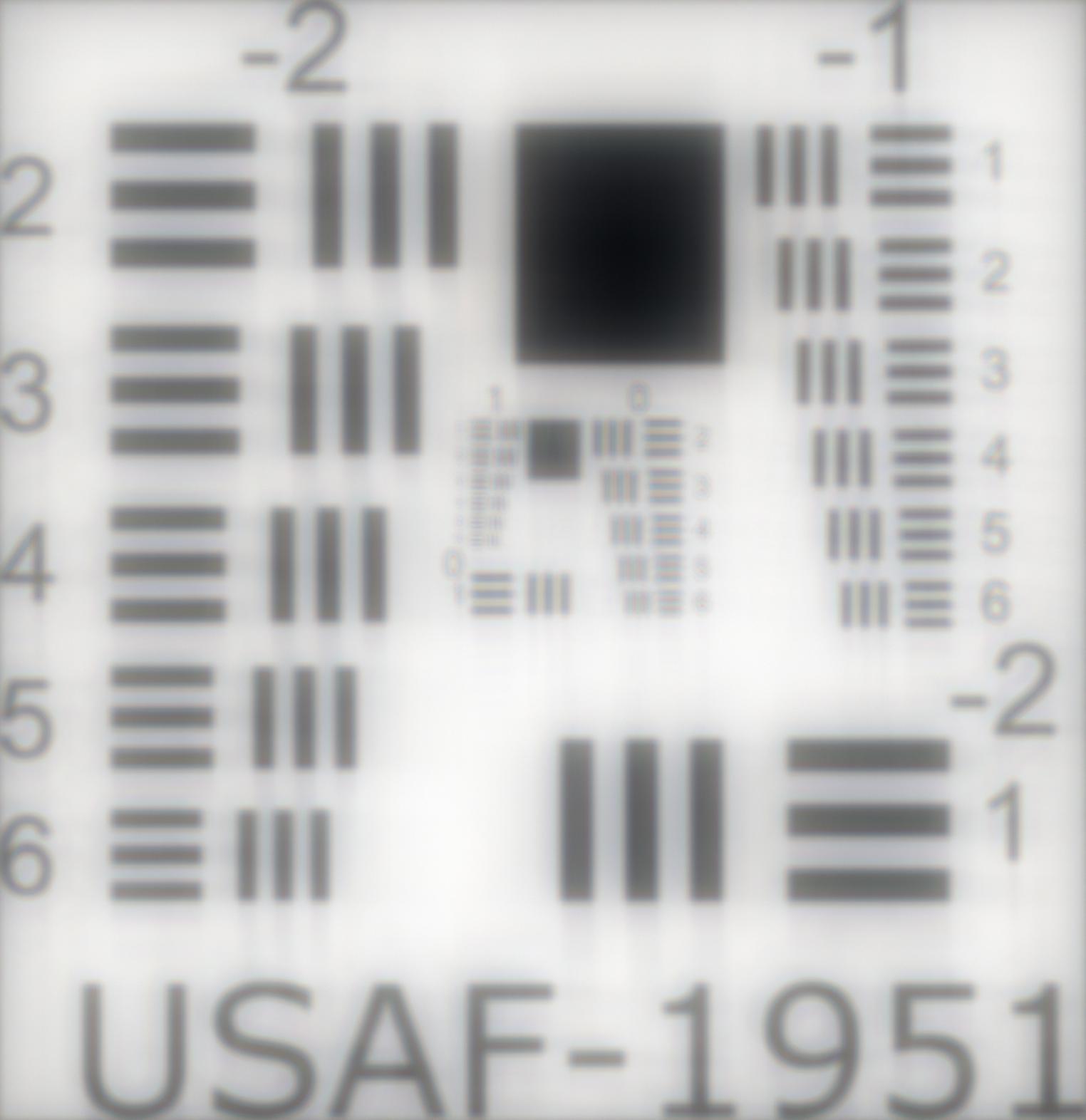}		
	\end{subfigure}
	\begin{subfigure}[t]{0.24\columnwidth}
		\centering
		\includegraphics[width=\columnwidth]{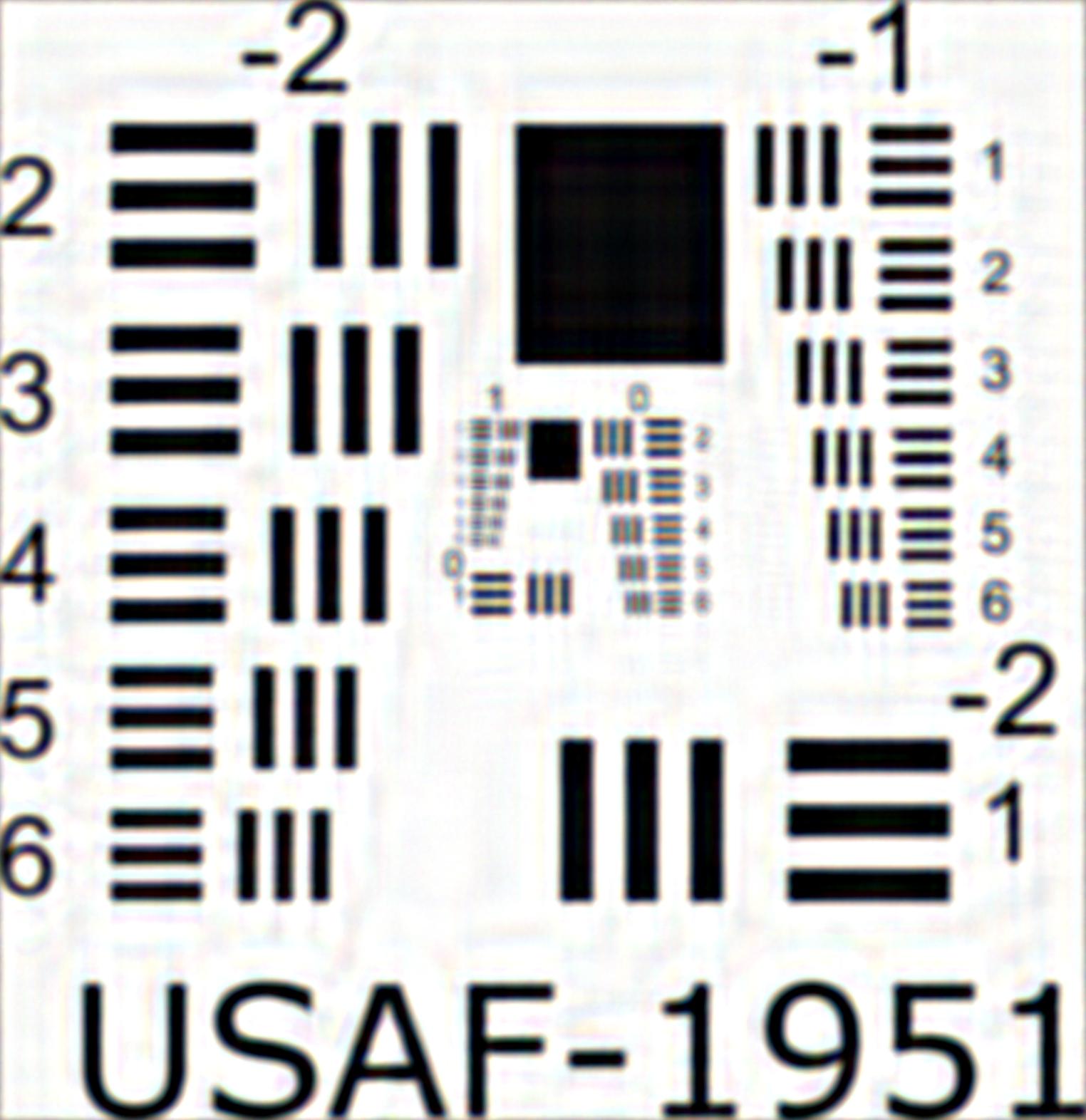}		
	\end{subfigure}
	\begin{subfigure}[t]{0.24\columnwidth}
		\centering
		\includegraphics[width=\columnwidth]{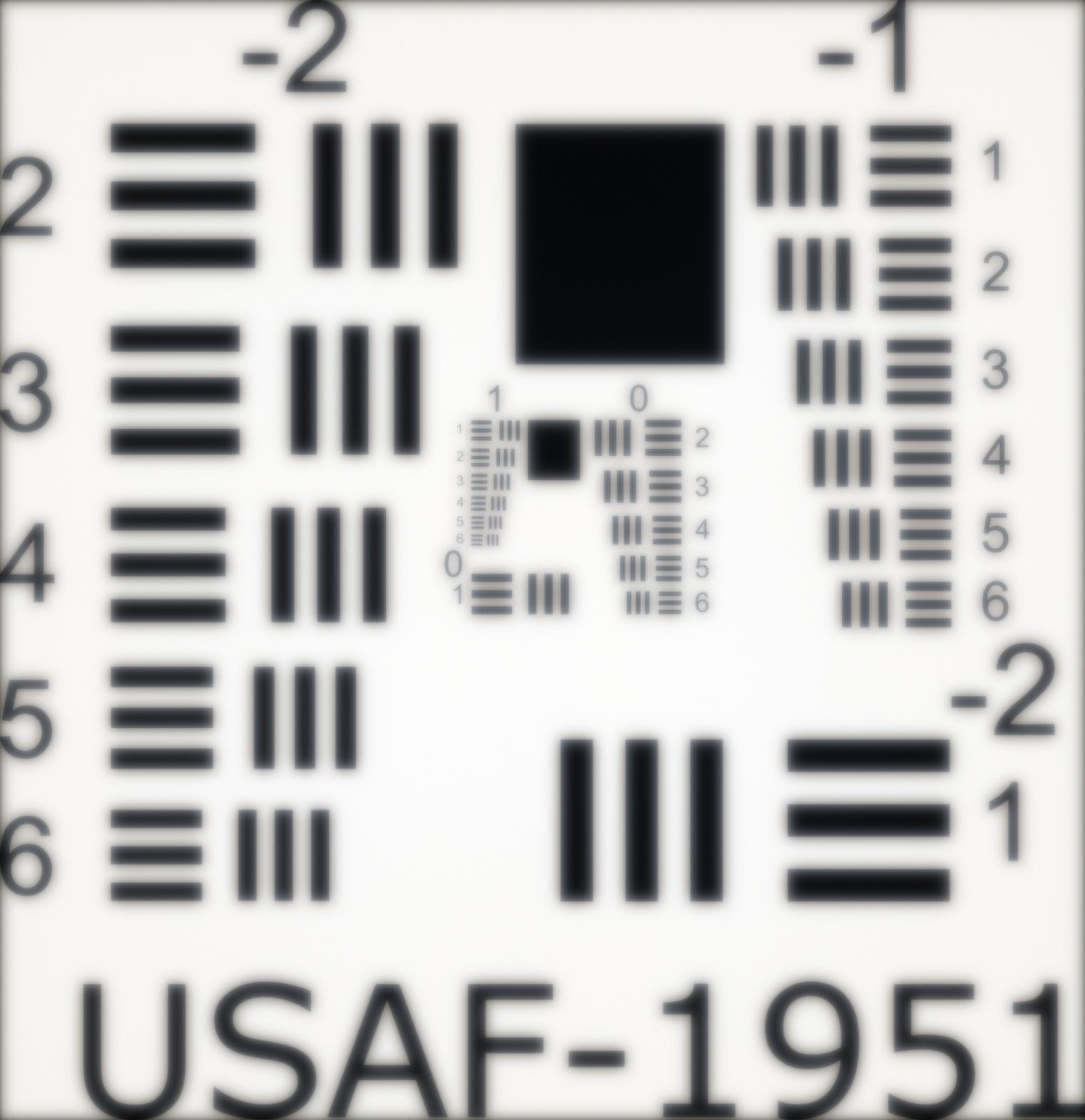}		
	\end{subfigure}
	\begin{subfigure}[t]{0.24\columnwidth}
		\centering
		\includegraphics[width=\columnwidth]{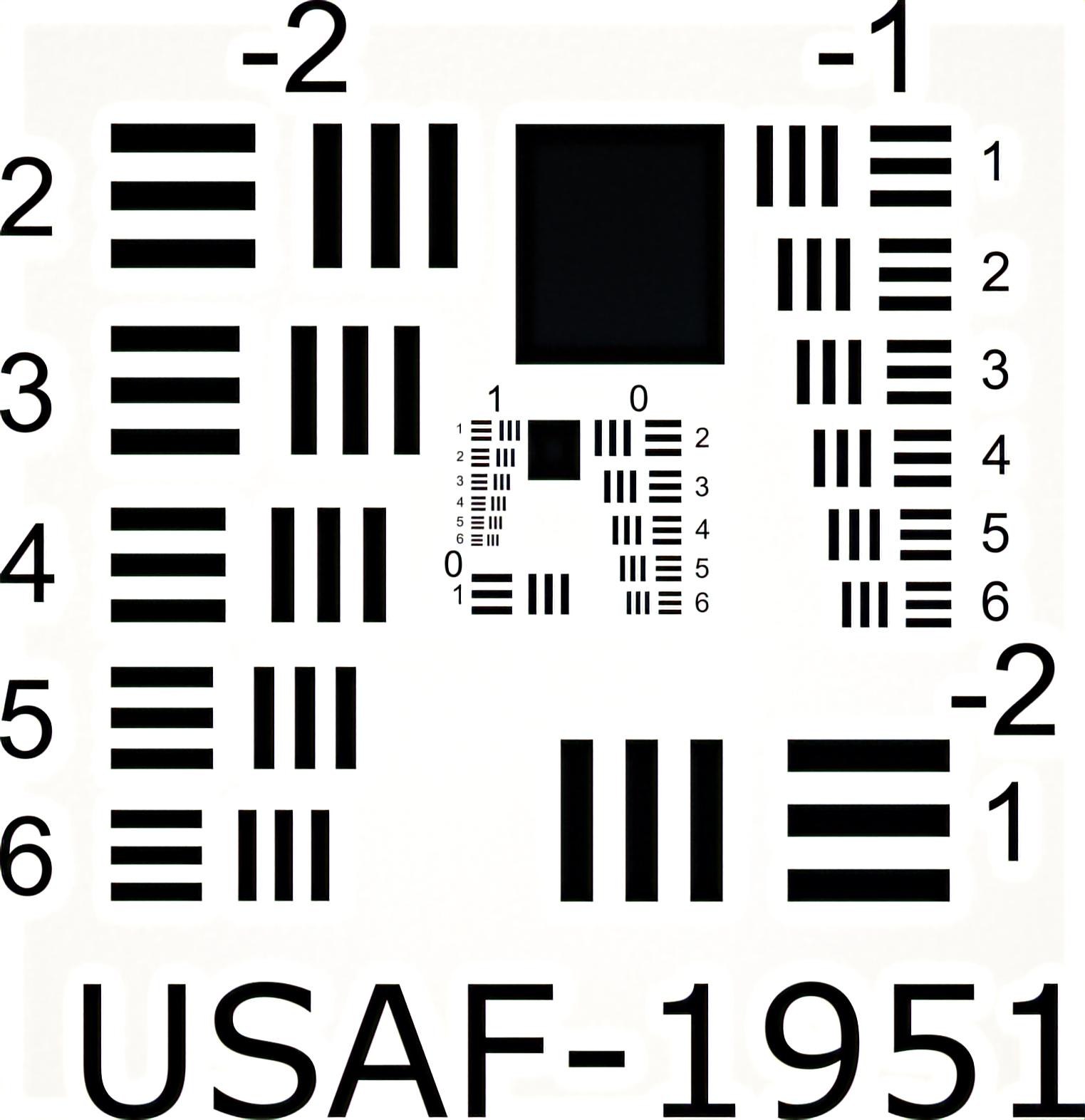}		
	\end{subfigure}
	\begin{subfigure}[t]{0.24\columnwidth}
		\centering
		\includegraphics[width=\columnwidth]{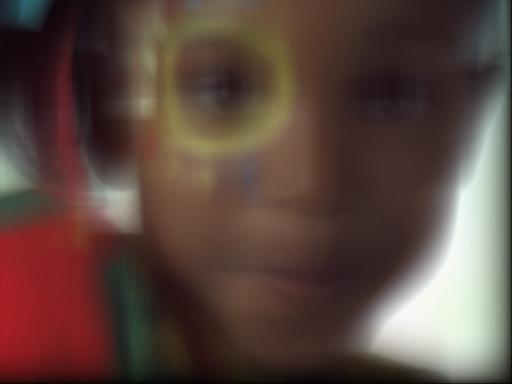}
        \put(-70,2){\rotatebox{90}{$z=\SI{0.8}{\meter}$}}
	\end{subfigure}
	\vspace{1mm}
	\begin{subfigure}[t]{0.24\columnwidth}
		\centering
		\includegraphics[width=\columnwidth]{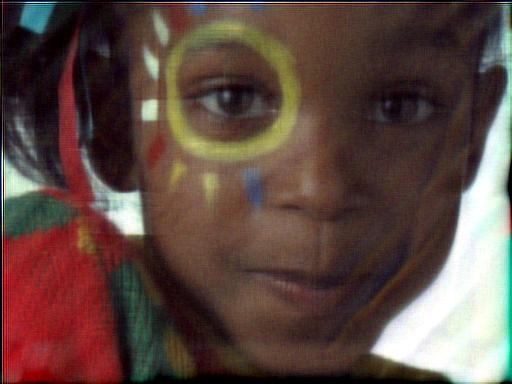}		
	\end{subfigure}
	\begin{subfigure}[t]{0.24\columnwidth}
		\centering
		\includegraphics[width=\columnwidth]{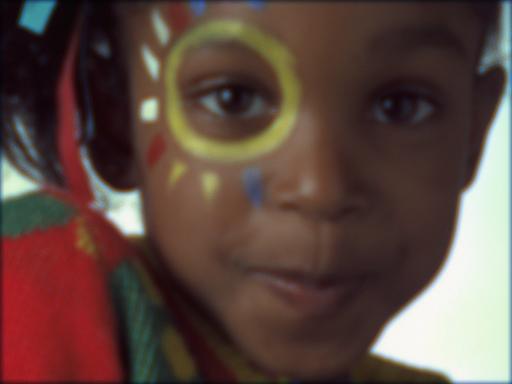}		
	\end{subfigure}
	\begin{subfigure}[t]{0.24\columnwidth}
		\centering
		\includegraphics[width=\columnwidth]{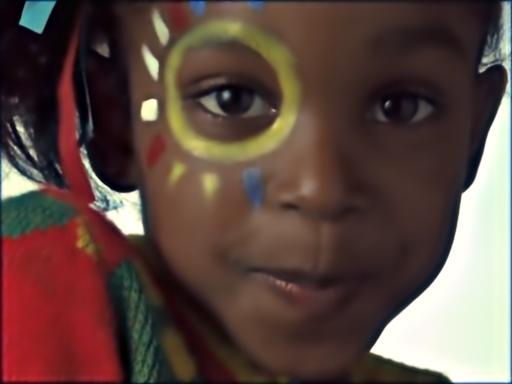}		
	\end{subfigure}
	\begin{subfigure}[t]{0.24\columnwidth}
		\centering
		\includegraphics[width=\columnwidth]{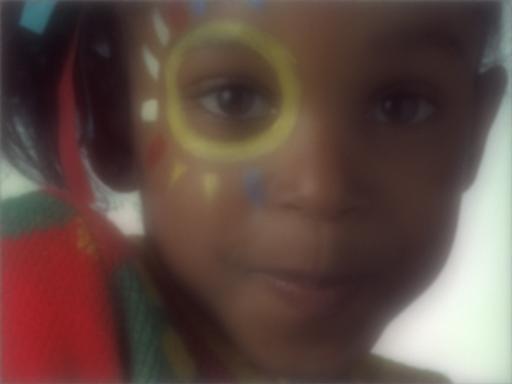}		
	\end{subfigure}
	\begin{subfigure}[t]{0.24\columnwidth}
		\centering
		\includegraphics[width=\columnwidth]{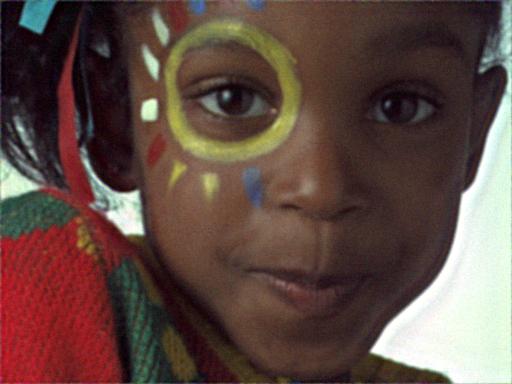}		
	\end{subfigure}
	\begin{subfigure}[t]{0.24\columnwidth}
		\centering
		\includegraphics[width=\columnwidth]{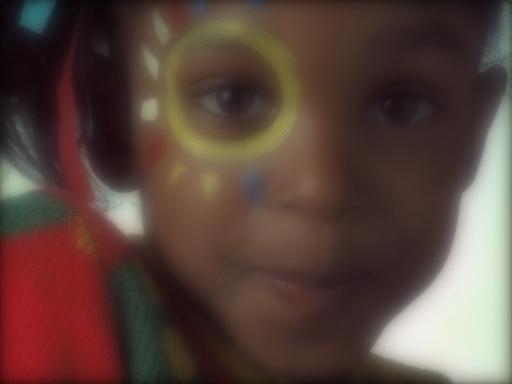}		
	\end{subfigure}
	\begin{subfigure}[t]{0.24\columnwidth}
		\centering
		\includegraphics[width=\columnwidth]{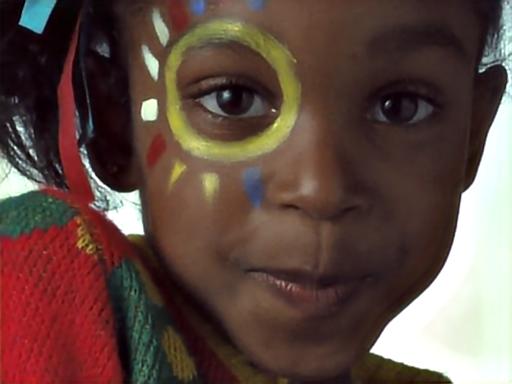}		
	\end{subfigure}
	\begin{subfigure}[t]{0.24\columnwidth}
		\centering
		\includegraphics[width=\columnwidth]{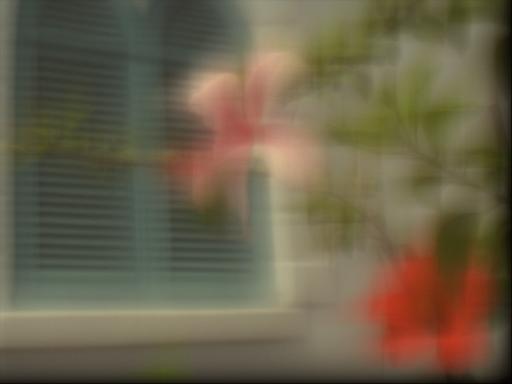}		
	\end{subfigure}
	\vspace{2mm}
	\begin{subfigure}[t]{0.24\columnwidth}
		\centering
		\includegraphics[width=\columnwidth]{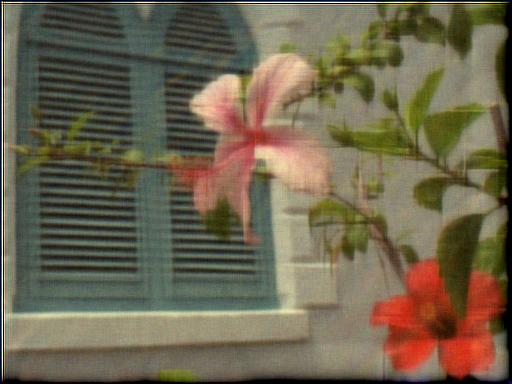}		
	\end{subfigure}
	\begin{subfigure}[t]{0.24\columnwidth}
		\centering
		\includegraphics[width=\columnwidth]{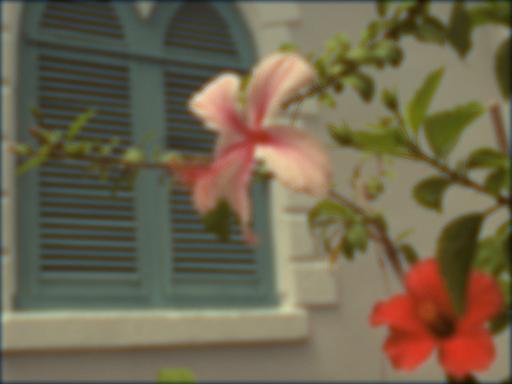}		
	\end{subfigure}
	\begin{subfigure}[t]{0.24\columnwidth}
		\centering
		\includegraphics[width=\columnwidth]{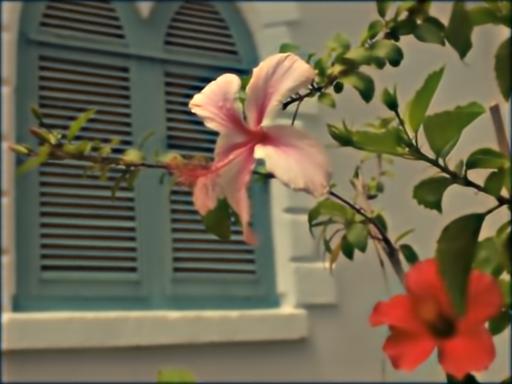}		
	\end{subfigure}
	\begin{subfigure}[t]{0.24\columnwidth}
		\centering
		\includegraphics[width=\columnwidth]{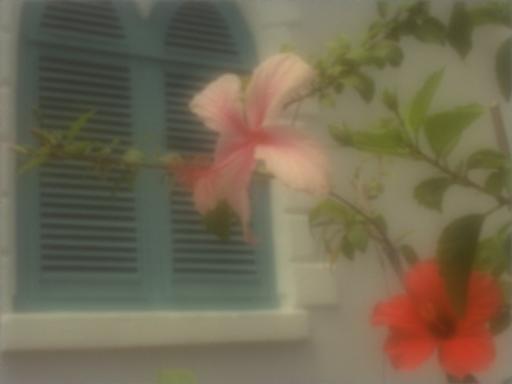}		
	\end{subfigure}
	\begin{subfigure}[t]{0.24\columnwidth}
		\centering
		\includegraphics[width=\columnwidth]{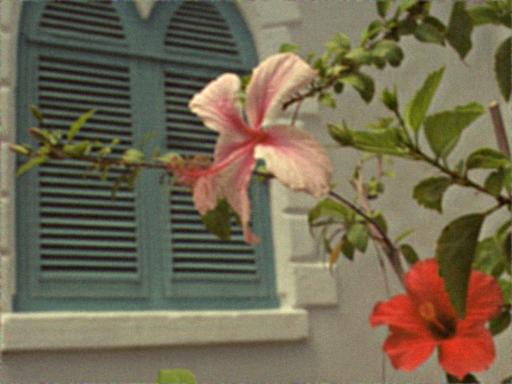}		
	\end{subfigure}
	\begin{subfigure}[t]{0.24\columnwidth}
		\centering
		\includegraphics[width=\columnwidth]{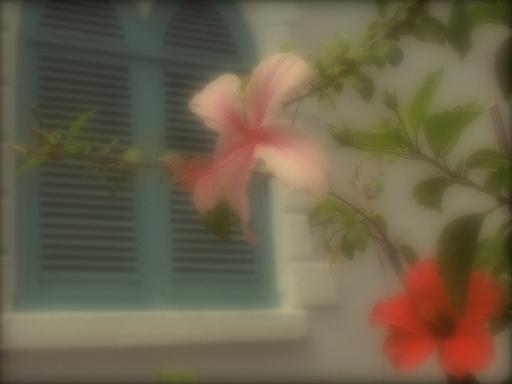}		
	\end{subfigure}
	\begin{subfigure}[t]{0.24\columnwidth}
		\centering
		\includegraphics[width=\columnwidth]{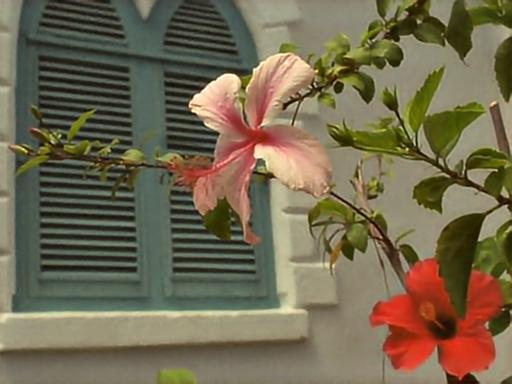}		
	\end{subfigure}
	\begin{subfigure}[t]{0.24\columnwidth}
		\centering
		\includegraphics[width=\columnwidth]{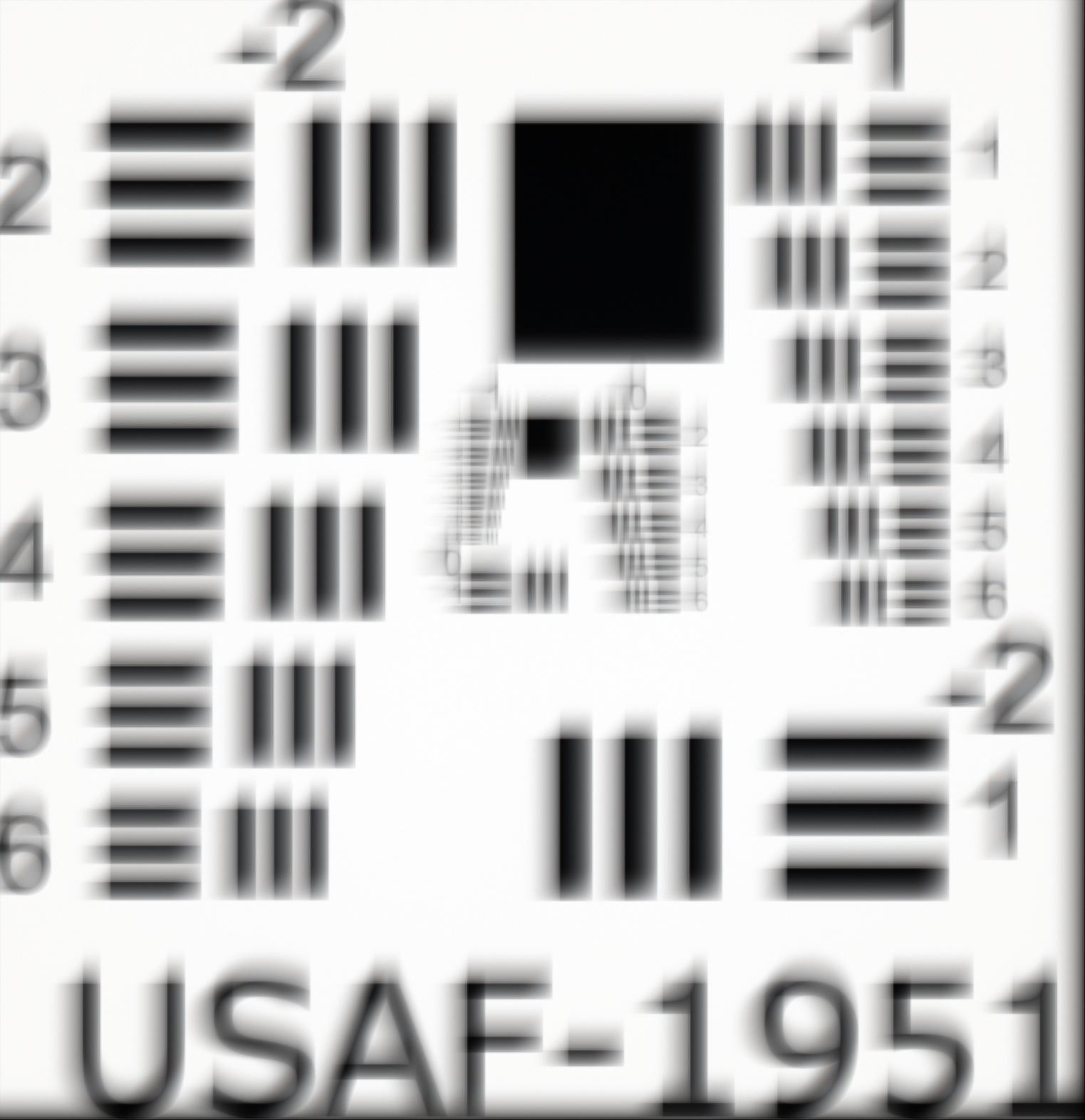}		
	\end{subfigure}
	\vspace{1mm}
	\begin{subfigure}[t]{0.24\columnwidth}
		\centering
		\includegraphics[width=\columnwidth]{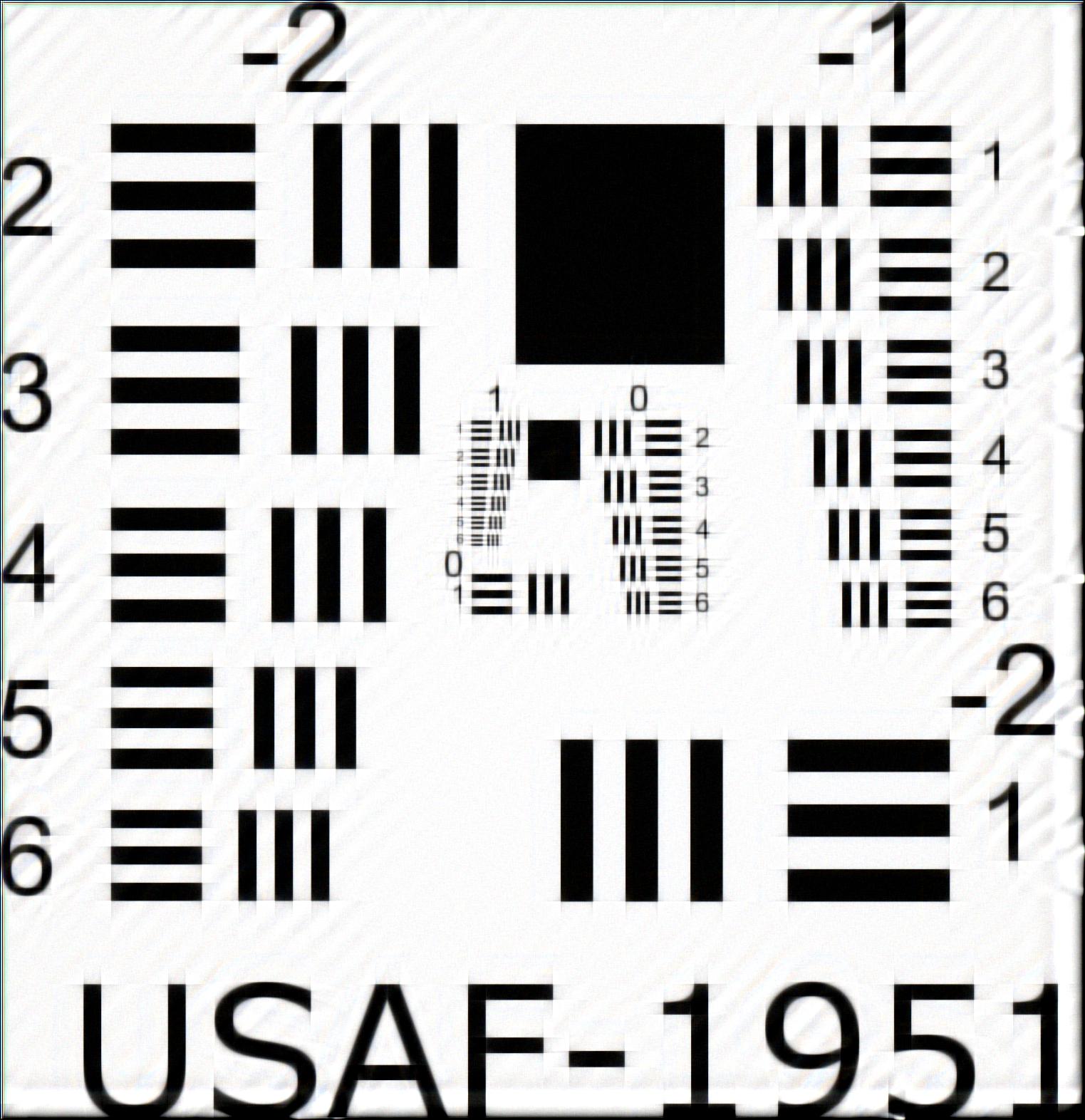}		
	\end{subfigure}
	\begin{subfigure}[t]{0.24\columnwidth}
		\centering
		\includegraphics[width=\columnwidth]{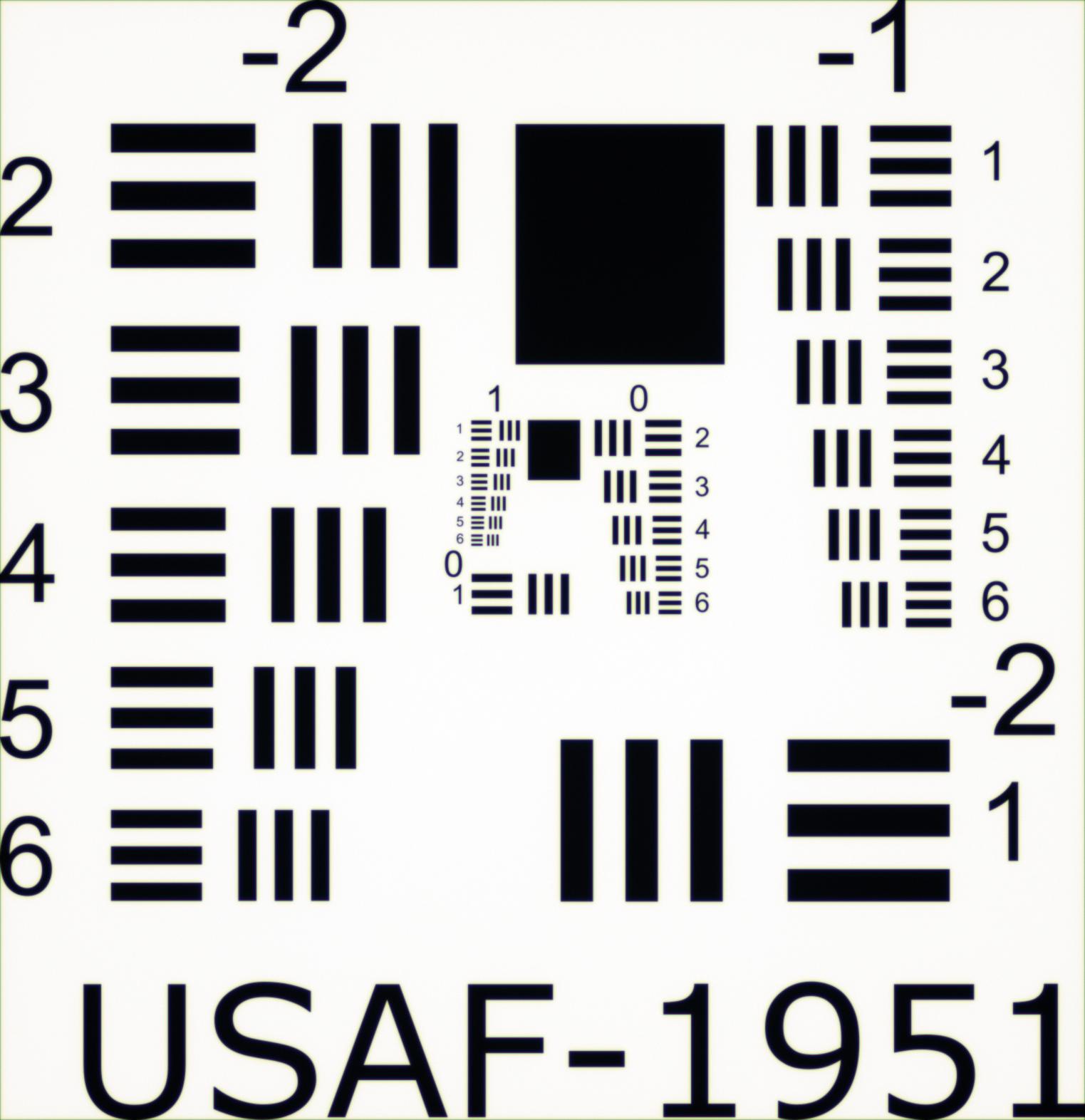}
	\end{subfigure}
	\begin{subfigure}[t]{0.24\columnwidth}
		\centering
		\includegraphics[width=\columnwidth]{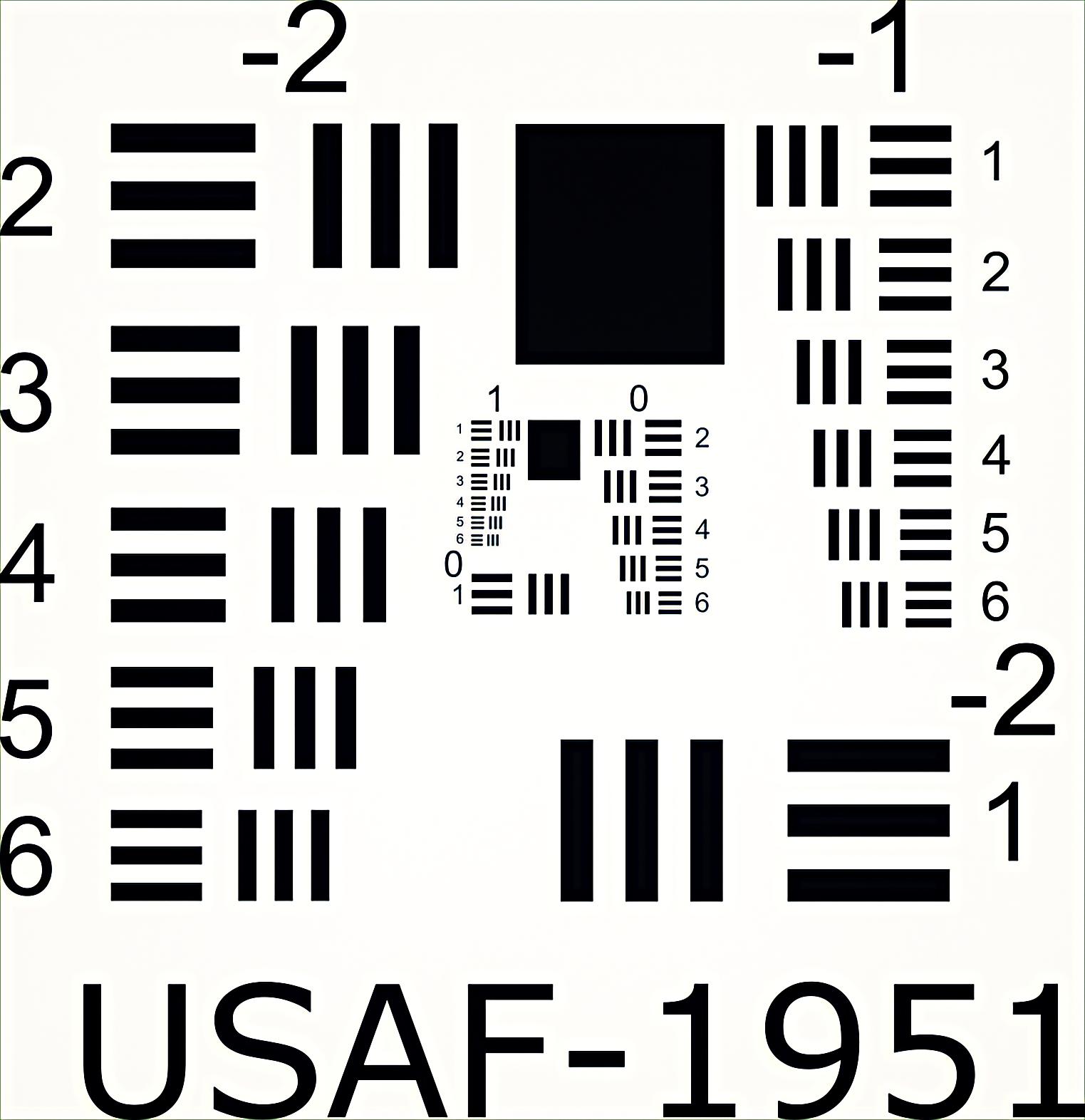}		
	\end{subfigure}
	\begin{subfigure}[t]{0.24\columnwidth}
		\centering
		\includegraphics[width=\columnwidth]{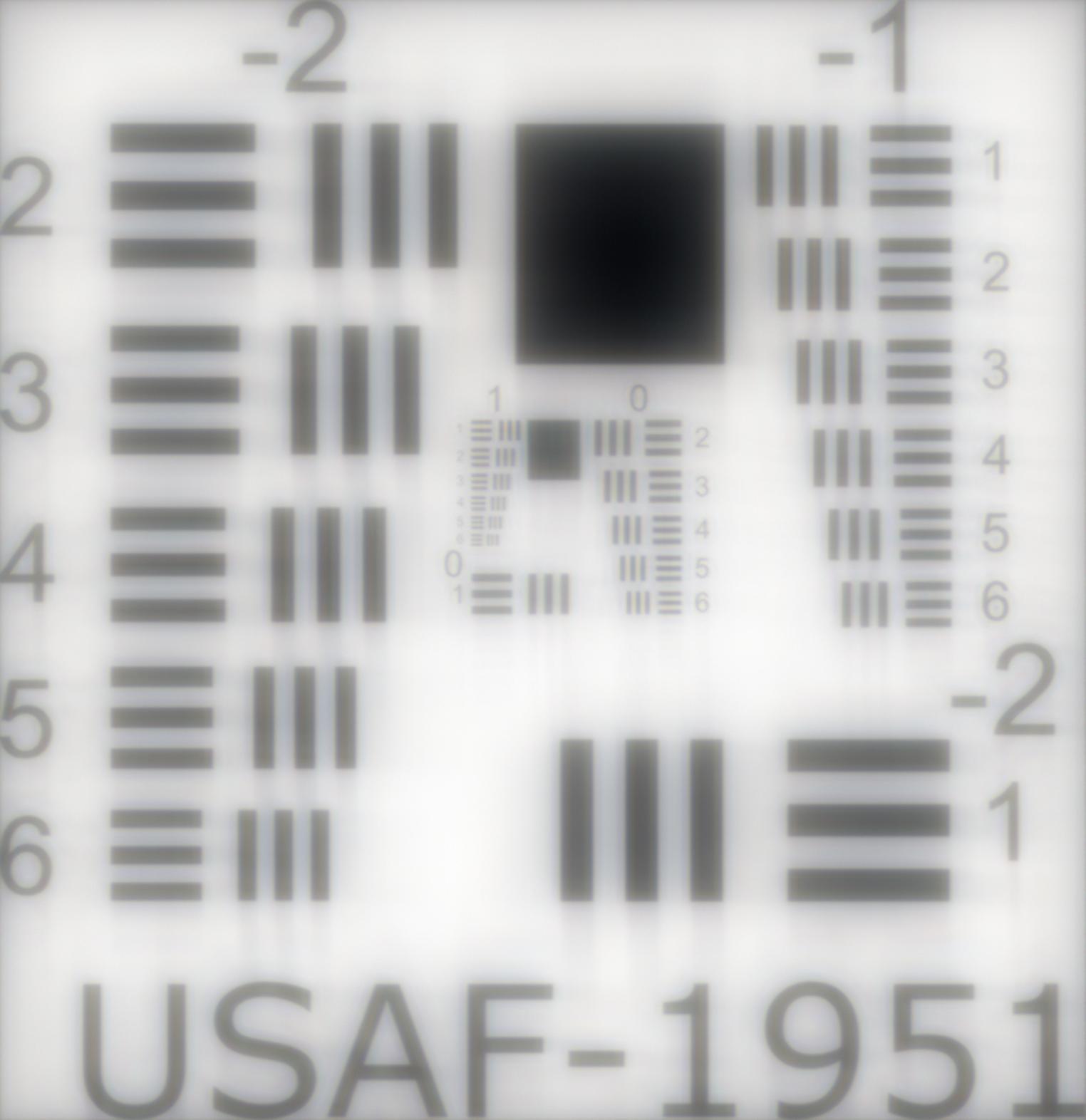}		
	\end{subfigure}
	\begin{subfigure}[t]{0.24\columnwidth}
		\centering
		\includegraphics[width=\columnwidth]{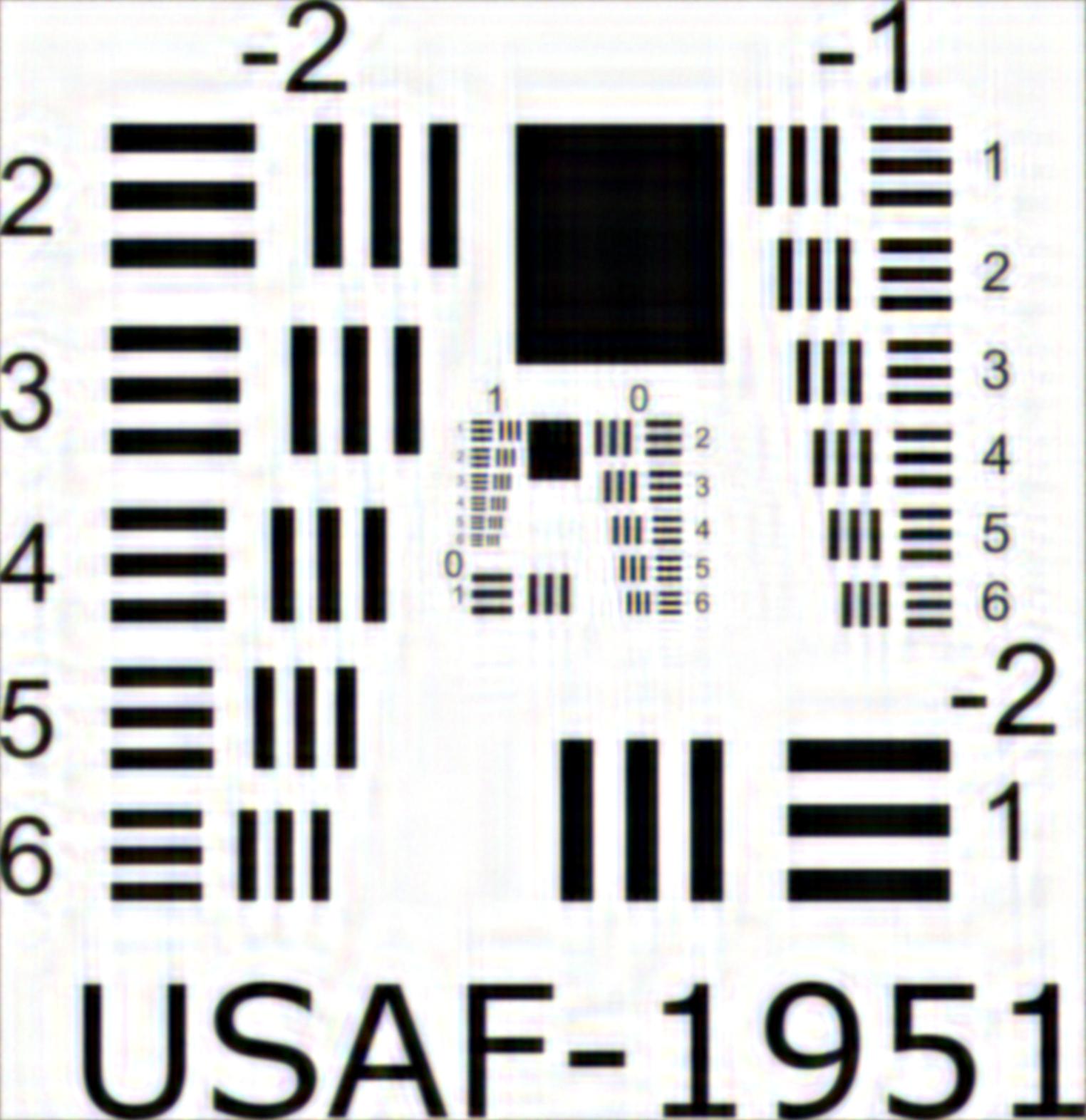}		
	\end{subfigure}
	\begin{subfigure}[t]{0.24\columnwidth}
		\centering
		\includegraphics[width=\columnwidth]{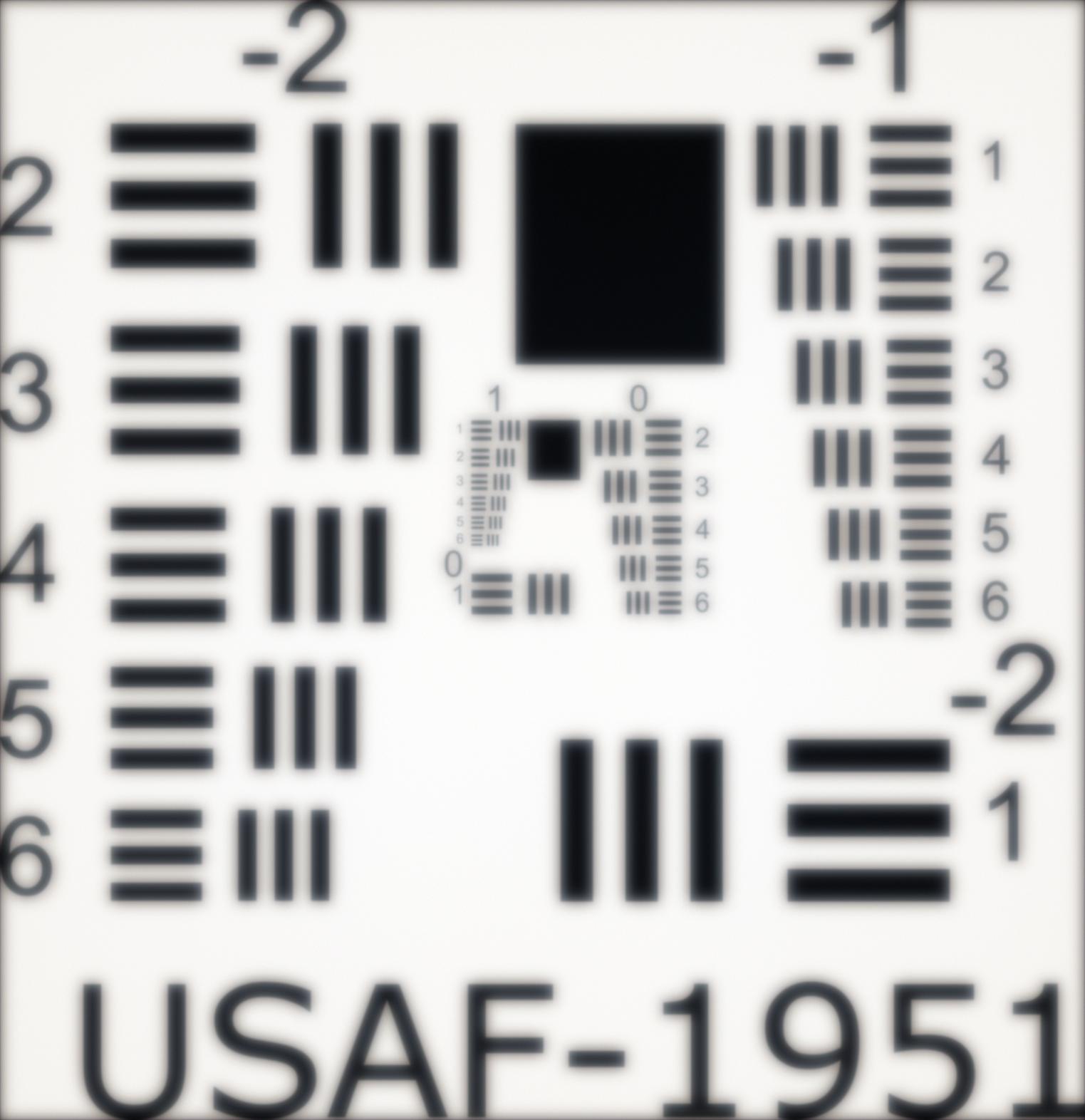}		
	\end{subfigure}
	\begin{subfigure}[t]{0.24\columnwidth}
		\centering
		\includegraphics[width=\columnwidth]{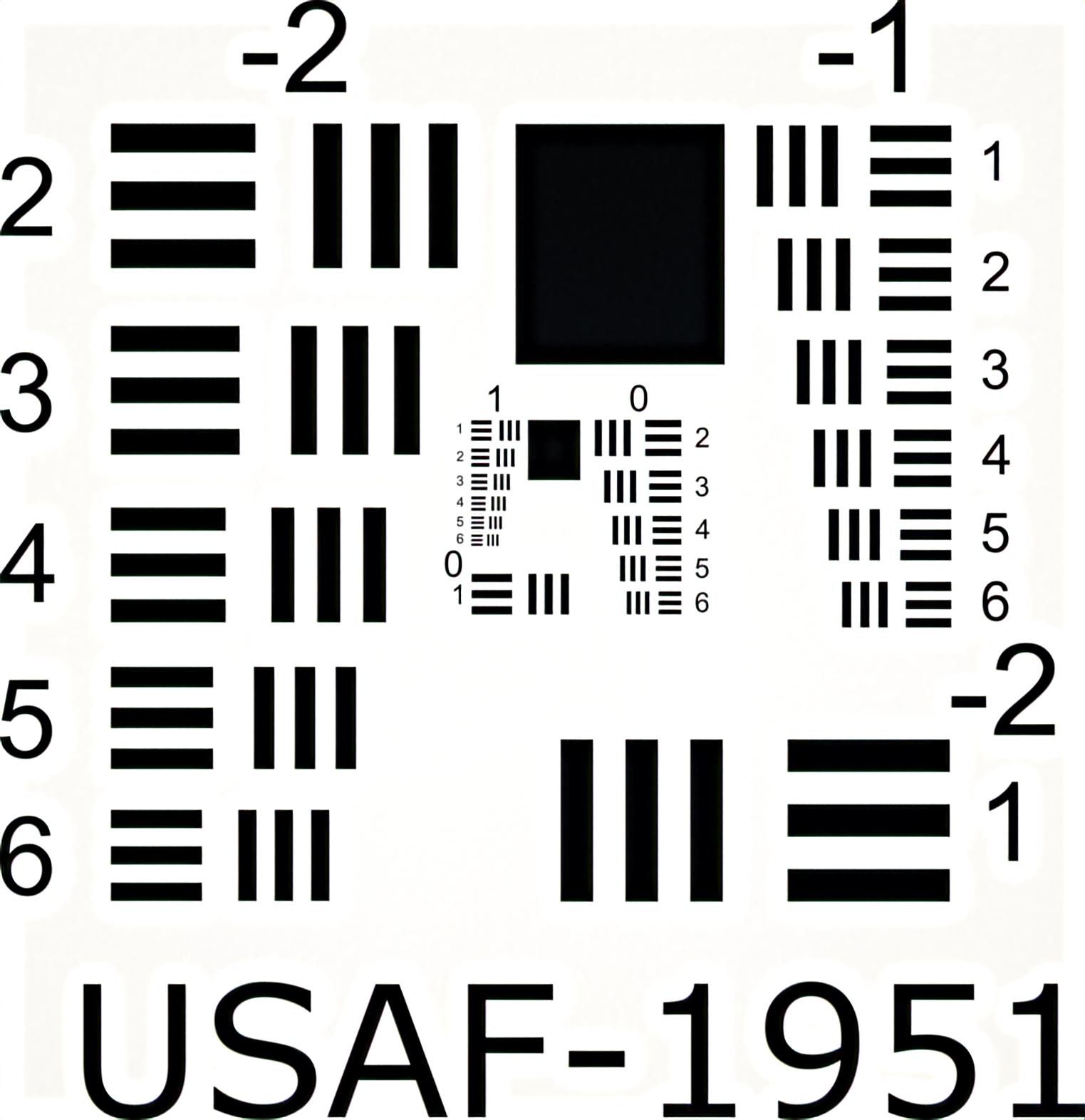}		
	\end{subfigure}
	\begin{subfigure}[t]{0.24\columnwidth}
		\centering
		\includegraphics[width=\columnwidth]{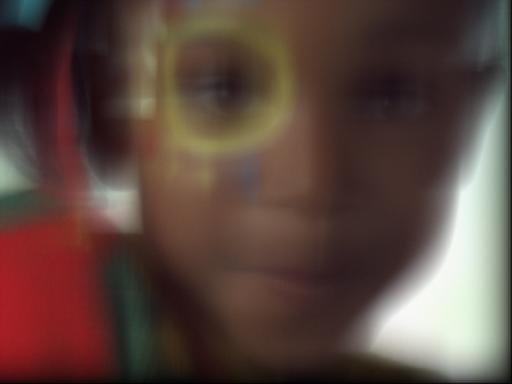}
        \put(-70,6){\rotatebox{90}{$z=\SI{1}{\meter}$}}		
	\end{subfigure}
	\vspace{1mm}
	\begin{subfigure}[t]{0.24\columnwidth}
		\centering
		\includegraphics[width=\columnwidth]{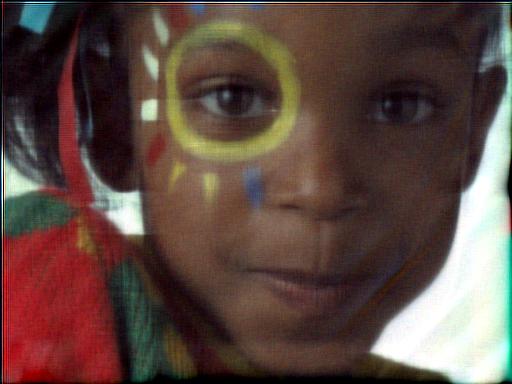}		
	\end{subfigure}
	\begin{subfigure}[t]{0.24\columnwidth}
		\centering
		\includegraphics[width=\columnwidth]{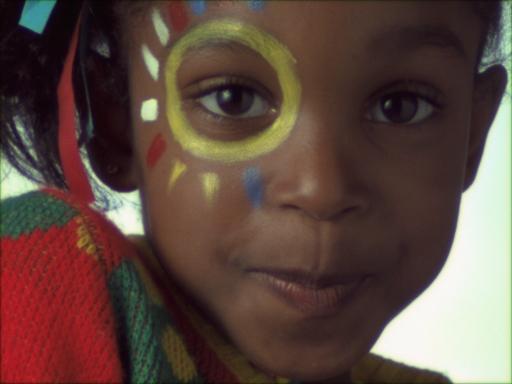}		
	\end{subfigure}
	\begin{subfigure}[t]{0.24\columnwidth}
		\centering
		\includegraphics[width=\columnwidth]{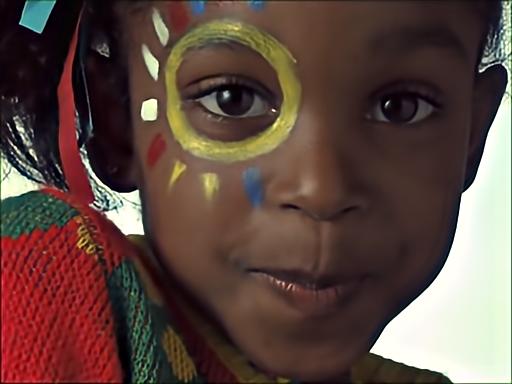}		
	\end{subfigure}
	\begin{subfigure}[t]{0.24\columnwidth}
		\centering
		\includegraphics[width=\columnwidth]{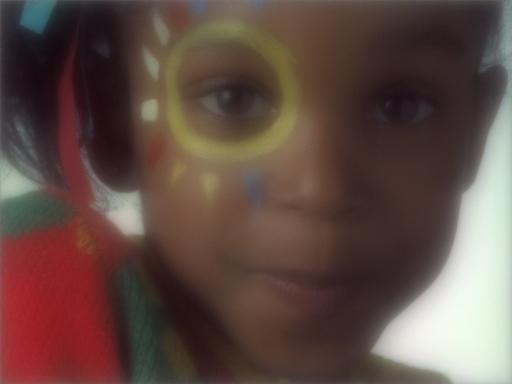}		
	\end{subfigure}
	\begin{subfigure}[t]{0.24\columnwidth}
		\centering
		\includegraphics[width=\columnwidth]{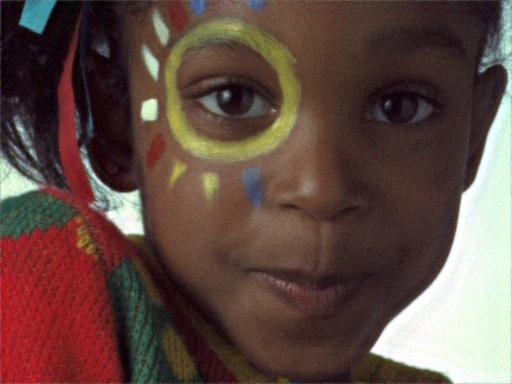}		
	\end{subfigure}
	\begin{subfigure}[t]{0.24\columnwidth}
		\centering
		\includegraphics[width=\columnwidth]{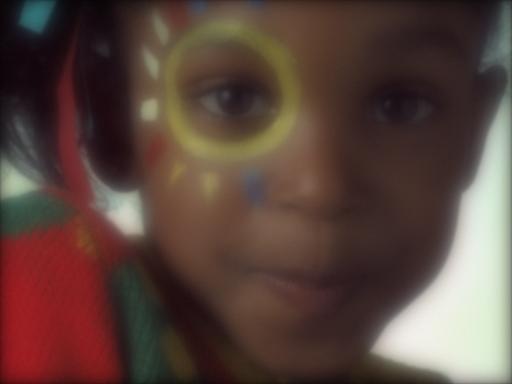}		
	\end{subfigure}
	\begin{subfigure}[t]{0.24\columnwidth}
		\centering
		\includegraphics[width=\columnwidth]{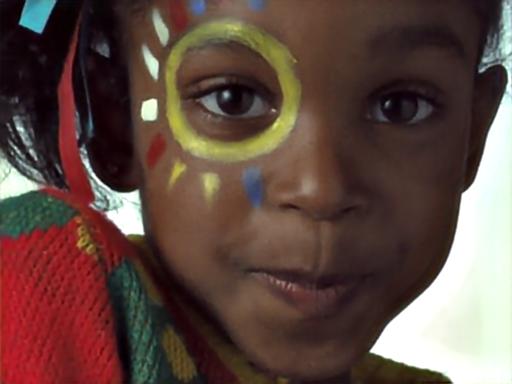}		
	\end{subfigure}
	\begin{subfigure}[t]{0.24\columnwidth}
		\centering
		\includegraphics[width=\columnwidth]{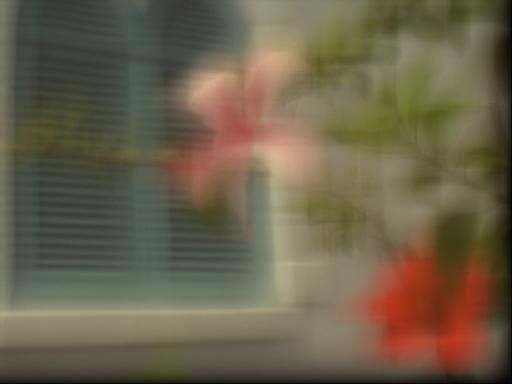}		
	\end{subfigure}
	\begin{subfigure}[t]{0.24\columnwidth}
		\centering
		\includegraphics[width=\columnwidth]{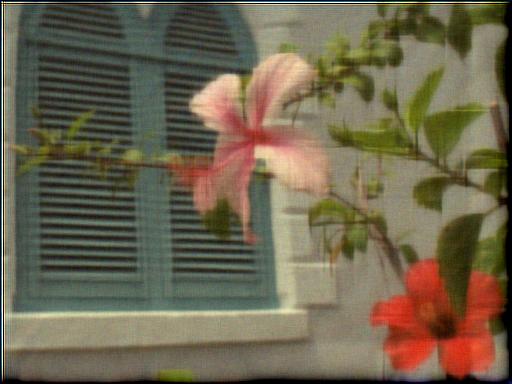}		
	\end{subfigure}
	\begin{subfigure}[t]{0.24\columnwidth}
		\centering
		\includegraphics[width=\columnwidth]{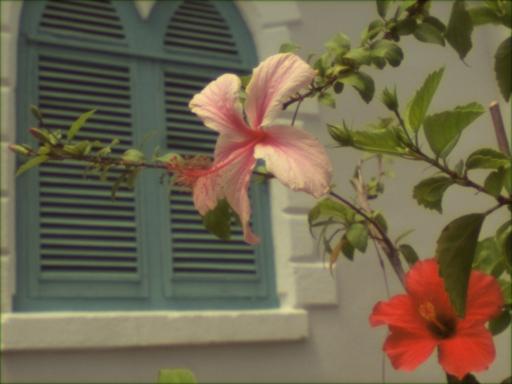}		
	\end{subfigure}
	\begin{subfigure}[t]{0.24\columnwidth}
		\centering
		\includegraphics[width=\columnwidth]{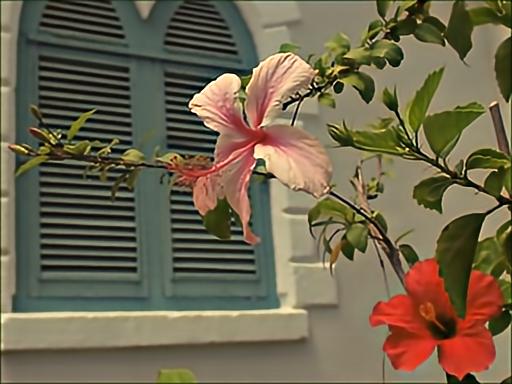}		
	\end{subfigure}
	\begin{subfigure}[t]{0.24\columnwidth}
		\centering
		\includegraphics[width=\columnwidth]{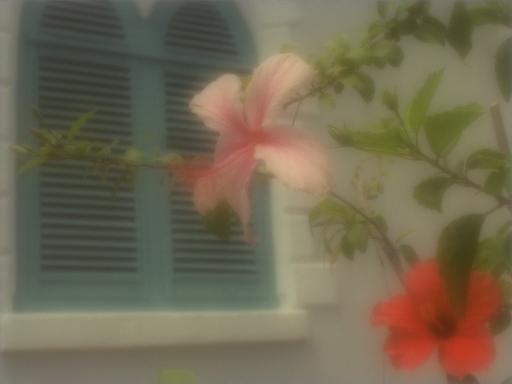}		
	\end{subfigure}
	\begin{subfigure}[t]{0.24\columnwidth}
		\centering
		\includegraphics[width=\columnwidth]{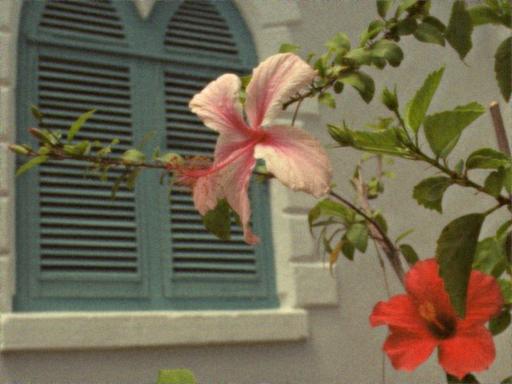}		
	\end{subfigure}
	\begin{subfigure}[t]{0.24\columnwidth}
		\centering
		\includegraphics[width=\columnwidth]{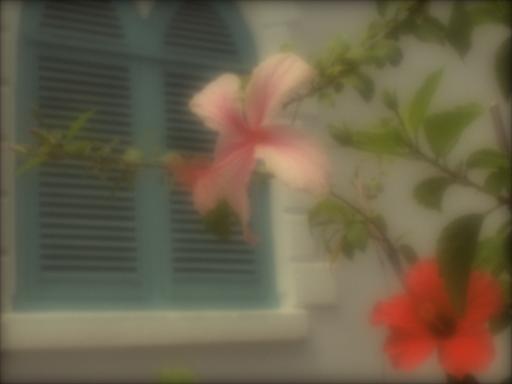}		
	\end{subfigure}
	\begin{subfigure}[t]{0.24\columnwidth}
		\centering
		\includegraphics[width=\columnwidth]{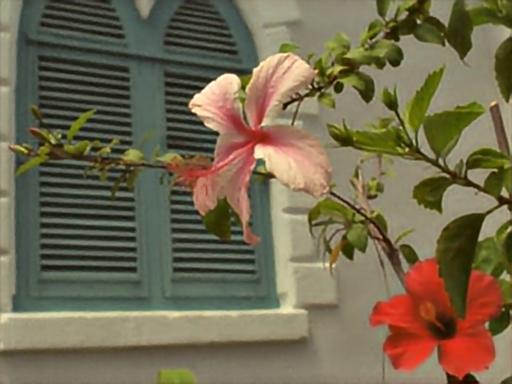}		
	\end{subfigure}
	\caption{Comparison of various simulated extended DoF imaging algorithms. From left to right: the cubic phase plate sensor image, the deblurring result with Wiener deconvolution, sensor image of Elmalem et. al. \cite{EDOFElmalem}, the deblurring result of their jointly-optimized post-processing CNN, the sensor image of Sitzmann et. al. \cite{EDOFSitzmann}, the result of their jointly-optimized Wiener deconvolution, the sensor image of the proposed method, and the output of the D-CNN.} 
	\label{fig:TestImage}
\end{figure*}

We provide the comparison of the proposed algorithm with the existing methods, as illustrated in Fig.~\ref{fig:TestImage}. In particular, the USAF test chart pattern and two natural images from Kodak natural image dataset are tested at three different depths. Three different approaches are considered for comparison. The cubic wavefront coding \cite{Cubic} is chosen as the conventional wavefront coding approach for DoF extension. The algorithms proposed by Elmalem et.al. \cite{EDOFElmalem} and Sitzmann et.al. \cite{EDOFSitzmann} are chosen as the more recent advanced methods based on the end-to-end learning framework. All the sensor images are simulated with $\sigma_s = 0.005$ and $\sigma_d = \SI{30}{\nm}$. The cubic phase-plate and the method of \cite{EDOFSitzmann} are deconvolved via Wiener filtering as done in the original works, while the post-processing of \cite{EDOFElmalem} is done with the network trained with their phase mask. The results demonstrate that the proposed method has the superior accuracy to the existing methods. In particular, the cubic wavefront coding \cite{Cubic} suffers from ringing artifacts, which are significant for the objects at \SI{0.5}{\meter}. Similar artifacts are visible for \cite{EDOFSitzmann}, though they are much less noticeable. This is because both methods utilize Wiener deconvolution as the post-processing, which is simply a fast Fourier transform (FFT)-based filter suffering from boundary value problem. Although this problem does not occur in \cite{EDOFElmalem}, their reconstruction results appear blurry for the objects at \SI{0.5}{\meter}. Please note, however, that \cite{EDOFElmalem} is optimized for a shallower defocus range of [0,8]. Their algorithm still performs well in the target range covering \SI{0.8}{\meter} and \SI{1}{\meter}.

 \begin{table}[h]
 \centering
 \begin{tabular}{|c|c|c|c|c|c|}
 \hline
 &  & \multicolumn{1}{c|}{\cite{Cubic}} & \multicolumn{1}{c|}{\cite{EDOFElmalem}} & \multicolumn{1}{c|}{\cite{EDOFSitzmann}} & \multicolumn{1}{c|}{Proposed}\\
 \hline
\parbox[t]{2mm}{\multirow{3}{*}{\rotatebox[origin=c]{90}{Scene 1}}} & $0.5m$ & 14.82 & 14.57 & 15.16 & \textbf{28.31} \\ 
 & $0.8m$ & 18.71 & 27.75 & 16.72 & \textbf{31.44} \\ 
 & $1m$ & 19.10 & 29.11 & 18.43 & \textbf{30.00} \\ 
\hline
\parbox[t]{2mm}{\multirow{3}{*}{\rotatebox[origin=c]{90}{Scene 2}}} & $0.5m$ & 20.18 & 20.12 & 25.45 & \textbf{30.98} \\ 
& $0.8m$ & 21.27 & 26.04 & 27.64 & \textbf{30.41} \\ 
 & $1m$ & 21.30 & 26.48 & 29.28 & \textbf{29.81} \\ 
\hline \parbox[t]{2mm}{\multirow{3}{*}{\rotatebox[origin=c]{90}{Scene 3}}} & $0.5m$ & 18.95 & 19.49 & 24.67 & \textbf{31.33} \\ 
& $0.8m$ & 19.98 & 26.20 & 27.43 & \textbf{31.07} \\ 
& $1m$ & 20.05 & 26.30 & 29.20 & \textbf{30.04} \\ 
\hline
\multicolumn{2}{|c|}{Avg. 24} & 17.77 & 21.13 & 25.64 & \textbf{26.26} \\ 
\hline
\multicolumn{2}{|c|}{Avg. 100} & 19.29 & 22.43 & 25.08 & \textbf{27.31} \\ 
\hline
\end{tabular}
\caption{Quantitative analysis of the DOF extension methods.}
\label{tbl:PSNR}
\end{table}

In addition to the qualitative results, we provide the quantitative analysis of the proposed method together with existing approaches. In particular, we derive the peak signal-to-noise-rations (PSNRs) for each test setup in Fig.~\ref{fig:TestImage}, as well as the average PSNRs for all images in the Kodak 24 data set and 100 test images of the BSDS500 data set \cite{BSDS500}. The results are given in Table~\ref{tbl:PSNR}. The average PSNR values are calculated by assigning a random depth to each image in the data set, uniformly chosen within the scene depth range of $[\SI{0.5}{\meter},\infty]$. The PSNR values are then averaged. The table demonstrates that the proposed method achieves better PSNR values, both for the individual images illustrated in Fig.~\ref{fig:TestImage}, and the average values in two different data sets. An important observation is that the USAF test chart image (Scene 1 in Table~\ref{tbl:PSNR}) is a failure case for \cite{EDOFSitzmann}. Although their method also targets a depth range of $[\SI{0.5}{\meter}, \infty)$ and demonstrates high-quality results for the natural images, it suffers from the ringing artifacts, blur and chromatic aberrations in the reconstructed test chart image. The superior performance of the proposed method, concerning both the demonstrated images and the average PSNR values of the utilized data sets, can be explained by a few critical factors. First, the optimization of DOE is not limited to a specific set of patterns (as in \cite{EDOFElmalem}); it is analytically defined based on the desired range of EDoF. Furthermore, as the imaging systems with hybrid refractive-diffractive optical elements are known to decrease the chromatic aberrations \cite{Ref-Diff}, it is natural to observe less artifacts (e.g., compared to \cite{EDOFSitzmann}) related to color aberrations (as especially seen in the USAF test chart ). Second, the proposed method utilizes CNN for the deblurring, which is expected to suppress the ringing artifacts and achieve better results 
than the Wiener filtering that is used in other compared methods. 

\subsubsection{Noise Analysis}

\begin{figure}[htbp]	
	\centering
	\begin{subfigure}[t]{0.22\columnwidth}
		\caption{$\sigma_s=0.005$}
		\centering
		\includegraphics[width=\columnwidth]{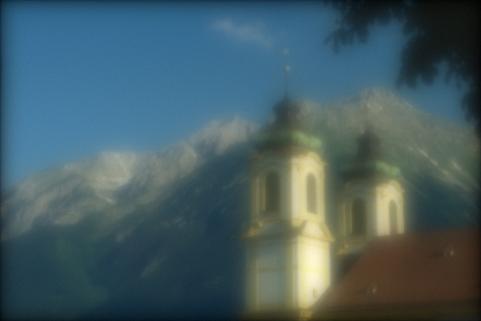}
	\end{subfigure}
	\vspace{1mm}
	\begin{subfigure}[t]{0.22\columnwidth}
		\caption{$\sigma_s=0.009$}
		\centering
		\includegraphics[width=\columnwidth]{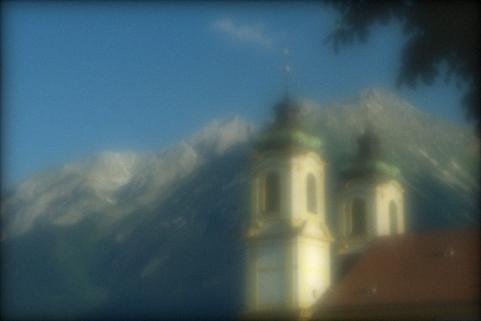}
	\end{subfigure}
	\begin{subfigure}[t]{0.22\columnwidth}
		\caption{$\sigma_s=0.015$}
		\centering
		\includegraphics[width=\columnwidth]{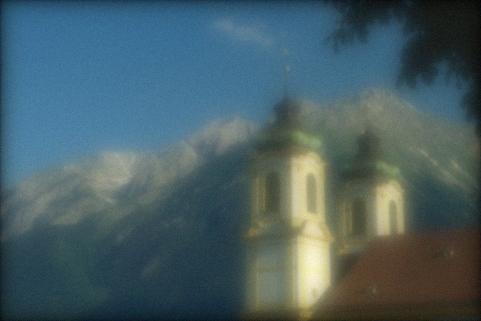}
	\end{subfigure}
	\begin{subfigure}[t]{0.22\columnwidth}
		\caption{$\sigma_s=0.02$}
		\centering
		\includegraphics[width=\columnwidth]{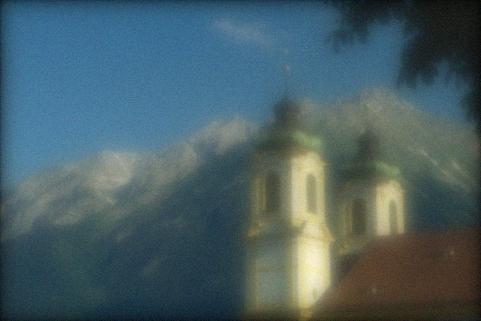}
	\end{subfigure}
	\begin{subfigure}[t]{0.22\columnwidth}
		\centering
		\includegraphics[width=\columnwidth]{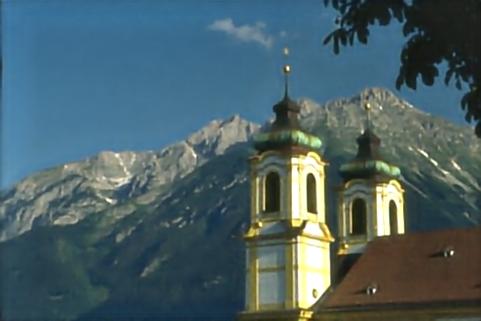}
		\caption{29.30}
		\label{sfig:noise5}
	\end{subfigure}
	\begin{subfigure}[t]{0.22\columnwidth}
		\centering
		\includegraphics[width=\columnwidth]{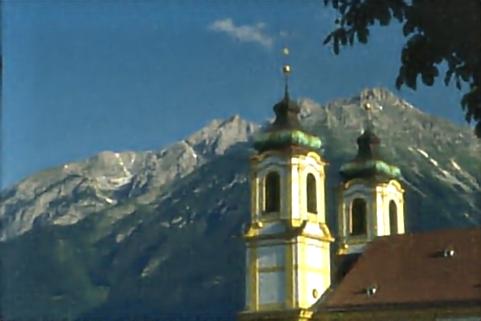}
		\caption{28.82}
		\label{sfig:noise9}
	\end{subfigure}
	\begin{subfigure}[t]{0.22\columnwidth}
		\centering
		\includegraphics[width=\columnwidth]{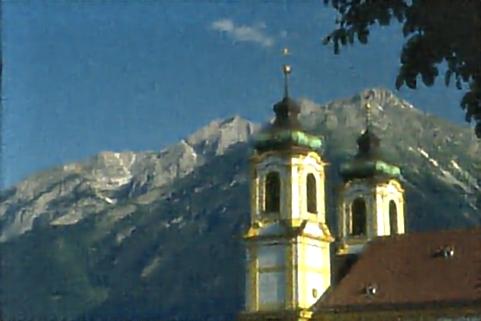}
		\caption{27.95}
		\label{sfig:noise15}
	\end{subfigure}
	\begin{subfigure}[t]{0.22\columnwidth}
		\centering
		\includegraphics[width=\columnwidth]{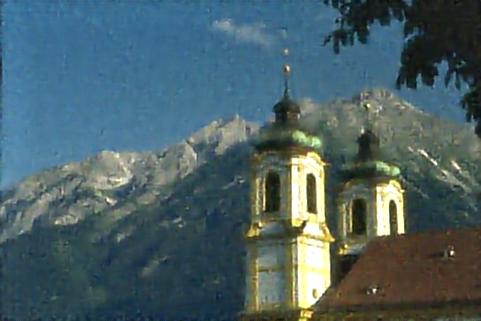}
		\caption{27.28}
		\label{sfig:noise20}
	\end{subfigure}
	\caption{Reconstruction results for increasing sensor read noise. Top: sensor outputs with different noise levels, bottom: post-processed images given with PSNR values.}
	\label{fig:Noise}
\end{figure}

The robustness of the proposed algorithm is tested against varying sensor image and DOE height map noise levels for both inside and outside the training noise limits. In particular, we evaluate the network output by increasing the sensor and the height-map noises respectively. The tested sensor noise levels are $\sigma_s=\{0.005,0.009.0.015,0.020\}$, where the height-map noise is assumed as $\sigma_d=\SI{30}{\nm}$. Fig.~\ref{fig:Noise} illustrates the results for a natural image located at $z=z_{f_G}$. As it can be inferred from the figure, the post-processing is robust within the noise limits introduced in the training, after which the image quality is observed to decrease. The robustness of the algorithm can be increased by broadening the noise limits during training; however this is expected to come at the cost of decrease in the overall image quality, as we have observed such effect in different sets of experiments. The noise limits therefore should be chosen in accordance with the sensor specifications; a sensor with small pixel size will introduce more noise. Please also note that no extra post-processing is applied to images in Fig. ~\ref{fig:Noise} for image denoising. Thus, the results can be actually improved by further utilizing denoising algorithms. 

\begin{figure}[htbp]	
	\centering
	\begin{subfigure}[t]{0.22\columnwidth}
		\caption{$\sigma_d=\SI{20}{\nm}$}
		\centering
		\includegraphics[width=\columnwidth]{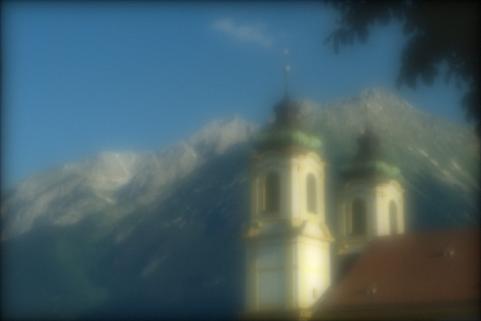}
	\end{subfigure}
	\vspace{1mm}
	\begin{subfigure}[t]{0.22\columnwidth}
		\caption{$\sigma_d=\SI{30}{\nm}$}
		\centering
		\includegraphics[width=\columnwidth]{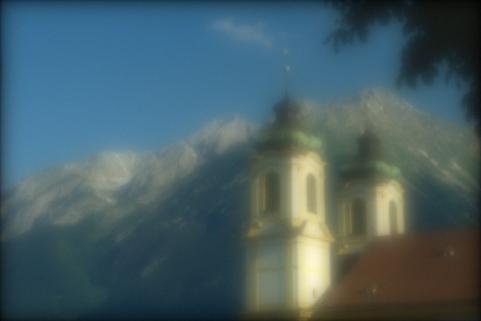}
	\end{subfigure}
	\begin{subfigure}[t]{0.22\columnwidth}
		\caption{$\sigma_d=\SI{40}{\nm}$}
		\centering
		\includegraphics[width=\columnwidth]{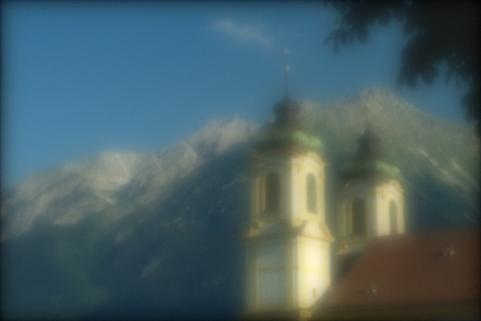}
	\end{subfigure}
	\begin{subfigure}[t]{0.22\columnwidth}
		\caption{$\sigma_d=\SI{50}{\nm}$}
		\centering
		\includegraphics[width=\columnwidth]{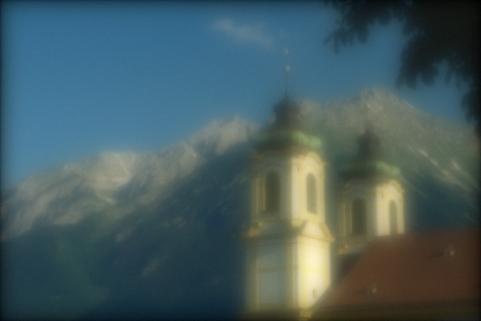}
	\end{subfigure}
	\begin{subfigure}[t]{0.22\columnwidth}
		\centering
		\includegraphics[width=\columnwidth]{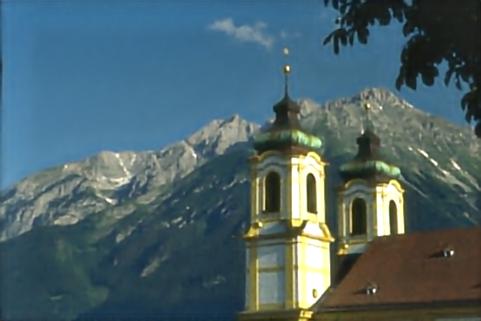}
		\caption{29.31}
		\label{sfig:hnoise20}
	\end{subfigure}
	\begin{subfigure}[t]{0.22\columnwidth}
		\centering
		\includegraphics[width=\columnwidth]{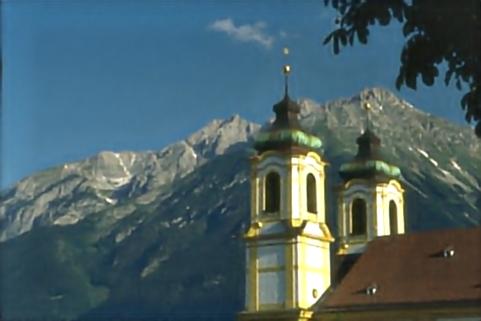}
		\caption{29.30}
		\label{sfig:hnoise30}
	\end{subfigure}
	\begin{subfigure}[t]{0.22\columnwidth}
		\centering
		\includegraphics[width=\columnwidth]{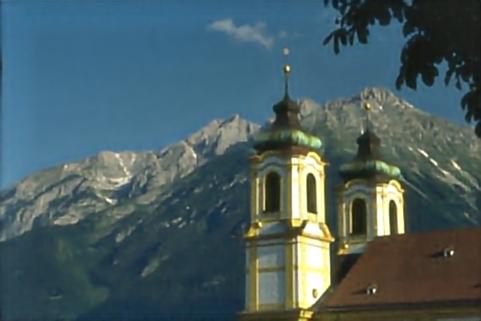}
		\caption{28.07}
		\label{sfig:hnoise40}
	\end{subfigure}
	\begin{subfigure}[t]{0.22\columnwidth}
		\centering
		\includegraphics[width=\columnwidth]{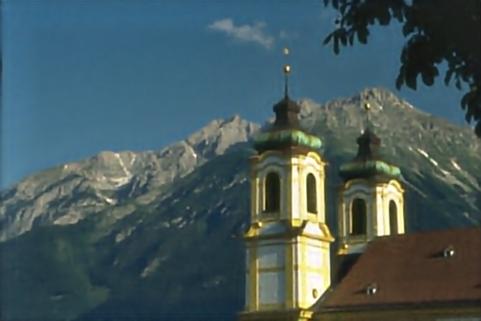}
		\caption{25.76}
		\label{sfig:hnoise50}
	\end{subfigure}
	\caption{Reconstruction results for increasing height-map noise. Top: sensor outputs with different noise levels, bottom: post-processed images given with PSNR values.}
	\label{fig:NoiseHmap}
\end{figure}

In the second noise analysis, the DOE height map fabrication noise levels of $\sigma_d=\{\SI{20}{\nm},\SI{30}{\nm},\SI{40}{\nm},\SI{50}{\nm}\}$ are tested, where the sensor noise is set as $\sigma_s=0.005$. Fig.~\ref{fig:NoiseHmap} shows the results for the same natural image located at $z=z_{f_G}$. We observe the effect of the noise in the phase mask as hazing in the image domain, which becomes significant at and above the training noise limit of $\sigma_d=\SI{40}{\nm}$. Additional dehazing algorithms may improve the results.

\subsubsection{Broadband Imaging}

\begin{figure}[htbp]
\hspace{-4mm}
\begin{subfigure}[t]{0.45\columnwidth}
    \centering
    \includegraphics[width=\columnwidth]{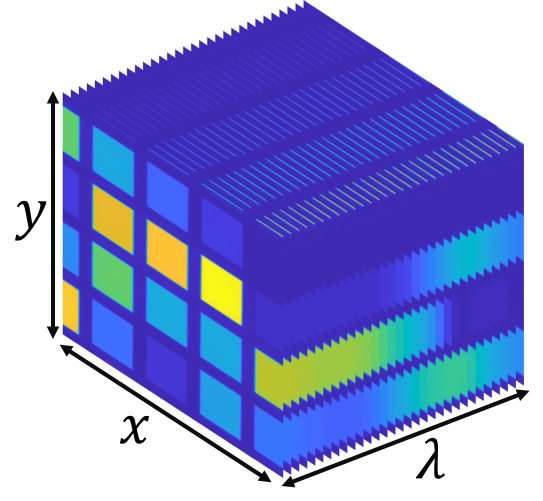}
\end{subfigure}
\begin{subfigure}[t]{0.45\columnwidth}
    \definecolor{myred}{rgb}{0.8078,0.1255,0.1608}%
\definecolor{mygreen}{rgb}{0.1608,0.8078,0.1255}%
\definecolor{myblue}{rgb}{0.1255,0.1608,0.8078}%

%
%
\begin{tikzpicture}

\begin{axis}[%
width=\columnwidth,
height=0.7\columnwidth,
at={(0\columnwidth,0\columnwidth)},
scale only axis,
xmin=400,
xmax=750,
xlabel style={font=\color{white!15!black}},
xlabel={Wavelength (\SI{}{\nm})},
ymin=0,
ymax=0.1,
ytick = {0,0.02,0.04,0.06,0.08,0.1},
yticklabels = {0,0.02,0.04,0.06,0.08,0.1},
axis background/.style={fill=white},
title style={font=\bfseries},
title={Sensor Spectral Response},
xmajorgrids,
ymajorgrids,
legend style={legend cell align=left, align=left, draw=white!15!black}
]
\addplot [color=myred, line width=1.4pt]
  table[row sep=crcr]{%
420	0.00943574259294206\\
430	0.00943574259294206\\
440	0.00943574259294206\\
450	0.0108511039818834\\
460	0.013681826759766\\
470	0.0160407624080015\\
480	0.018399698056237\\
490	0.0212304208341196\\
500	0.022645782223061\\
510	0.0235893564823552\\
520	0.0259482921305907\\
530	0.0283072277788262\\
540	0.0306661634270617\\
550	0.0306661634270617\\
560	0.0306661634270617\\
570	0.0283072277788262\\
580	0.0330250990752972\\
590	0.0566144555576524\\
600	0.0872806189847141\\
610	0.0990752972258917\\
620	0.0896395546329496\\
630	0.073127005095301\\
640	0.0589733912058879\\
650	0.0471787129647103\\
660	0.0353840347235327\\
670	0.0259482921305907\\
680	0.0212304208341196\\
690	0.0141536138894131\\
700	0.0117946782411776\\
710	0.00943574259294206\\
720	0.00783166635214191\\
};
\addlegendentry{R}

\addplot [color=mygreen, line width=1.4pt]
  table[row sep=crcr]{%
420	0.0081888246628131\\
430	0.00915221579961464\\
440	0.0108381502890173\\
450	0.0120423892100193\\
460	0.016859344894027\\
470	0.0264932562620424\\
480	0.0385356454720617\\
490	0.0433526011560694\\
500	0.0529865125240848\\
510	0.0674373795761079\\
520	0.0842967244701349\\
530	0.0939306358381503\\
540	0.0963391136801541\\
550	0.0915221579961464\\
560	0.0842967244701349\\
570	0.0722543352601156\\
580	0.0602119460500963\\
590	0.0469653179190751\\
600	0.0313102119460501\\
610	0.0180635838150289\\
620	0.00963391136801541\\
630	0.00602119460500963\\
640	0.00481695568400771\\
650	0.00361271676300578\\
660	0.00240847784200385\\
670	0.00240847784200385\\
680	0.00240847784200385\\
690	0.00240847784200385\\
700	0.00120423892100193\\
710	0\\
720	0\\
};
\addlegendentry{G}

\addplot [color=myblue, line width=1.4pt]
  table[row sep=crcr]{%
420	0.0564826700898588\\
430	0.061617458279846\\
440	0.0718870346598203\\
450	0.0821566110397946\\
460	0.086007702182285\\
470	0.0911424903722721\\
480	0.0949935815147625\\
490	0.0937098844672657\\
500	0.0847240051347882\\
510	0.0718870346598203\\
520	0.0577663671373556\\
530	0.0410783055198973\\
540	0.0282413350449294\\
550	0.0192554557124519\\
560	0.0128369704749679\\
570	0.0102695763799743\\
580	0.00898587933247753\\
590	0.00770218228498074\\
600	0.00641848523748395\\
610	0.00513478818998716\\
620	0.00385109114249037\\
630	0.00256739409499358\\
640	0.00128369704749679\\
650	0\\
660	0\\
670	0\\
680	0\\
690	0\\
700	0\\
710	0\\
720	0\\
};
\addlegendentry{B}

\end{axis}

\end{tikzpicture}%
\end{subfigure}
\caption{Broadband imaging simulation setup. Left: A sample hyper-spectral image (Data set from \cite{HSDataset}). Right: Example sensor spectral response (Adopted from \cite{KodakSensor}).}
\label{fig:BroadbandImaging}
\end{figure}
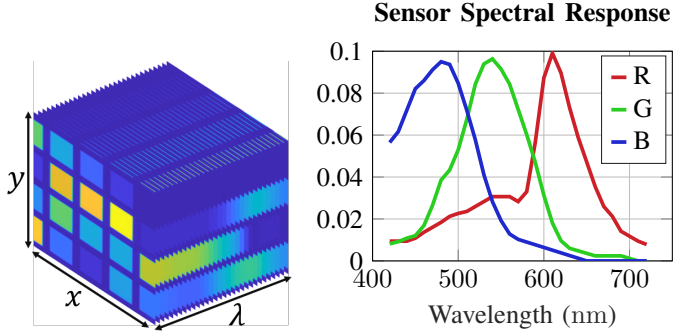

\begin{figure}[htbp]	
	\centering
	\begin{subfigure}[t]{0.48\columnwidth}
	    \caption{Colorchart}
	    \centering
        \includegraphics[width=0.50\columnwidth]{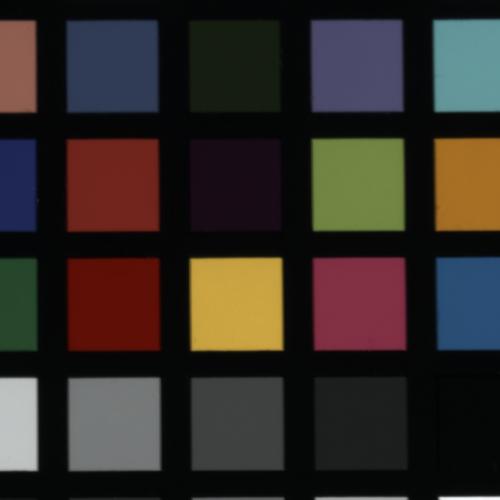}
        \label{sfig:gtColorchart}
	\end{subfigure}
    \begin{subfigure}[t]{0.48\columnwidth}
        \caption{Butterfly}
        \centering
        \includegraphics[width=0.50\columnwidth]{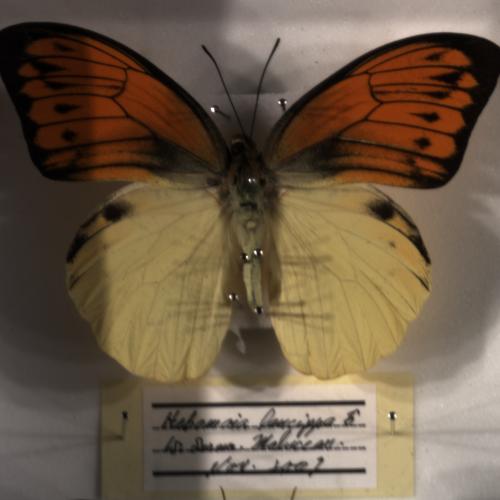}
        \label{sfig:gtButterfly}
    \end{subfigure}
    \begin{subfigure}[t]{0.24\columnwidth}
		\caption{Sensor}
		\centering
		\includegraphics[width=\columnwidth]{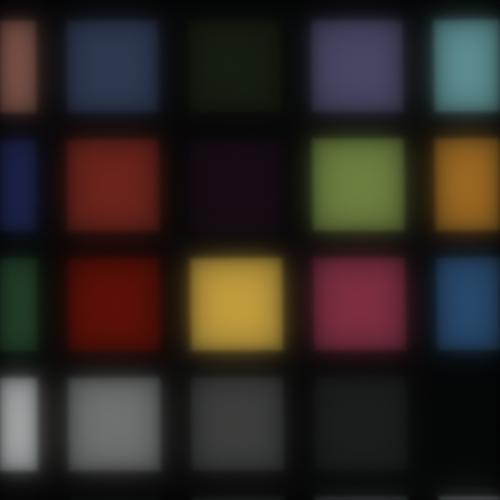}
        \put(-70,8){\rotatebox{90}{$z=\SI{0.5}{\meter}$}}
		\caption{PSNR}
	\end{subfigure}
	\begin{subfigure}[t]{0.24\columnwidth}
		\caption{Output}
		\centering
		\includegraphics[width=\columnwidth]{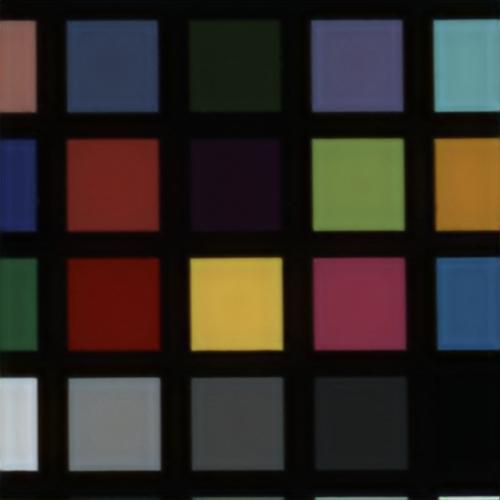}
		\caption{37.09}
	\end{subfigure}
	\begin{subfigure}[t]{0.24\columnwidth}
	    \caption{Sensor}
		\centering
		\includegraphics[width=\columnwidth]{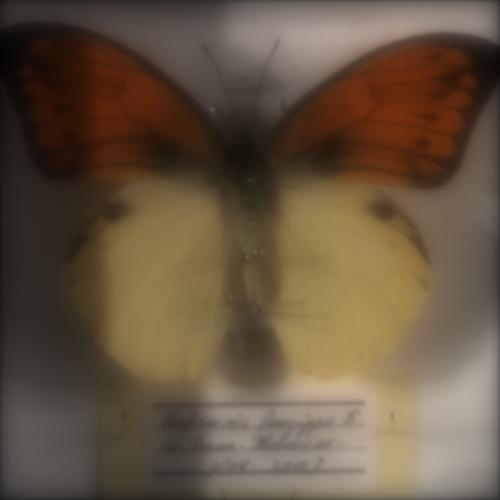}
	\end{subfigure}
	\begin{subfigure}[t]{0.24\columnwidth}
		\caption{Output}
		\centering
		\includegraphics[width=\columnwidth]{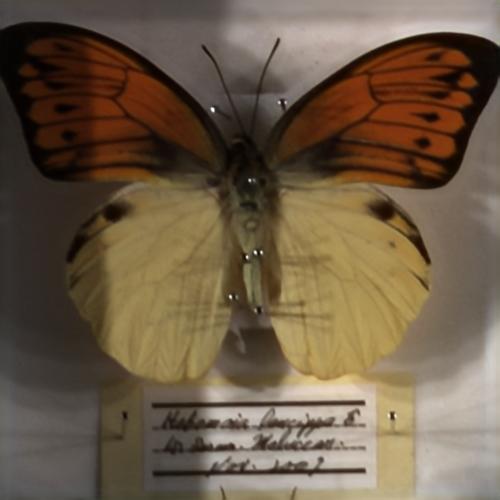}	
		\caption{36.55}
	\end{subfigure}
	\begin{subfigure}[t]{0.24\columnwidth}
		\centering
		\includegraphics[width=\columnwidth]{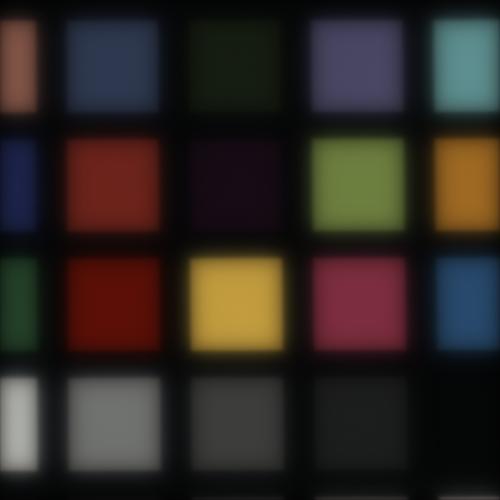}
        \put(-70,10){\rotatebox{90}{$z=\SI{1}{\meter}$}}
	\end{subfigure}
	\begin{subfigure}[t]{0.24\columnwidth}
		\centering
		\includegraphics[width=\columnwidth]{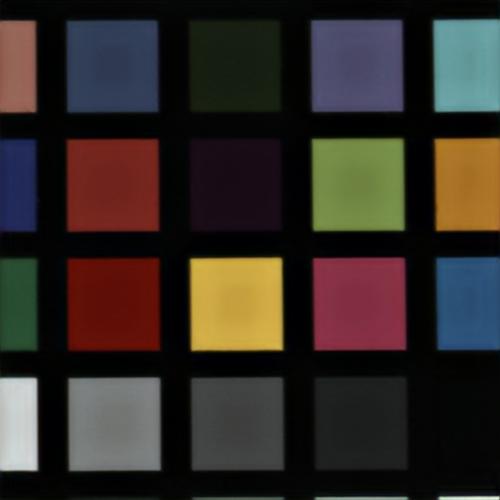}		
	    \caption{35.31}
	\end{subfigure}
	\begin{subfigure}[t]{0.24\columnwidth}
		\centering
		\includegraphics[width=\columnwidth]{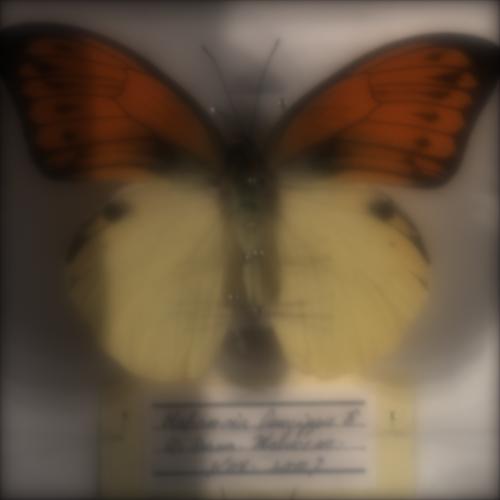}
	\end{subfigure}
	\begin{subfigure}[t]{0.24\columnwidth}
		\centering
		\includegraphics[width=\columnwidth]{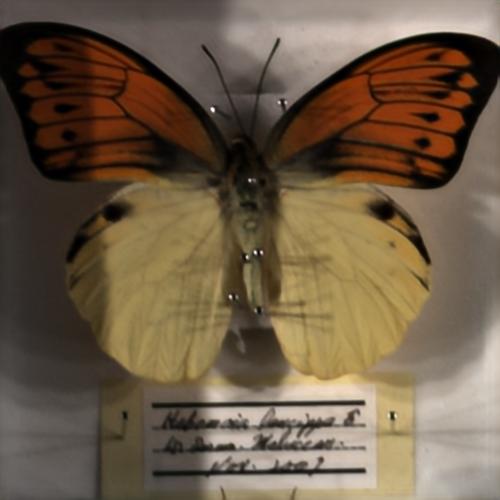}
		\caption{35.12}
	\end{subfigure}
	\begin{subfigure}[t]{0.24\columnwidth}
		\centering
		\includegraphics[width=\columnwidth]{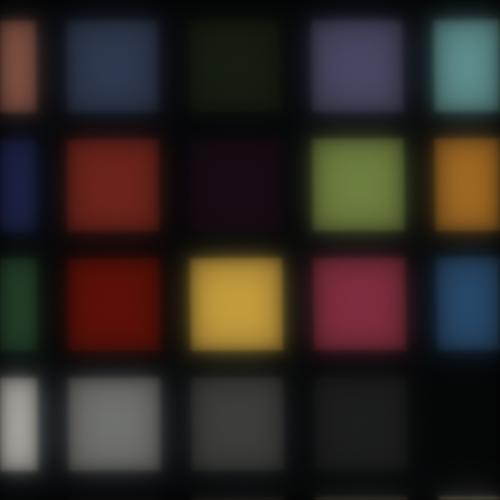}
        \put(-68,14){\rotatebox{90}{$z=\infty$}}		
	\end{subfigure}
	\begin{subfigure}[t]{0.24\columnwidth}
		\centering
		\includegraphics[width=\columnwidth]{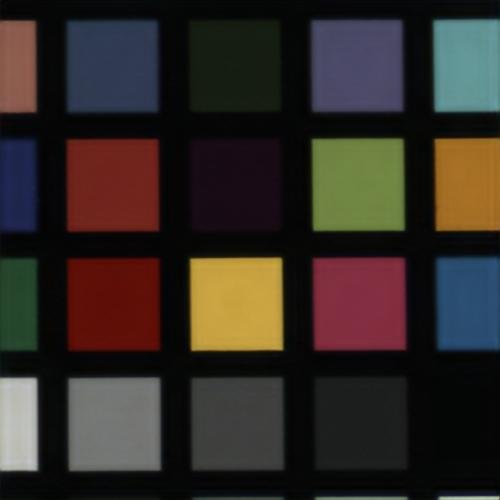}
		\caption{36.46}
	\end{subfigure}
	\begin{subfigure}[t]{0.24\columnwidth}
		\centering
		\includegraphics[width=\columnwidth]{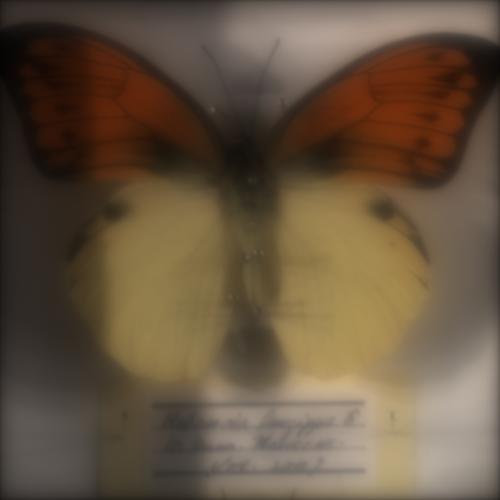}
	\end{subfigure}
	\begin{subfigure}[t]{0.24\columnwidth}
		\centering
		\includegraphics[width=\columnwidth]{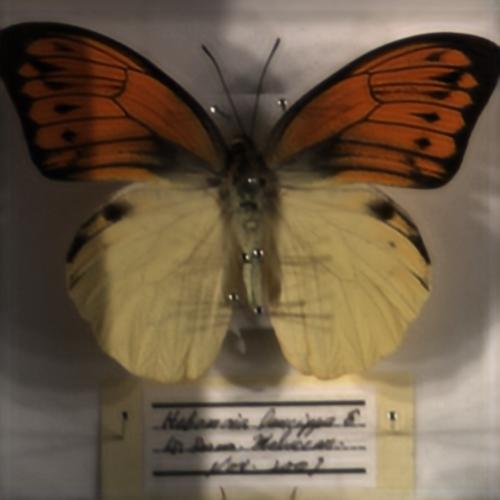}
		\caption{35.77}
	\end{subfigure}
	\caption{Broadband imaging results for two hyper-spectral data. Top row: Ground-truth color images are obtained by taking the weighted average of each color channel by the sensor spectral response (see Fig.~\ref{fig:BroadbandImaging}). Second to last rows: The sensor images and the network outputs for object depths of $\SI{0.5}{\meter},\SI{1}{\meter}$, and $\infty$, respectively.} 
	\label{fig:TestImageMS}
\end{figure}

Although the method is trained and tested above for three distinct wavelengths standing for the red, green, and blue channels of the sensor, in reality a sensor pixel integrates over a continuous spectrum range, based on its spectral response. Therefore, here we test the broadband imaging performance of the algorithm to investigate possible artifacts (e.g., chromatic aberrations) introduced due to the discrete wavelength assumption in the imaging model. For this purpose, a hyperspectral image data set \cite{HSDataset} is used as a source for broadband test images, in which each image has 31 equally spaced bands from \SI{420}{\nm} to \SI{720}{\nm} (a hyperspectral image is represented as a data cube as illustrated in Fig.~\ref{fig:BroadbandImaging}, left). In the camera model, the sensor image is first found for each channel by convolving the wavelength-dependent PSF with the corresponding channel of the hyperspectral image. The RGB sensor image is then formed by weighted average of each channel based on the spectral response shown in Fig.~\ref{fig:BroadbandImaging}, right. The reconstruction results are shown in Fig.~\ref{fig:TestImageMS} for two hyperspectral test images at three different depths. We report the PSNR values of each case, comparing the network output with the ground-truth color image, which is defined as the weighted average of the original hyperspectral image with the sensor response. The proposed method is observed to perform well also in the broadband imaging scenario, providing achromatic DoF extension.

\subsubsection{3D Scene}

\begin{figure}[htbp]	
	\centering
	\begin{subfigure}[t]{0.45\columnwidth}
		\caption{Image}
		\centering
		\includegraphics[width=\columnwidth]{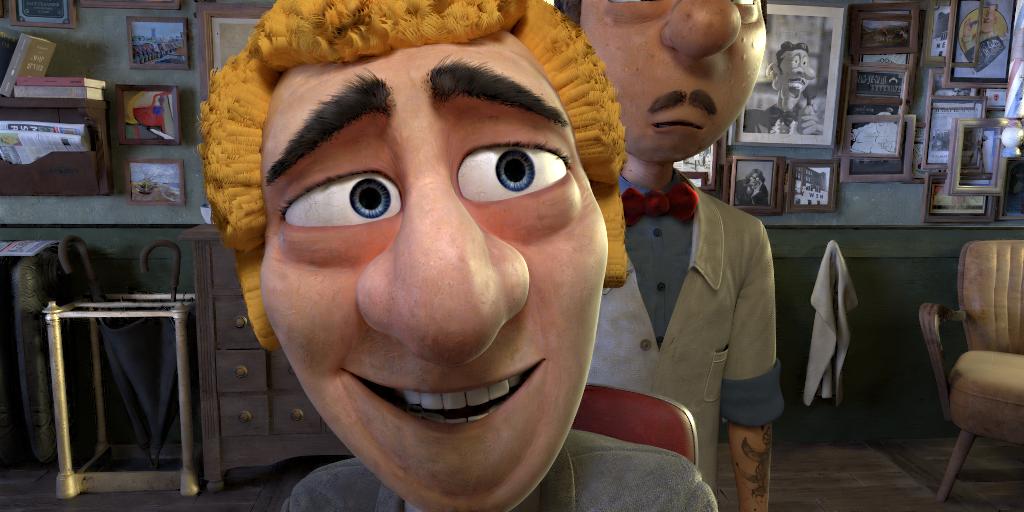}
	\end{subfigure}
	\begin{subfigure}[t]{0.45\columnwidth}
	    \caption{Depth}
		\centering
		\includegraphics[width=\columnwidth]{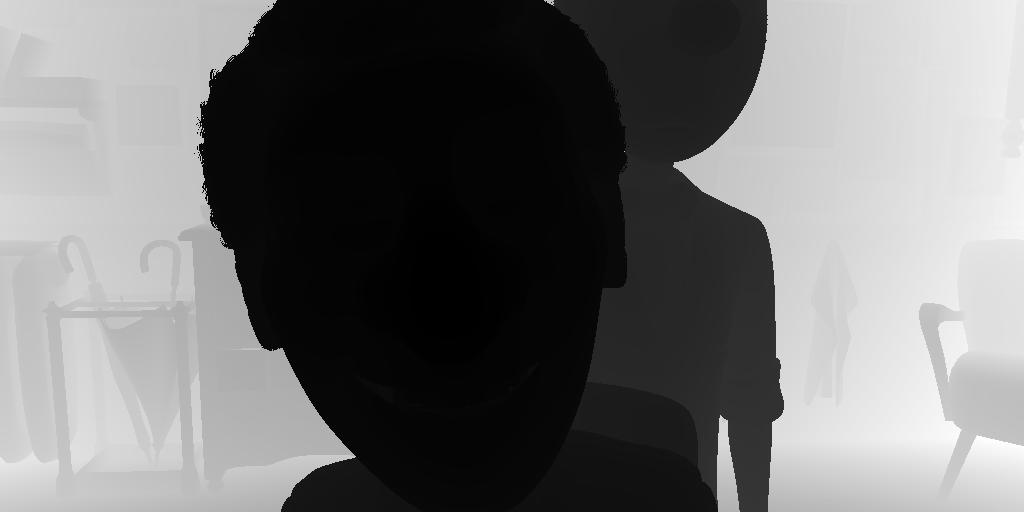}
		\caption{$ z \in (\SI{0.5}{\meter},\SI{4.46}{\meter})$}
	\end{subfigure}
	\begin{subfigure}[t]{0.45\columnwidth}
		\caption{Sensor}
		\centering
		\includegraphics[width=\columnwidth]{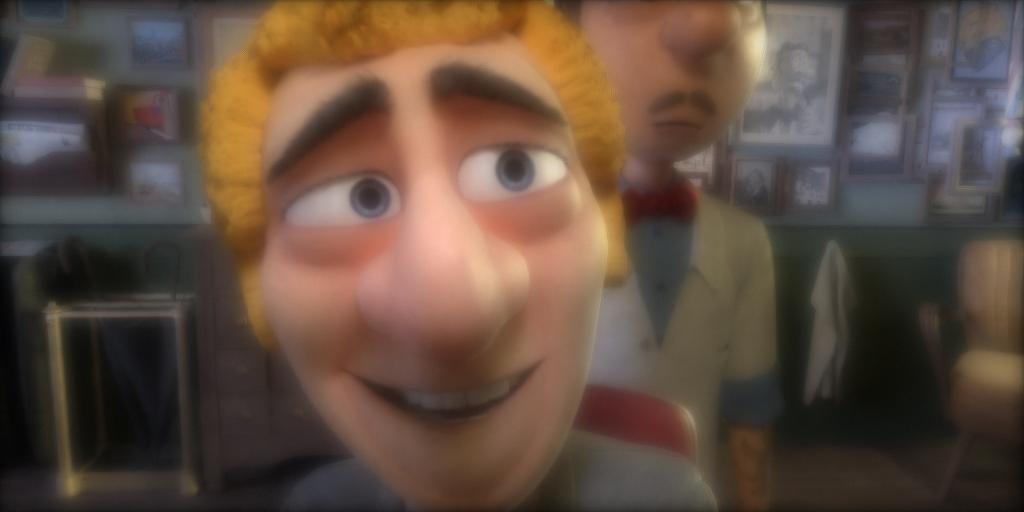}
	\end{subfigure}
	\begin{subfigure}[t]{0.45\columnwidth}
		\centering
		\caption{Output}
		\includegraphics[width=\columnwidth]{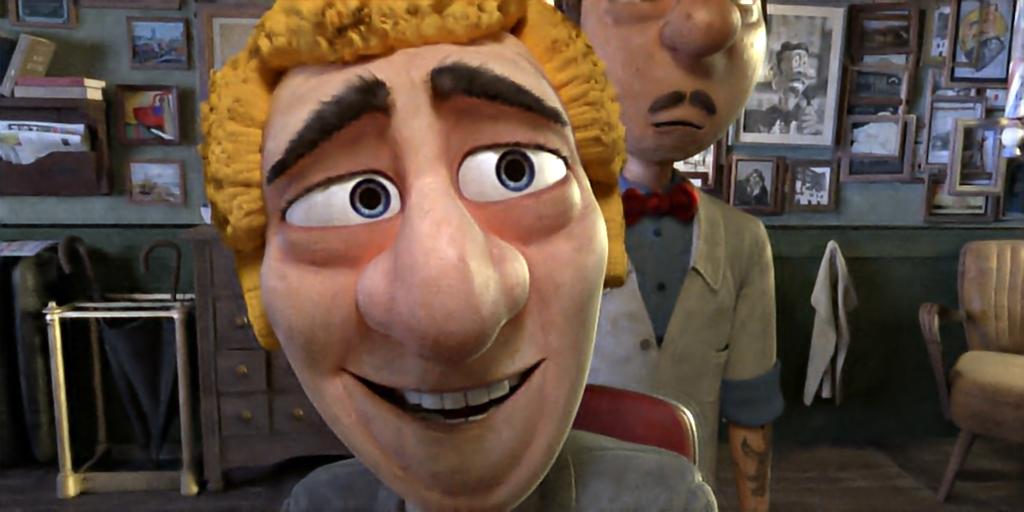}
	    \caption{PSNR=30.76}
	\end{subfigure}
	\caption{Reconstruction results with synthetic 3D scene.}
	\label{fig:Test3D}
\end{figure}

During training, we simulate the sensor images by assuming planar objects at constant depth. Such approach simplifies the forward model to a single convolution between the image and the PSF calculated at the corresponding depth, which in turn speeds up the training. A more rigorous imaging model, however, should incorporate the pixel-wise depth map as well, from which the sensor image is obtained through the depth-dependent convolution via Eq.~\ref{eq:Forwardmodel} and~\ref{eq:SensorImage}. To test our approach on a more realistic scene model, we utilize TAU-Agent data set \cite{DepthEstPhase}, which consists of synthetic sharp images together with accurate pixel-wise depth maps created in Blender. Fig.~\ref{fig:Test3D} shows an example image from the data set with its depth map. The scene depth range is shifted with respect to the original so that the closest object is located at \SI{0.5}{\meter}, where the furthest object is at \SI{4.46}{\meter}. The depth values are quantized with \SI{1}{\mm} steps, which corresponds to 475 different depth values within the scene. The simulated sensor image and the output of the post-processing are presented in the figure. The reconstructed image appears free from ringing artifacts, which are likely to occur in most conventional extended DoF imaging methods.

\section{Experimental Results}
\label{sec:Experiments}

\begin{figure*}[t]
\includegraphics[width=\textwidth]{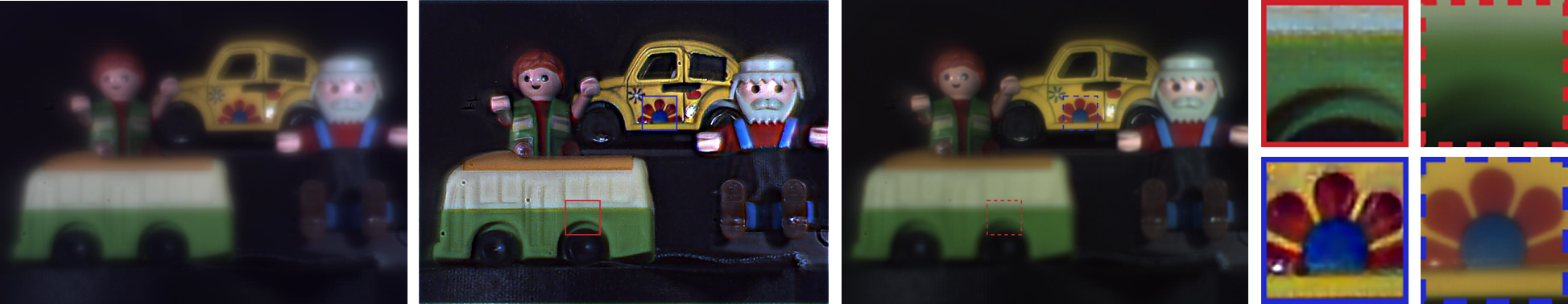}
\caption{Experimental results. From left to right: sensor image taken with the proposed implementation, output of the D-CNN on the sensor image taken with the proposed method, sensor image taken with the refractive lens-only, zoomed-in images of the network output and the sensor image of refractive lens-only, corresponding to the defocused and focused regions.}
\label{fig:Experiment3D}
\end{figure*}

\begin{figure}[htbp]
\centering
\includegraphics[width=0.9\columnwidth]{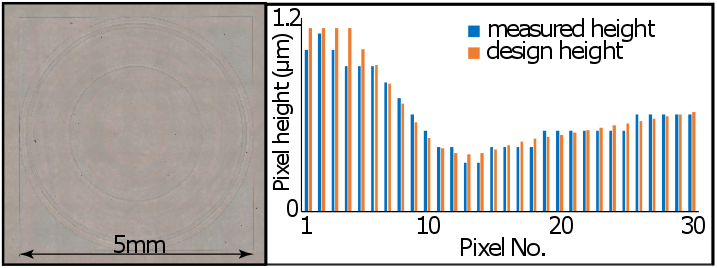}
\caption{Fabrication of the DOE. Left: optical micrograph of the fabricated DOE. The size of the device is approximately $\SI{5}{\mm} \times \SI{5}{\mm}$. Right: to estimate the fabrication error, the heights of 30 randomly selected pixels were measured. By comparing to the design values, we estimate the average and standard deviation of height-map error as \SI{44}{\nm} and \SI{49.5}{\nm}, respectively.}
\label{fig:Hmapfabrication}
\end{figure}

The EDoF capabilities of the proposed approach is tested through a DOE that is fabricated onto a positive-tone photoresist, S1813 (Microchem, GmbH) \cite{S1813} using grayscale lithography. The optical micro-graph of the fabricated DOE is shown in Fig.~\ref{fig:Hmapfabrication}. By exposing the photoresist with a spatially modulated exposure (based upon a previous calibration step), one can generate 3D structures after development \cite{Meem2020,meem2020multi}. The photopolymer is spun at 1000 rpm for \SI{60}{\second}, then baked on a hot-plate at \SI{110}{\celsius} for \SI{30}{\minute} in convection oven. After exposure, it is developed in AZ developer 1:1 \cite{Microchemicals} for \SI{40}{\second}, rinsed with DI water and dried with N2. A \SI{50.8}{\mm}-diameter soda-lime glass wafer of thickness \SI{0.55}{\mm} is used as the support substrate. The minimum feature width, maximum feature depth and number of gray-levels in this design are \SI{3}{\um}, \SI{1.2}{\um} and \SI{98}{\um}, respectively. To estimate the fabrication error, we measure the height of 30 randomly selected pixels and compare them with the corresponding design heights, shown in Fig.~\ref{fig:Hmapfabrication}. The average and standard deviation of height error are found to be \SI{44}{\nm} and \SI{49.5}{\nm}, respectively.

\begin{figure}[htbp]
    \centering
    \includegraphics[clip, trim=105 125 55 120, width=\columnwidth]{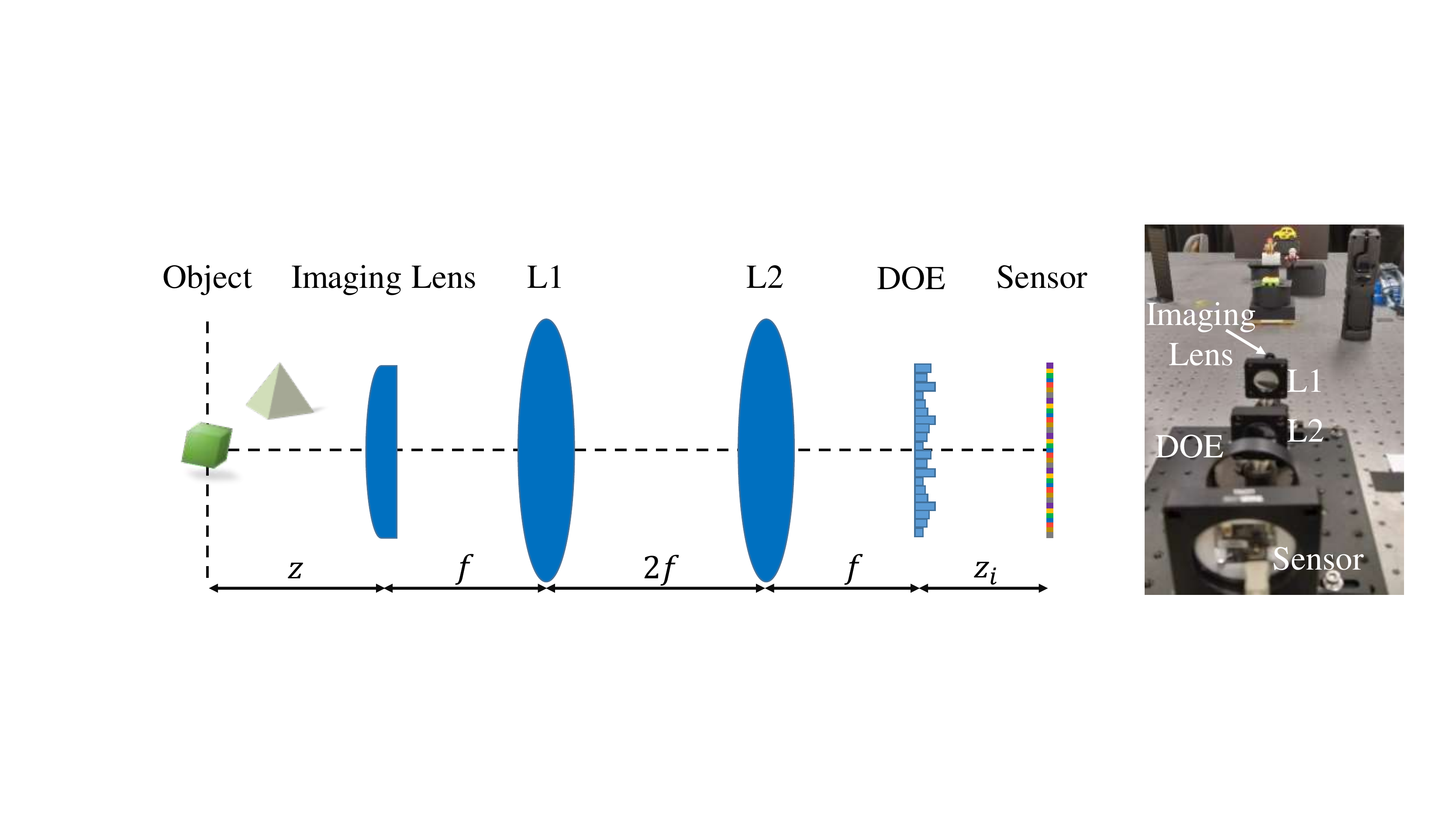}
    \caption{Experimental setup. Left: for imaging, the imaging lens and the DOE was placed in the front and back focal plane of a “4f-system”. Right: Photograph of the setup. }
    \label{fig:Setup}
\end{figure}

We set a benchtop optical setup to test our EDoF camera in real world imagery. The schematic of the setup is shown in Fig.~\ref{fig:Setup}, left. We employ a “4f-system” using two identical lenses (AC254-050-A, Thorlabs), L1 and L2, where the imaging lens (\SI{6}{\mm} Dia $\times$ \SI{36}{\mm} FL VIS \SI{0}{\degree} Coated, UV Plano-Convex Lens, Edmund Optics) is placed at the front focal plane of L1 and the DOE is placed at the back focal plane of L2. First, the DOE is removed and a green object is placed \SI{1.3}{\meter} away from the imaging lens, an image sensor (DFM 72BUC02-ML, The Imaging Source) is placed behind the back focal plane of L2. The sensor distance is adjusted until a sharp image of the green object is formed. The distance between the back focal plane and the sensor plane is found to be ~\SI{36.9}{\mm}, as consistent with the design. Keeping the sensor-to-DOE distance fixed, different objects are placed within the depth range $z \in [\SI{0.5}{\meter},\SI{1.3}{\meter}]$ from the imaging lens, corresponding to the defocus range of $\Psi \in [0,\Psi_{max}]$, and their corresponding images are captured on the image sensor. All objects are illuminated with a white LED flash light. Two sets of images are taken for each case, one with the DOE and a second without the DOE. Fig.~\ref{fig:Experiment3D} illustrates the comparative results, where the green bus is at around \SI{0.5}{\meter} and yellow car is at around \SI{1.3}{\meter}. As it is illustrated through the zoomed-in images in the figure, the proposed method is capable of reconstructing the details within the defocused regions of the refractive lens, without significantly sacrificing from the quality at the focused region. In addition, our end-to-end approach seems to address the spherical aberrations in the refractive lens as well. However, we still observe a few artifacts in the network output. The ringing artifacts due to the deconvolution are slightly visible at the edges of the textures. Even though the network addresses most of the halo artifacts observed in the sensor image, such artifacts are still visible at the edges between the objects and the background in the network output. Please also note that the standard deviation of the DOE fabrication error is found to be \SI{49.5}{\nm}, which also causes degradation in the image quality, as discussed in Sec.~\ref{sec:Simulations}. Some of the above-mentioned artifacts can be addressed by designing a network more robust to the DOE fabrication noise. Nevertheless, the experimental results demonstrate the proposed end-to-end system is a promising solution for the EDoF imaging.

\section{Conclusion}
\label{sec:Conclusion}

In the presented work, we propose an EDoF imaging system co-optimizing the optics and the signal post-processing. In particular, by designing a differentiable imaging model and incorporating it into a CNN, we design an end-to-end optimization framework. The optical model consists of a refractive lens and a DOE-based phase modulating element, of which the phase profile is an optimization parameter. The hybrid refractive-diffractive imaging system has two main advantages compared to the existing approaches. First, the main refractive lens restricts the depth-dependent PSF sizes that helps us to define the optimal search space with an analytical expression. Second, the color aberration is minimized due to the color dispersions of refractive and diffractive elements being opposite to each other. The reconstruction results through rigorous simulations as well as real-world experiments demonstrate that the proposed algorithm improves the state of the art in broadband EDoF imaging, achieving good results in wide EDoF range.

As future work, we would like to explore the potential of the co-design of optics and signal post-(pre-)processing scheme in different camera (display) applications. This is expected to lead to optimal task-specific computational camera and display designs, such as hyperspectral camera or light field display. 


%

\appendix[PSF Modelling as a Network Layer]


\section*{Forward}

Let us consider a point source located at distance $z$ from the camera plane on the optical axis of the lens. Under monochromatic illumination with the wavelength $\lambda$, the wave field in front of the camera plane due to such point is approximated as \cite{Goodman} 
\begin{equation}
U_{0_{\lambda,z}}^-(s,t) = \frac{\exp(-jkz)}{-j\lambda z}\exp\bigg[\frac{jk}{2z}(s^2+t^2) \bigg].
\label{eq:U0minus}
\end{equation}
At the camera plane, the incoming wave is modulated by the transmittance function of the lens and the DOE. The field right after the camera plane is 
\begin{equation}
U_{0_{\lambda,z}}^+(s,t) = U_0^-(s,t)A(s,t)\exp(j\Phi_\lambda(s,t))\exp(j\Phi^l_\lambda(s,t)),  
\label{eq:U0plus}
\end{equation}
with $A(s,t)$ being the lens amplitude, $\Phi_\lambda(s,t)$ and $\Phi^l_\lambda(s,t)$ being the phase modulation by the mask and the lens, respectively. The modified wave field $U_{0_{\lambda,z}}^+(s,t)$ then propagates the distance $z_i$ to the sensor plane, from where the intensity is recorded. The wave propagation can be expressed using the Fresnel diffraction model as \cite{Goodman}
\begin{equation}
\begin{split}
U_{z_i}(x,y) = & \frac{\exp(jkz_i)}{j\lambda z_i}\exp\bigg[\frac{jk}{2z_i}(x^2+y^2) \bigg] \\
& \times \mathcal{F} \bigg\{U_{0_{\lambda,z}}^+(s,t) \exp\bigg[\frac{jk}{2z_i}(s^2+t^2) \bigg] \bigg\} \biggl |_{\bigl(\frac{x}{\lambda z_i},\frac{y}{\lambda z_i}\bigr)},
\end{split}
\label{eq:Uzi}
\end{equation}
where $\mathcal{F}\{.\}$ is the Fourier transform operator. Finally, the incoherent PSF is $h_{\lambda,z}(x,y) = |U_{z_i}(x,y)|^2$. 

A plano-convex spherical refractive lens with the central thickness $d_0$ and radius of curvature $R$ has a thickness function, $d^l(s,t)$, in the form of 
\begin{equation}
    d^l(s,t) = d_0 - (R - \sqrt{R^2 - (s^2+t^2)}).
    \label{eq:spherical}
\end{equation}
In the thin lens model, where we assume $s^2+t^2 \ll R^2$, the thickness function is approximated using $\sqrt{R^2 - (s^2+t^2)} \approx R - (s^2+t^2)/2R$ as
\begin{equation}
    d^l(s,t) \approx d_0 - (s^2+t^2)/2R.
    \label{eq:paraboliclens}
\end{equation}
Following Eq.~\ref{eq:dtophi}~ and~\ref{eq:paraboliclens}, we derive the phase modification through the lens, $\Phi^l_\lambda(s,t)$ as
\begin{equation}
    \Phi^l_\lambda(s,t) =  -k (n^l_\lambda - 1) (R - \sqrt{R^2 - (s^2+t^2)}) \approx - \frac{k}{2f_\lambda}(s^2+t^2),
    \label{lensphase}
\end{equation}
where $f_\lambda = R/(n^l_\lambda-1)$. The constant term $d_0$ is omitted as its effect is negligible. Using the thin lens model and Eq.~\ref{eq:U0minus}~-~\ref{eq:Uzi}, we derive the incoherent PSF expression of Eq.~\ref{eq:PSF}.

\section*{Backward}

In this section, we derive the partial derivatives of the error with respect to the individual elements of the phase mask. Let us first define the DOE over a discrete finite 2D grid of size $M \times N$, and denote the discrete samples of the phase delay as $\Phi_{m,n} = \Phi_\lambda[m,n]$, $\forall m \in [0,M-1], \forall n \in [0,N-1]$. Similarly, the samples of the discrete defocused pupil function is $Q_{m,n} = Q_{\lambda,z}[m,n]$. Let us denote the indices of the discrete PSF with $k \in [0,M-1], l \in [0,N-1]$, that is, $h_{k,l} = h_{\lambda,z}[k,l]$. In the backward pass, the partial derivative of the error with respect to the PSF, $\partial \mathcal{L}/\partial h$, and the defocused pupil function, $Q$, are assumed to be known. Using the chain rule
\begin{equation}
   \frac{\partial \mathcal{L}}{\partial \Phi_{m,n}} = \sum_{k,l} \frac{\partial \mathcal{L}}{\partial h_{k,l}}\frac{\partial h_{k,l}}{\partial \Phi_{m,n}}.
\label{eq:partialEphi}
\end{equation}
Based on Eq.~\ref{eq:PSF}, the discrete PSF can be expressed as 
\begin{equation}
    h_{k,l} = \hat{Q}_{k,l}\hat{Q}^\ast_{k,l},
\label{eq:discretePSF}
\end{equation}
where $\hat{Q}$ is the discrete Fourier transform of $Q$,
\begin{equation}
    \hat{Q}_{k,l} = \sum_{m,n} Q_{m,n} e^{-j2\pi(\frac{k}{M}m+\frac{l}{N}n)},    
\label{eq:DFQ}
\end{equation}
and $\ast$ denotes complex conjugate. Using the product rule on Eq.~\ref{eq:discretePSF}
\begin{equation}
    \frac{\partial h_{k,l}}{\partial\Phi_{m,n}} = \frac{\partial \hat{Q}_{k,l}}{\partial\Phi_{m,n}}\hat{Q}^\ast_{k,l}  + \frac{\partial \hat{Q}^\ast_{k,l}}{\partial\Phi_{m,n}}\hat{Q}_{k,l}.
    \label{eq:productrule}
\end{equation}
Please note that the first and the second terms in the right hand side of Eq.~\ref{eq:productrule} are conjugate variables. Since the sum of any complex conjugate pairs is $z+z^\ast = 2\operatorname{Re}(z)$, deriving the first term is sufficient. The partial derivative $\partial \hat{Q}_{k,l} / \partial\Phi_{m,n}$ can be calculated using the chain rule as
\begin{equation}
    \frac{\partial \hat{Q}_{k,l}}{\partial\Phi_{m,n}} = \sum_{p,q} \frac{\partial \hat{Q}_{k,l}}{\partial Q_{p,q}} \frac{\partial Q_{p,q}}{\partial\Phi_{m,n}}.
\label{eq:partialFQphi}
\end{equation}
Considering Eq.~\ref{eq:Q} in discrete form, one can notice that the relation between $Q$ and $\Phi$ is element-wise, making the Jacobian matrix diagonal. That is, $\partial Q_{p,q} / \partial\Phi_{m,n} = jQ_{p,q}$ for $p = m, q = n$, and 0 elsewhere. Replacing the derivative of Eq.~\ref{eq:DFQ} into Eq.~\ref{eq:partialFQphi} then, we conclude that
\begin{equation}
    \frac{\partial \hat{Q}_{k,l}}{\partial\Phi_{m,n}} = jQ_{m,n} e^{-j2\pi(\frac{k}{M}m+\frac{l}{N}n)}.
\label{eq:partialFQphi2}
\end{equation}
Placing Eq.~\ref{eq:productrule} and \ref{eq:partialFQphi2} into Eq.~\ref{eq:partialEphi},
\begin{equation}
    \frac{\partial \mathcal{L}}{\partial \Phi_{m,n}} = 2 \operatorname{Re}\biggl( \sum_{k,l} \frac{\partial \mathcal{L}}{\partial h_{k,l}} jQ_{m,n} e^{-j2\pi(\frac{k}{M}m+\frac{l}{N}n)} \hat{Q}^\ast_{k,l} \biggr).
    \label{eq:partialEphi2}
\end{equation}
Please note that we use the fact that $\partial \mathcal{L}/\partial h_{k,l}$ is real. It is straightforward to notice that the conjugate of Eq.~\ref{eq:partialEphi2} corresponds to the inverse discrete Fourier transform. We obtain in the matrix form that
\begin{equation}
    \frac{\partial \mathcal{L}}{\partial \mathbf{\Phi}} = 2MN\operatorname{Im}\biggl(F^{-1}\biggl( \frac{\partial \mathcal{L}}{\partial \mathbf{h}} \circ \mathbf{\hat{Q}} \biggr) \circ \mathbf{Q}^\ast \biggr),
    \label{eq:partialEphi3}
\end{equation}
where $F^{-1}(.)$ is the inverse discrete Fourier transform and $\circ$ represents the Hadamard (element-wise) multiplication.

\section*{Acknowledgment}
The authors would like to thank Shay Elmalem for sharing their implementation. Ugur Akpinar is supported by the graduate school funding of Tampere University.
\ifCLASSOPTIONcaptionsoff
  \newpage
\fi



\bibliographystyle{IEEEtran}
%

\bibliography{refs.bib}

\begin{thebibliography}{10}
\providecommand{\url}[1]{#1}
\csname url@samestyle\endcsname
\providecommand{\newblock}{\relax}
\providecommand{\bibinfo}[2]{#2}
\providecommand{\BIBentrySTDinterwordspacing}{\spaceskip=0pt\relax}
\providecommand{\BIBentryALTinterwordstretchfactor}{4}
\providecommand{\BIBentryALTinterwordspacing}{\spaceskip=\fontdimen2\font plus
\BIBentryALTinterwordstretchfactor\fontdimen3\font minus
  \fontdimen4\font\relax}
\providecommand{\BIBforeignlanguage}[2]{{%
\expandafter\ifx\csname l@#1\endcsname\relax
\typeout{** WARNING: IEEEtran.bst: No hyphenation pattern has been}%
\typeout{** loaded for the language `#1'. Using the pattern for}%
\typeout{** the default language instead.}%
\else
\language=\csname l@#1\endcsname
\fi
#2}}
\providecommand{\BIBdecl}{\relax}
\BIBdecl

\bibitem{EDOFMicroscopy}
S.~Liu and H.~Hua, ``Extended depth-of-field microscopic imaging with a
  variable focus microscope objective,'' \emph{Optics express}, vol.~19, no.~1,
  pp. 353--362, 2011.

\bibitem{DeblurringEM}
M.~A. Figueiredo and R.~D. Nowak, ``An em algorithm for wavelet-based image
  restoration,'' \emph{IEEE Transactions on Image Processing}, vol.~12, no.~8,
  pp. 906--916, 2003.

\bibitem{DeblurringVariational}
A.~C. Likas and N.~P. Galatsanos, ``A variational approach for bayesian blind
  image deconvolution,'' \emph{IEEE transactions on signal processing},
  vol.~52, no.~8, pp. 2222--2233, 2004.

\bibitem{BlindDefDeblur}
M.~S. Almeida and L.~B. Almeida, ``Blind and semi-blind deblurring of natural
  images,'' \emph{IEEE Transactions on Image Processing}, vol.~19, no.~1, pp.
  36--52, 2009.

\bibitem{DeblurringSingleImage}
\BIBentryALTinterwordspacing
X.~Zhang, R.~Wang, X.~Jiang, W.~Wang, and W.~Gao, ``{Spatially variant defocus
  blur map estimation and deblurring from a single image},'' \emph{Journal of
  Visual Communication and Image Representation}, vol.~35, pp. 257--264, 2016.
  [Online]. Available: \url{http://dx.doi.org/10.1016/j.jvcir.2016.01.002}
\BIBentrySTDinterwordspacing

\bibitem{GraphBlindDec}
Y.~Bai, G.~Cheung, X.~Liu, and W.~Gao, ``Graph-based blind image deblurring
  from a single photograph,'' \emph{IEEE Transactions on Image Processing},
  vol.~28, no.~3, pp. 1404--1418, 2018.

\bibitem{LevinCA}
A.~Levin, R.~Fergus, F.~Durand, and W.~T. Freeman, ``Image and depth from a
  conventional camera with a coded aperture,'' \emph{ACM transactions on
  graphics (TOG)}, vol.~26, no.~3, p.~70, 2007.

\bibitem{ZhouCA}
C.~Zhou and S.~Nayar, ``What are good apertures for defocus deblurring?'' in
  \emph{2009 IEEE international conference on computational photography
  (ICCP)}.\hskip 1em plus 0.5em minus 0.4em\relax IEEE, 2009, pp. 1--8.

\bibitem{ZhouCAPairs}
C.~Zhou, S.~Lin, and S.~K. Nayar, ``{Coded aperture pairs for depth from
  defocus and defocus deblurring},'' \emph{International Journal of Computer
  Vision}, vol.~93, no.~1, pp. 53--72, 2011.

\bibitem{AnnularCA}
W.~T. Welford, ``{Use of Annular Apertures to Increase Focal Depth},''
  \emph{Journal of the Optical Society of America}, vol.~50, no.~8, p. 749, aug
  1960.

\bibitem{DappledCA}
A.~Veeraraghavan, R.~Raskar, A.~Agrawal, A.~Mohan, and J.~Tumblin, ``Dappled
  photography: Mask enhanced cameras for heterodyned light fields and coded
  aperture refocusing,'' in \emph{ACM transactions on graphics (TOG)}, vol.~26,
  no.~3.\hskip 1em plus 0.5em minus 0.4em\relax ACM, 2007, p.~69.

\bibitem{MasiaCA}
B.~Masia, L.~Presa, A.~Corrales, and D.~Gutierrez, ``Perceptually optimized
  coded apertures for defocus deblurring,'' in \emph{Computer Graphics Forum},
  vol.~31, no.~6.\hskip 1em plus 0.5em minus 0.4em\relax Wiley Online Library,
  2012, pp. 1867--1879.

\bibitem{Cubic}
E.~R. Dowski and W.~T. Cathey, ``Extended depth of field through wave-front
  coding,'' \emph{Applied optics}, vol.~34, no.~11, pp. 1859--1866, 1995.

\bibitem{tangent}
S.~Chen, Z.~Fan \emph{et~al.}, ``Optimized asymmetrical tangent phase mask to
  obtain defocus invariant modulation transfer function in incoherent imaging
  systems,'' \emph{Optics letters}, vol.~39, no.~7, pp. 2171--2174, 2014.

\bibitem{sinusoidal}
H.~Zhao and Y.~Li, ``Optimized sinusoidal phase mask to extend the depth of
  field of an incoherent imaging system,'' \emph{Optics letters}, vol.~35,
  no.~2, pp. 267--269, 2010.

\bibitem{logarithmic}
S.~S. Sherif, W.~T. Cathey, and E.~R. Dowski, ``Phase plate to extend the depth
  of field of incoherent hybrid imaging systems,'' \emph{Applied optics},
  vol.~43, no.~13, pp. 2709--2721, 2004.

\bibitem{Ref-Diff}
T.~Stone and N.~George, ``Hybrid diffractive-refractive lenses and achromats,''
  \emph{Applied Optics}, vol.~27, no.~14, pp. 2960--2971, 1988.

\bibitem{SpectralSweepEDoF}
O.~Cossairt and S.~Nayar, ``{Spectral Focal Sweep: Extended depth of field from
  chromatic aberrations},'' \emph{2010 IEEE International Conference on
  Computational Photography, ICCP 2010}, pp. 1--8, 2010.

\bibitem{EDoFMobilePhone}
H.-Y. Sung, S.~S. Yang, and H.~Chang, ``{Design of mobile phone lens with
  extended depth of field based on point-spread function focus invariance},''
  \emph{Novel Optical Systems Design and Optimization XI}, vol. 7061, p.
  706107, 2008.

\bibitem{DefPlenoptic}
R.~Ng, M.~Levoy, M.~Br{\'e}dif, G.~Duval, M.~Horowitz, P.~Hanrahan
  \emph{et~al.}, ``Light field photography with a hand-held plenoptic camera.''

\bibitem{FocusedPlenoptic}
A.~Lumsdaine and T.~Georgiev, ``The focused plenoptic camera,'' in \emph{2009
  IEEE International Conference on Computational Photography (ICCP)}.\hskip 1em
  plus 0.5em minus 0.4em\relax IEEE, 2009, pp. 1--8.

\bibitem{EDOFPlenoptic}
C.~Perwass and L.~Wietzke, ``Single lens 3d-camera with extended
  depth-of-field,'' in \emph{Human Vision and Electronic Imaging XVII}, vol.
  8291.\hskip 1em plus 0.5em minus 0.4em\relax International Society for Optics
  and Photonics, 2012, p. 829108.

\bibitem{Latticefocal}
A.~Levin, S.~W. Hasinoff, P.~Green, F.~Durand, and W.~T. Freeman, ``4d
  frequency analysis of computational cameras for depth of field extension,''
  in \emph{ACM Transactions on Graphics (TOG)}, vol.~28, no.~3.\hskip 1em plus
  0.5em minus 0.4em\relax ACM, 2009, p.~97.

\bibitem{LFDiffuser}
O.~Cossairt, C.~Zhou, and S.~Nayar, ``Diffusion coded photography for extended
  depth of field,'' \emph{ACM Transactions on Graphics (TOG)}, vol.~29, no.~4,
  p.~31, 2010.

\bibitem{FourierSlice}
R.~Ng, ``Fourier slice photography,'' in \emph{ACM transactions on graphics
  (TOG)}, vol.~24, no.~3.\hskip 1em plus 0.5em minus 0.4em\relax ACM, 2005, pp.
  735--744.

\bibitem{EDOFAkpinar}
U.~{Akpinar}, E.~{Sahin}, and A.~{Gotchev}, ``Learning optimal phase-coded
  aperture for depth of field extension,'' in \emph{2019 IEEE International
  Conference on Image Processing (ICIP)}, Sep. 2019, pp. 4315--4319.

\bibitem{EDOFSitzmann}
V.~Sitzmann, S.~Diamond, Y.~Peng, X.~Dun, S.~Boyd, W.~Heidrich, F.~Heide, and
  G.~Wetzstein, ``{End-to-end optimization of optics and image processing for
  achromatic extended depth of field and super-resolution imaging},'' \emph{ACM
  Transactions on Graphics}, vol.~37, no.~4, 2018.

\bibitem{EDOFElmalem}
S.~Elmalem, R.~Giryes, and E.~Marom, ``{Learned phase coded aperture for the
  benefit of depth of field extension},'' \emph{Optics Express}, vol.~26,
  no.~12, p. 15316, 2018.

\bibitem{DepthEstPhase}
H.~Haim, S.~Elmalem, R.~Giryes, A.~M. Bronstein, and E.~Marom, ``{Depth
  Estimation From a Single Image Using Deep Learned Phase Coded Mask},''
  \emph{IEEE Transactions on Computational Imaging}, vol.~4, no.~3, pp.
  298--310, 2018.

\bibitem{CADepth}
P.~A. Shedligeri, S.~Mohan, and K.~Mitra, ``{Data driven coded aperture design
  for depth recovery},'' \emph{Proceedings - International Conference on Image
  Processing, ICIP}, vol. 2017-Septe, pp. 56--60, 2018.

\bibitem{LFCALearning}
Y.~Inagaki, Y.~Kobayashi, K.~Takahashi, T.~Fujii, and H.~Nagahara, ``Learning
  to capture light fields through a coded aperture camera,'' in
  \emph{Proceedings of the European Conference on Computer Vision (ECCV)},
  2018, pp. 418--434.

\bibitem{HSCNNJoint}
L.~Wang, T.~Zhang, Y.~Fu, and H.~Huang, ``{HyperReconNet: Joint Coded Aperture
  Optimization and Image Reconstruction for Compressive Hyperspectral
  Imaging},'' \emph{IEEE Transactions on Image Processing}, vol.~28, no.~5, pp.
  2257--2270, 2019.

\bibitem{OpticalCNN}
\BIBentryALTinterwordspacing
J.~Chang, V.~Sitzmann, X.~Dun, W.~Heidrich, and G.~Wetzstein, ``{Hybrid
  optical-electronic convolutional neural networks with optimized diffractive
  optics for image classification},'' \emph{Scientific Reports}, vol.~8, no.~1,
  pp. 1--10, 2018. [Online]. Available:
  \url{http://dx.doi.org/10.1038/s41598-018-30619-y}
\BIBentrySTDinterwordspacing

\bibitem{Goodman}
J.~W. Goodman, \emph{Introduction to Fourier Optics}.\hskip 1em plus 0.5em
  minus 0.4em\relax Roberts and Company Publishers, 2005.

\bibitem{regularization}
H.~Son and S.~Lee, ``Fast non-blind deconvolution via regularized residual
  networks with long/short skip-connections,'' in \emph{2017 IEEE International
  Conference on Computational Photography (ICCP)}, May 2017, pp. 1--10.

\bibitem{DebCNN}
K.~Zhang, W.~Zuo, S.~Gu, and L.~Zhang, ``Learning deep {CNN} denoiser prior for
  image restoration,'' in \emph{IEEE Conference on Computer Vision and Pattern
  Recognition}, 2017, pp. 3929--3938.

\bibitem{DebCNN2}
J.~Jiao, W.-C. Tu, S.~He, and R.~W. Lau, ``Formresnet: formatted residual
  learning for image restoration,'' in \emph{Computer Vision and Pattern
  Recognition Workshops (CVPRW), 2017 IEEE Conference on}.\hskip 1em plus 0.5em
  minus 0.4em\relax IEEE, 2017, pp. 1034--1042.

\bibitem{L1loss}
H.~Zhao, O.~Gallo, I.~Frosio, and J.~Kautz, ``Loss functions for image
  restoration with neural networks,'' \emph{IEEE Transactions on Computational
  Imaging}, vol.~3, no.~1, pp. 47--57, 2017.

\bibitem{INRIADataset}
H.~Jegou, M.~Douze, and C.~Schmid, ``Hamming embedding and weak geometric
  consistency for large scale image search,'' in \emph{European conference on
  computer vision}.\hskip 1em plus 0.5em minus 0.4em\relax Springer, 2008, pp.
  304--317.

\bibitem{AdamOptimizer}
D.~P. Kingma and J.~Ba, ``Adam: A method for stochastic optimization,''
  \emph{arXiv preprint arXiv:1412.6980}, 2014.

\bibitem{BSDS500}
\BIBentryALTinterwordspacing
P.~Arbelaez, M.~Maire, C.~Fowlkes, and J.~Malik, ``Contour detection and
  hierarchical image segmentation,'' \emph{IEEE Transactions on Pattern
  Analysis Machine Intelligence}, vol.~33, no.~5, pp. 898--916, may 2011.
  [Online]. Available: \url{http://dx.doi.org/10.1109/TPAMI.2010.161}
\BIBentrySTDinterwordspacing

\bibitem{HSDataset}
Y.~Monno, S.~Kikuchi, M.~Tanaka, and M.~Okutomi, ``A practical one-shot
  multispectral imaging system using a single image sensor,'' \emph{IEEE
  Transactions on Image Processing}, vol.~24, no.~10, pp. 3048--3059, 2015.

\bibitem{KodakSensor}
\BIBentryALTinterwordspacing
Kodak. (2007) Kaf-10500 image sensor. [Online]. Available:
  \url{https://www.datasheets360.com/pdf/4613689109339751409}
\BIBentrySTDinterwordspacing

\bibitem{S1813}
\BIBentryALTinterwordspacing
Dow. (2014, march) Microposit s1800 g2 series photoresists for microlithography
  applications. [Online]. Available:
  \url{https://kayakuam.com//wp-content/uploads/2019/09/S1800-G2.pdf}
\BIBentrySTDinterwordspacing

\bibitem{Meem2020}
\BIBentryALTinterwordspacing
M.~Meem, S.~Banerji, C.~Pies, T.~Oberbiermann, A.~Majumder,
  B.~Sensale-Rodriguez, and R.~Menon, ``Large-area, high-numerical-aperture
  multi-level diffractive lens via inverse design,'' \emph{Optica}, vol.~7,
  no.~3, pp. 252--253, Mar 2020. [Online]. Available:
  \url{http://www.osapublishing.org/optica/abstract.cfm?URI=optica-7-3-252}
\BIBentrySTDinterwordspacing

\bibitem{meem2020multi}
M.~Meem, A.~Majumder, and R.~Menon, ``Multi-plane, multi-band image projection
  via broadband diffractive optics,'' \emph{Applied Optics}, vol.~59, no.~1,
  pp. 38--44, 2020.

\bibitem{Microchemicals}
\BIBentryALTinterwordspacing
Microchemicals. Az developer. [Online]. Available:
  \url{https://www.microchemicals.com/downloads.html}
\BIBentrySTDinterwordspacing

\end{thebibliography}


%








\end{document}